\documentclass[fleqn,usenatbib,useAMS]{mnras}
\pdfoutput=1 

\usepackage{amssymb}
\usepackage{amstext}
\usepackage{amsmath}
\usepackage{epsf}
\usepackage{graphicx}
\usepackage{graphics}
\usepackage{epsfig}
\usepackage{wrapfig}
\usepackage{listings}
\usepackage{float}
\usepackage{esint}
\usepackage{natbib}
\usepackage{ifthen}
\usepackage{hyperref}

\usepackage{xcolor}
\usepackage[normalem]{ulem} 
\newcommand\hl{\bgroup\markoverwith
  {\textcolor{yellow}{\rule[-.5ex]{2pt}{2.5ex}}}\ULon}

\newcommand{\eg}{e.g.,~}
\newcommand{\ie}{i.e.,~}
\input{epsf}

\begin{document}

\title[Paper 1]{The features of the Cosmic Web unveiled by the flip-flop field}

\author[ Shandarin \& Medvedev]
	{Sergei F. Shandarin,  \thanks{E-mail: sergei@ku.edu}
	Mikhail V. Medvedev \\
	Department of Physics and Astronomy, University of Kansas, Lawrence, KS 66045, USA}

 \maketitle
\begin{abstract}
Understanding of the observed structure in the universe can be reached only in the theoretical framework 
of dark matter.
N-body simulations are indispensable for the analysis of the formation and evolution of the dark matter web.
Two primary fields - density and velocity fields - are used in most of studies.  However dark matter provides two additional 
fields which are unique for collisionless media only. These are the multi-stream field in Eulerian space and
flip-flop field in Lagrangian space. The flip-flop field  represents the number of   
sign reversals of an elementary volume of each collisionless fluid element. 
This field can be estimated  by counting the sign reversals of the Jacobian at each particle at every time step of the simulation.
The Jacobian is evaluated by numerical differentiation of  the Lagrangian submanifold, 
\ie the three-dimensional dark matter sheet in the six-dimensional space formed by three Lagrangian and three Eulerian coordinates.
We present the results of the statistical study of the  evolution of the flip-flop field from $z=50$ to the present time $z=0$.
A number of statistical characteristics  show  that the pattern of the flip-flop field remains remarkably stable from $z \approx 30$
to the present time.
As a result the flip-flop field evaluated at $z=0$  stores a wealth of information about the 
 dynamical history of the dark matter web. 
In particular one of the most intriguing properties of the flip-flop is a unique capability to preserve  the information about 
the merging history of dark matter haloes.

\end{abstract}


\begingroup
\let\clearpage\relax
\endgroup

\section{Introduction} \label{sec:intro}

Modern redshift surveys such as 2dF Galaxy Redshift Survey  \footnote{http://msowww.anu.edu.au/2dFGRS/ }  
and the Sloan Digital Sky Survey  \footnote{http://www.sdss.org/} 
as well as  others reveal the wealth  of structures in the spatial distribution of galaxies. 
A useful abstraction helping to comprehend the complexity of the structure has been provided by the skeleton of the web introduced 
by the adhesion approximation \citep{Gurbatov_etal:85,Gurbatov_etal:89, Gurbatov_etal:12, Hidding_etal:12a,Hidding_etal:12b}. 
Its geometrical version designs 
a tiling of  three-dimensional space by irregular three-dimensional tiles associated with voids. 
The faces of the tiles are
associated with walls/pancakes, the edges -- with filaments and the vertices -- with the haloes. 
The geometrical model obviously highly simplifies  the dark matter (hereafter DM) web  especially its interior structure since 
it does not trace the details in  the complex distribution of mass inside the walls, filaments
and haloes.  However the skeleton looks qualitatively quite realistic revealing the multiscale nature of the web
\citep{Kofman_etal:92}.
For instance, it indicated  the presence of substructures in voids for the first time. 

Historically,  haloes have attracted the most of attention in theoretical studies of the large-scale structure formation.  
From the observational point of view,
haloes are most closely related to galaxies, galaxy groups and clusters of galaxies, which provide the bulk of
information about the structures in the universe. However, direct modeling of galaxy formation
based on fundamental laws of physics is precluded by enormous complexity of the physical processes involved. 
In addition to the gravitational coupling with dark matter (hereafter DM) structure baryons participate in extremely complex
hydrodynamical and thermal processes  including star formation and the stellar wind feedback, shocks and supernovae explosions, 
gas accretion onto black holes in active galactic nuclei and the feedback via relativistic jets to name just some of them. 

Hence various semi-empirical models of galaxy formation have been suggested, see \eg \cite{Angulo_etal:13} and
references therein. In particular, it has been argued that galaxies
are formed in the host DM haloes of corresponding masses. The DM haloes themselves  are formed in a
chain of mergers of smaller DM haloes which may start from tiny 
haloes of planet masses   ~\citep{Diemand_etal:05}. When two or more haloes merge 
their remnants may survive for a long time as subhaloes and/or streams within the resultant halo. 
Therefore, DM haloes are likely to have a nesting structure
where each  subhalo may include a number of  even smaller subhaloes down to the smallest haloes allowed 
by the initial power spectrum \citep{Diemand_etal:05,Ghigna_etal:98}. 

Dark matter structure results from the gravitational  growth of the initial  Gaussian perturbations of density. 
It is shaped by the  nonlinear collisionless gravitational dynamics.  Being very complex it is  still considerably
simpler than baryonic physics therefore it is more feasible to build the model of the DM web based on fundamental physical laws
with a minimal number of  heuristic assumptions. Cosmological N-body simulations play indispensable role in the studies of the DM web
providing the most reliable data on the evolution of the web in highly nonlinear regime. However identifying the basic elements
of the web (i.e. haloes, filaments, walls, and voids) represents a difficult problem even in pure DM N-body 
simulations where all dynamical information is readily available, see \eg \citet{Colberg_etal:08,Knebe_etal_structures:13,Cautun_etal:14}.

In early cosmological $N$-body simulations the haloes were loosely defined as compact concentrations of the simulation
particles in configuration space. A particularly popular simple technique used for this purpose and called 
`friends of friends' (FOF) algorithm was adopted from percolation analysis \citep{Zeldovich_etal:82,Shandarin:83, Davis_etal:85}.
According to this method one firstly finds all `friends' of each particle by linking every particle in the simulation 
with all neighbors  separated by less than a chosen distance -- the linking length. 
Then applying the criterion: a friend of my friend is my friend,  one can identify all groups of particles consisting of  friends. 
Choosing a particular value of the linking length  (often $\sim20\%$ of the mean particle separation, \cite{Davis_etal:85})
one can select a particular set of groups and call them haloes. 
A number of improved versions of FOF have been developed:  
\citet{Couchman_Carlberg:92,Suginohara_Suto:92,vanKampen:95,Summers:95,Klypin_etal:99,Okamoto_Habe:99}
just to mention a few.
Other more sophisticated methods that  identify
both haloes and subhaloes have been  suggested as well, for review  see   \citet{Knebe_etal_structures:13}
and references therein. 
Some of them rely only on the  particle positions, other also  use the phase space information.
 The methods using only the configuration space information regardless of their sophistication
 may suffer from projection effect that causes dynamically distinct structures in phase space temporally to overlap 
 in configuration space.  
 \subsection{One-dimensional example}
\begin{figure}  
\includegraphics[scale=0.33]{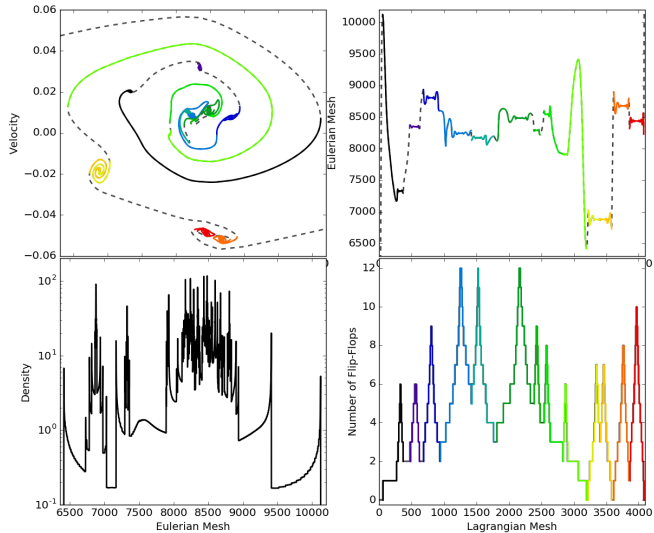}       
\caption{Illustration of a one-dimensional halo simulated from random but smooth initial
condition. Top left panel: the phase space. Individual subhaloes are shown by different colors.
Bottom left panel: density distribution in Eulerian space. Bottom right panel: the flip-flop field in Lagrangian
space. Top right panel: Lagrangian submanifold $x = x(q)$. Colors in the panels on the right correspond to
the colors in the top left panel. All panels correspond to the same instant of time.}
\label{fig:1d_example}
\end{figure}

 Let us consider a one dimensional example shown in Fig. \ref{fig:1d_example}. 
 It shows the phase space of a halo simulated in a one-dimensional
universe from some random but smooth initial condition. 
One can see a complicated substructure consisting of a number of subhaloes and streams
shown in phase space  by different colors. 
Using  only the coordinates of particles means that the complicated
phase space curve must be projected on the horizontal.
It is obvious from the figure that identifying individual subhaloes would be impossible
even in a simple one-dimensional model.
 For instance the pair
 of subhaloes in red at the bottom of the figure is currently projected on the center of the halo and thus cannot be identified as separate subhaloes
 using particle coordinates only.
 But at a later time it will take place of  the yellow subhalo on the left and thus may be identified as a separate pair of subhaloes.
 It becomes even more challenging in three-dimensional case, see \eg  \citet{Knebe_etal_structures:13, 
Hoffmann_etal:14} and reference therein.

From Fig. \ref{fig:1d_example} one can also see that a halo as well as subhaloes can be naturally defined as the regions in Eulerian space 
where the number of streams is greater than one \citep{Shandarin_etal:12, Ramachandra_Shandarin:15,Ramachandra_Shandarin:16}. 
However, this approach also is not free of the contamination effects due to projection effects.
 Using all dynamical information available in phase space  helps in solving this problem, however
 this is complicated by the fact that phase space is not a metric space, see e.g.  \cite{Ascasibar_Binney:05}.  
 Evaluating  distances in phase space requires additional parameter with the dimensions of time. 
 Unfortunately the time parameter  is not universal for the whole halo. 
 For illustration, consider again a simple example shown in Fig. \ref{fig:1d_example}. 
 The spiraling time of two red subhaloes  shown on the bottom of the figure is mostly determined by the density due to 
 the subhaloes themselves $\tau_{\rm sh}  \propto \rho_{\rm sh}^{-1/2}$ 
 rather than by the total density dominated by the central part of the halo.
 This is because they spend a relatively short time in the spatial vicinity of the center. 
 They live outside the central region of the halo the most of  time since they move with lower speed there.
 In the outskirts of the main halo their dynamical time is determined primarily by their own densities.
 Thus finding the relevant local time  requires  identification of a subhalo in phase space which in turn requires the knowledge 
 of the characteristic time of the same subhalo for making the corresponding patch of phase space metric.
 
One way overcoming this circular reasoning problem was suggest by \cite{Ascasibar_Binney:05}.
It might be also possible to develop  some iterative technique  but we have tried  a new completely different approach.

In order to outline the  main idea of this new method we begin with  a closer examination of the one-dimensional 
example  introduced above.
Let us follow along the phase space curve in the top left panel of Fig. \ref{fig:1d_example} starting from the top point 
of the spiral on the left boundary of the box through the bottom point on the right boundary of the box. 
Along this path, the initial (Lagrangian) coordinates $q_{\rm i}$ of the particles, which are in essence their IDs 
and thus are immutable, increase 
monotonically while their final (Eulerian) coordinates $x_{\rm i}$ are not monotonic.
This  is also seen in the top right panel of Fig. \ref{fig:1d_example} showing the Lagrangian submanifold of the halo which is 
the curve $x = x(q;t)$.  In other words there are fluid elements with $x_{\rm i+1} < x_{\rm i}$ while $q_{\rm i+1} > q_{\rm i}$. 
Later on the Eulerian coordinates of these particles will swap again when they pass the caustic on the left hand side
of the top left panel of Fig. \ref{fig:1d_example}.
We will dub every swap of the Eulerian coordinates of the two neighboring (in Lagrangian space) particles on the curve as a flip-flop. 
The analog of this phenomenon in a multi-dimensional space is a formal change of the sign
of the volume of a fluid element when it turns inside out.  The  volume of a fluid element is a continuous function
of time, therefore between the states of the volume with different signs it must be zero. 
At this moment in three-dimensional space it collapses into a piece of a two-dimensional surface called a caustic and its density
 becomes infinite.
The caustic concept is of cause a mathematical abstraction useful only when the discreteness effects are negligible.

The total number of flip-flops experienced by every fluid element by the time corresponding to the top left panel
 is shown in the bottom right  panel of Fig. \ref{fig:1d_example}. 
We will refer to it as a flip-flop field.
Colors show individual peaks of the flip-flop field in Lagrangian coordinates. 
The correspondence of the flip-flop peaks in Lagrangian space to  individual subhaloes
in the phase space shown in  Fig. \ref{fig:1d_example} is remarkable. 
Note that the tidal streams and haloes are also easily, unambiguously and robustly identified 
via the flip-flop field,  cf. the bottom right  and top left panels of Fig. \ref{fig:1d_example}.
Using the flip-flop field in Lagrangian space is 
the key idea of a new approach to the analysis  of DM haloes.

As far as we know the flip-flop phenomenon was firstly mentioned in cosmological context by \citet{Zeldovich_aap:70}.
Later on it was used by \citet{Arnold_etal:82} in the study of generic caustics 
and also discussed in the review by \citet{Shandarin_Zeldovich:89}.
Recently the flip-flop phenomenon was used in the studies of a few specific problems \cite{Shandarin_Medvedev:14}. 
Here we briefly describe major differences between these studies and the  approach discussed in this paper. 
\citet{Vogelsberger_White:11} invoked it 
for the  study of 'the properties of fine-grained phase-space streams and their  associated caustics'
by 'integrating the geodesic deviation equation' in cosmological N-body simulations.
Although the counts of caustics was used for computing some one-point statistics  the concept
of a field in Lagrangian space was not even mentioned. 
Another  method called "Origami"  was suggested in ~\cite{Neyrinck:12} and ~\cite{Neyrinck_etal:13}
It was used for exploring the connectivity of streams in Lagrangian space. The authors   used the results
for approximating the  boundaries of haloes, filaments and walls. In this approach a binary field
with two values +1 and -1 corresponding to positive and negative parities of local Lagrangian volumes
was introduced. The values of our flip-flop field can be any non-negative integer and
its peaks play the key role in our study.  
 
From dynamical point of view a subhalo can be described as a set of particles
participating in oscillatory motions about some center which in turn orbits about the center of the halo or about the center of a larger subhalo.
The both amplitudes of the oscillations of a subhalo in configuration and velocity spaces  are significantly smaller than
the corresponding  sizes of the halo. The characteristic times of the oscillations are also significantly shorter than 
the corresponding time of the orbiting of the subhalo around the center of the halo. 
In a simple one-dimensional halo without subhaloes the number of  flip-flops becomes a counter of  full orbits: two flip-flops  per a full orbit.
It is worth mentioning that in  such a halo the periods of  orbits becomes shorter as the particle gets closer to the center because the mean density
within smaller orbits is higher  than that within the larger orbits since the period is proportional to $<\rho>^{-1/2}$. 
The major goal of this paper is to investigate the properties of the flip-flop field an three-dimensional 
N-body simulation and explore its potential usefulness for identifying the DM web, \ie haloes, filaments, walls/pancakes, voids 
and their substructures.
\subsection{Flip-flop field in two dimensions}
Now  we consider a two-dimensional example which may help to  bridge the visualization gap between 1D in 3D.
\begin{figure}
	\centering
	\centerline{\includegraphics[scale=0.19]{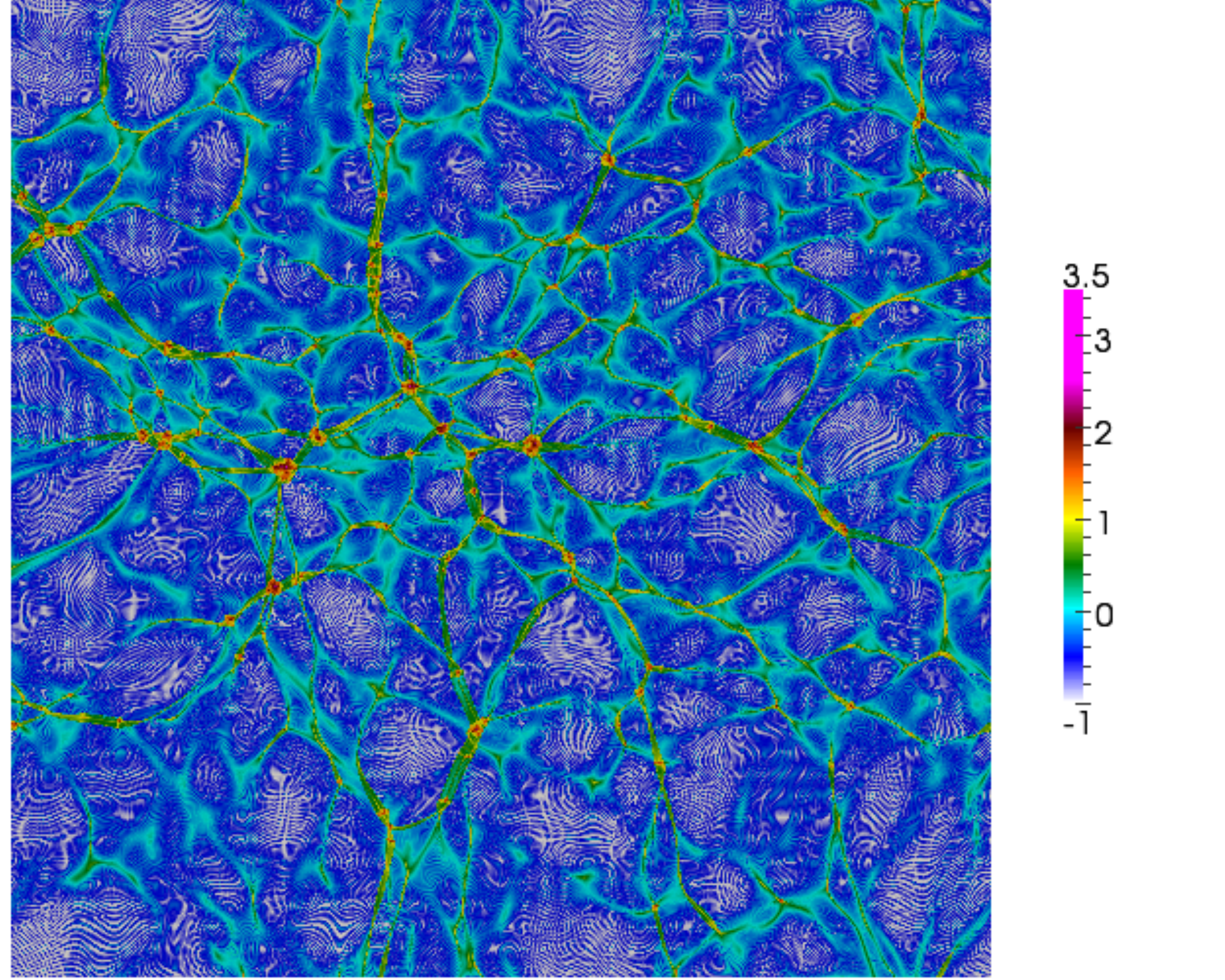}
	                     \includegraphics[scale=0.19]{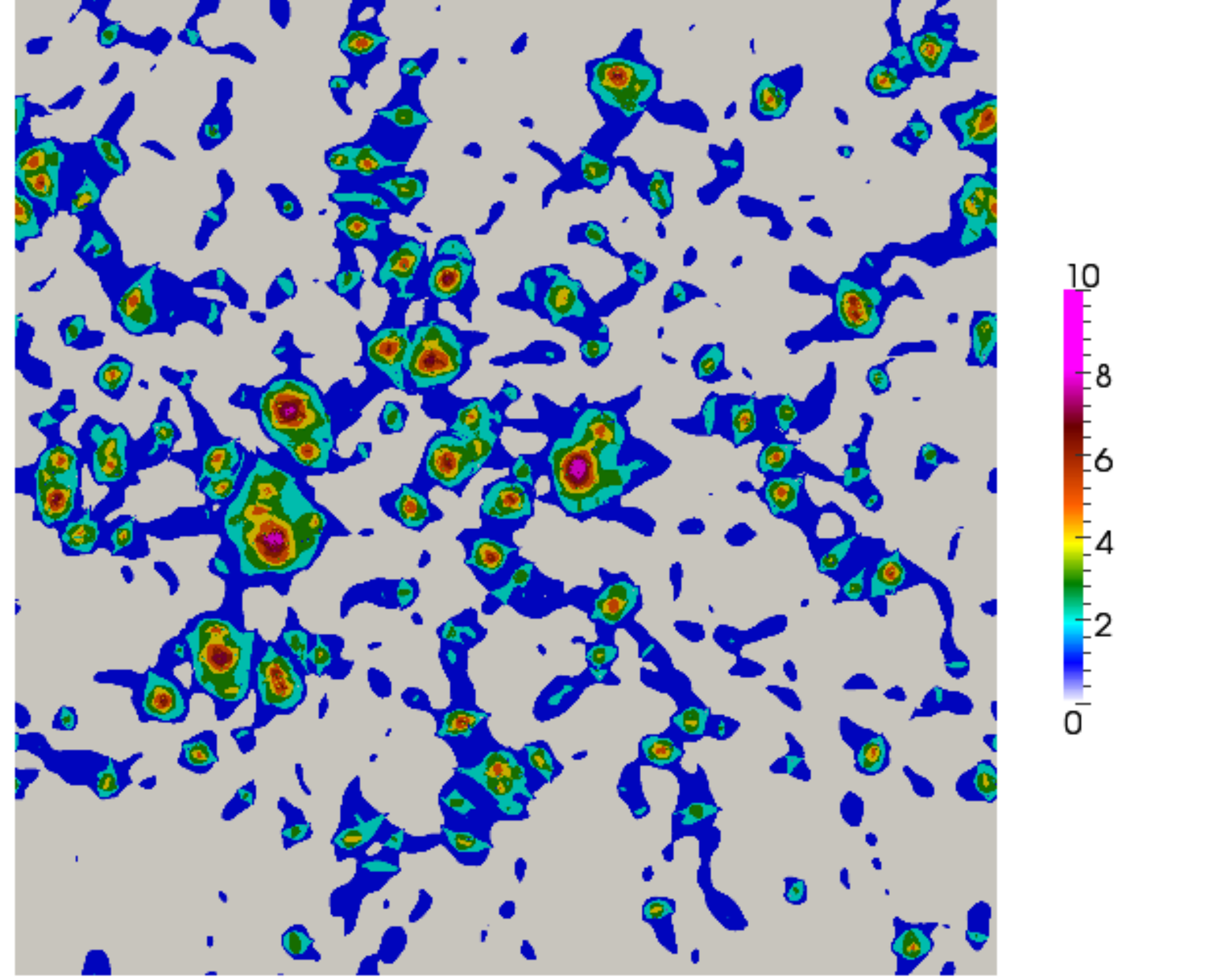}}
	\centerline{\includegraphics[scale=0.19]{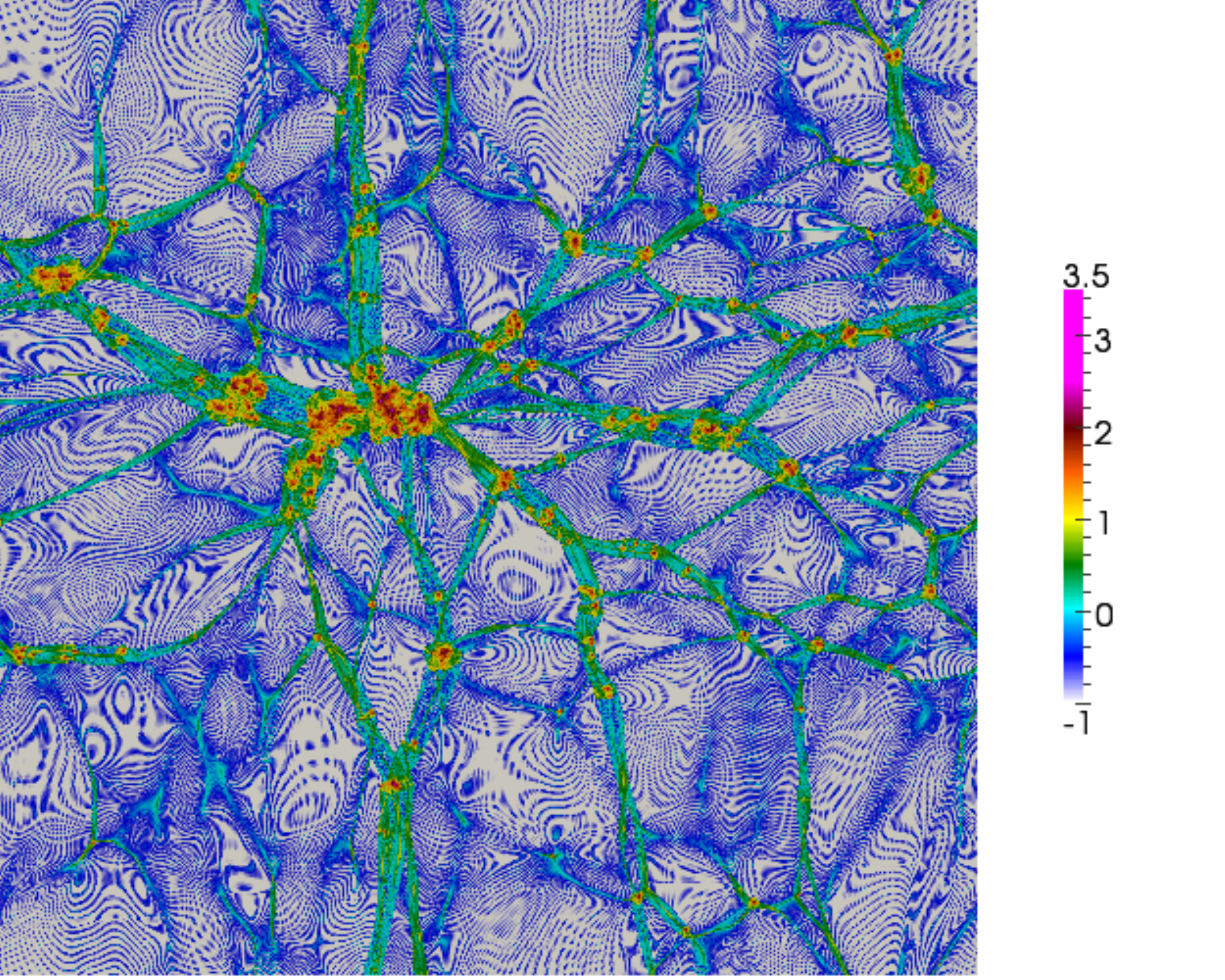}
	                     \includegraphics[scale=0.19]{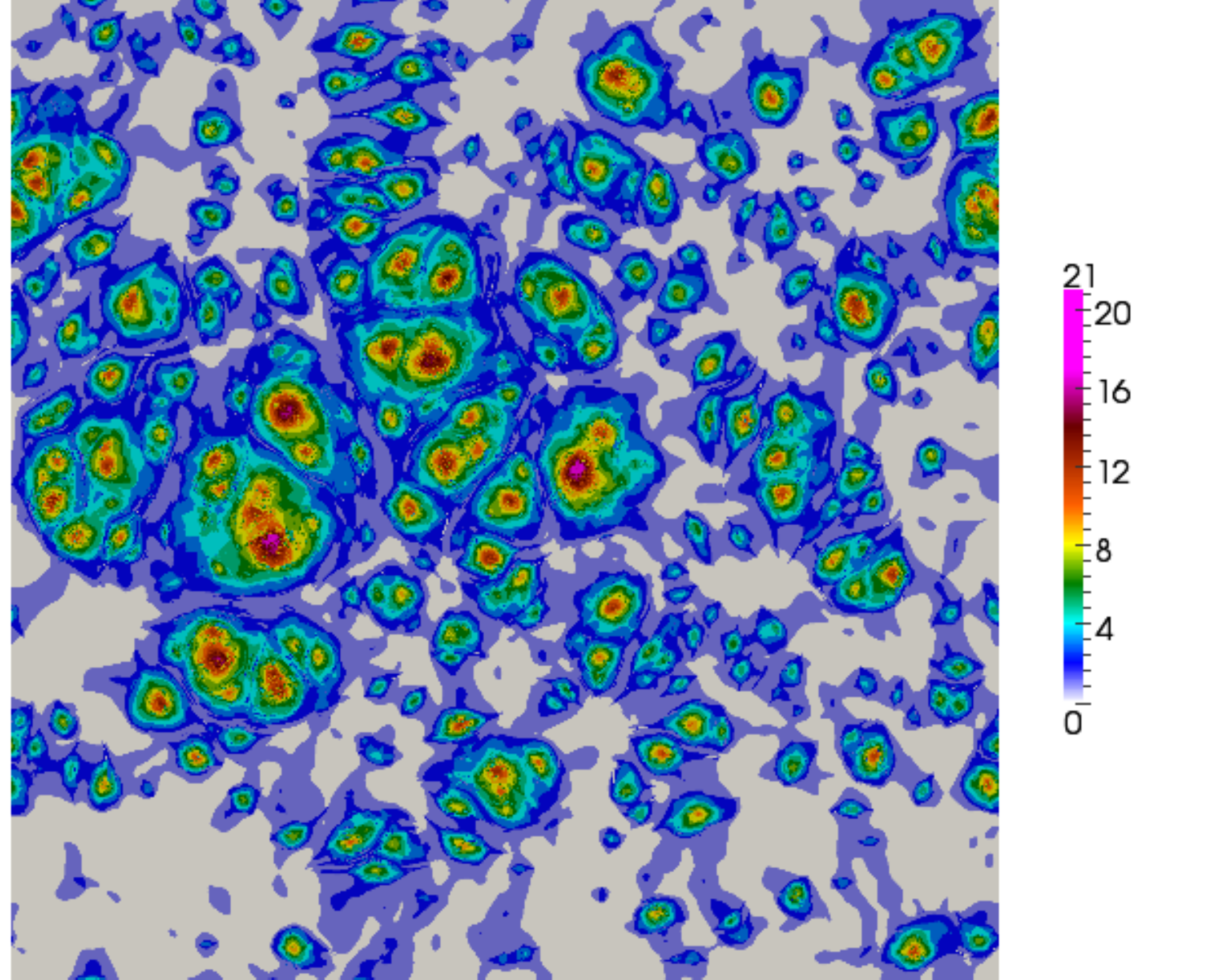}}
	\centerline{\includegraphics[scale=0.19]{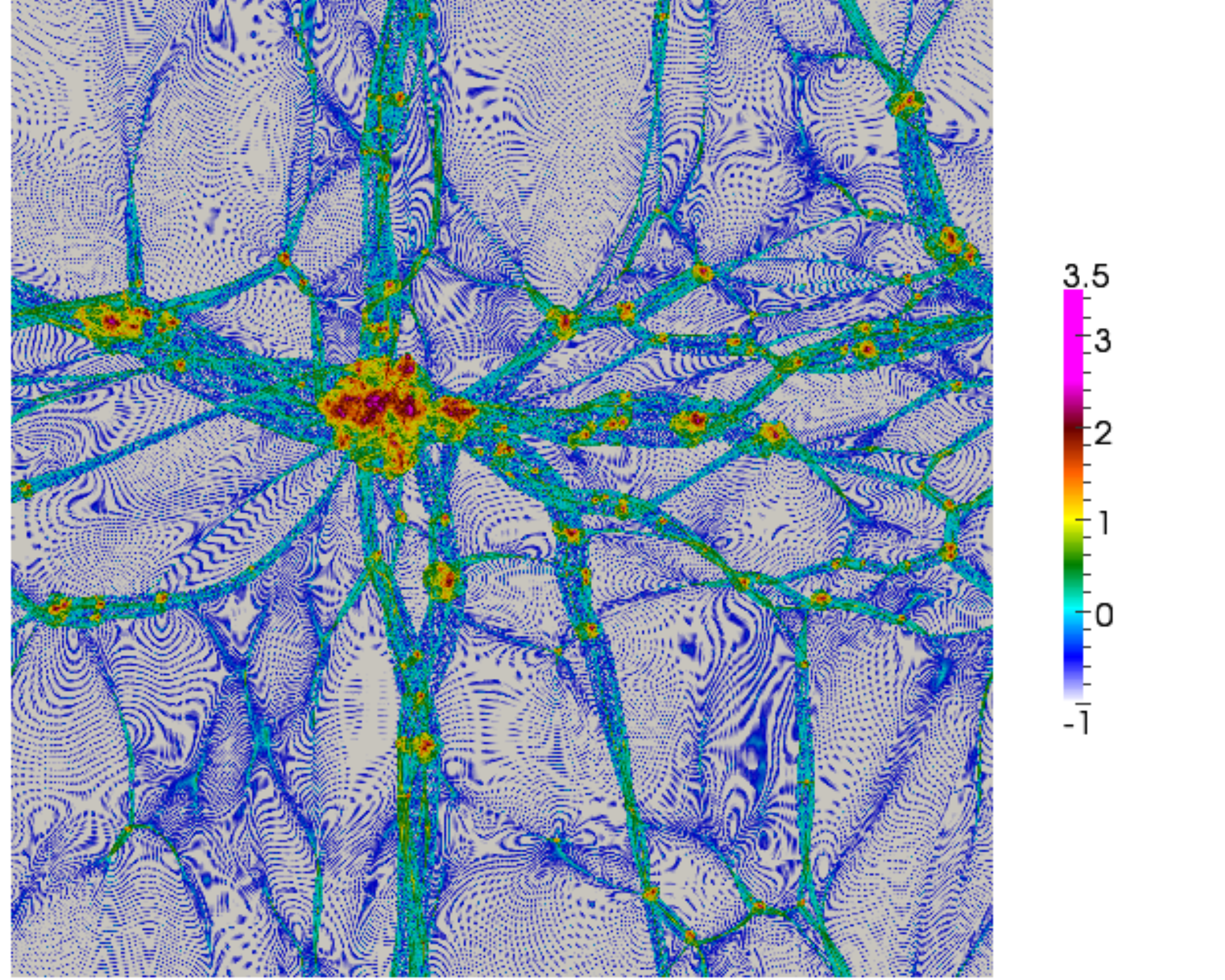}
	                     \includegraphics[scale=0.19]{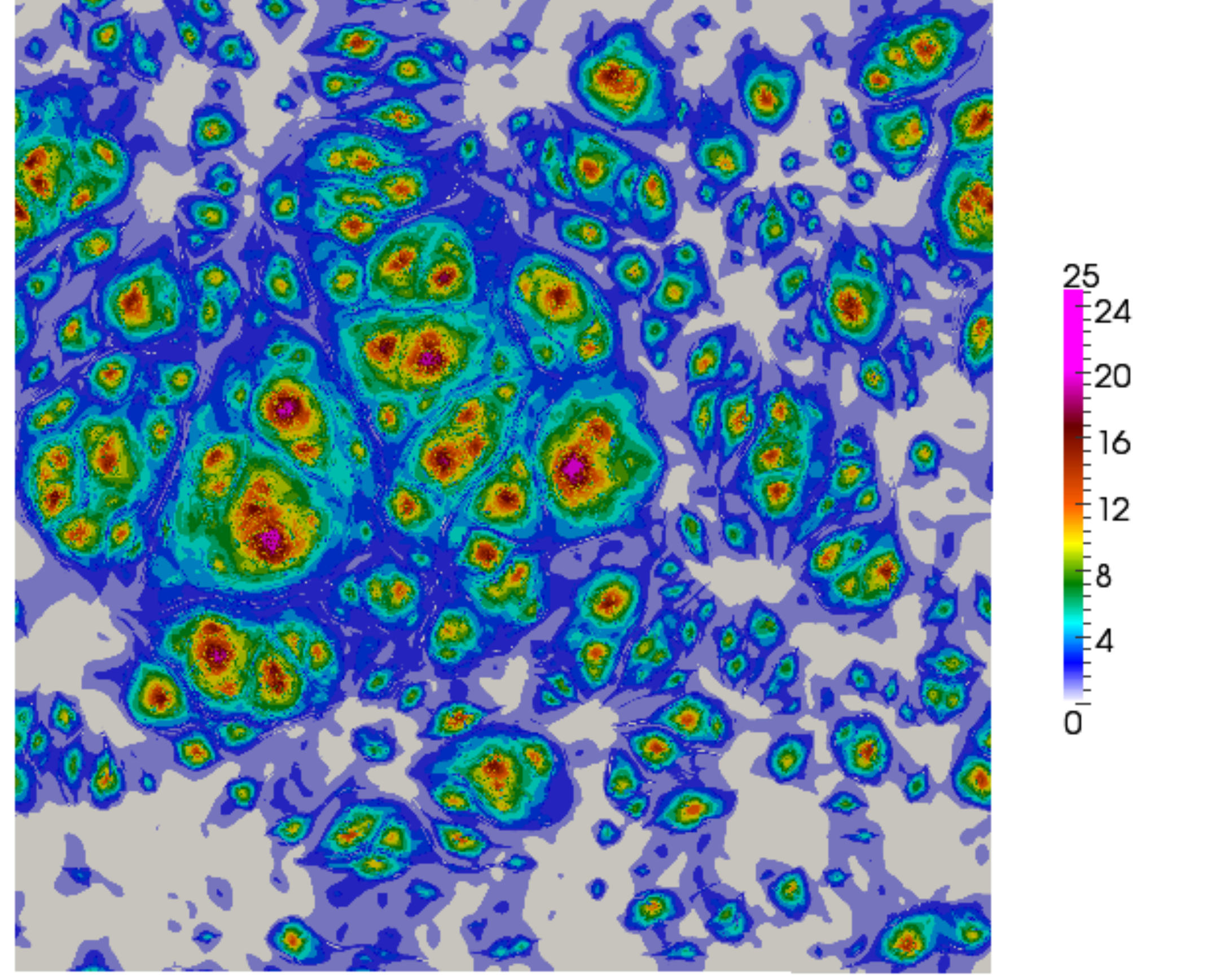}}
	\centerline{\includegraphics[scale=0.19]{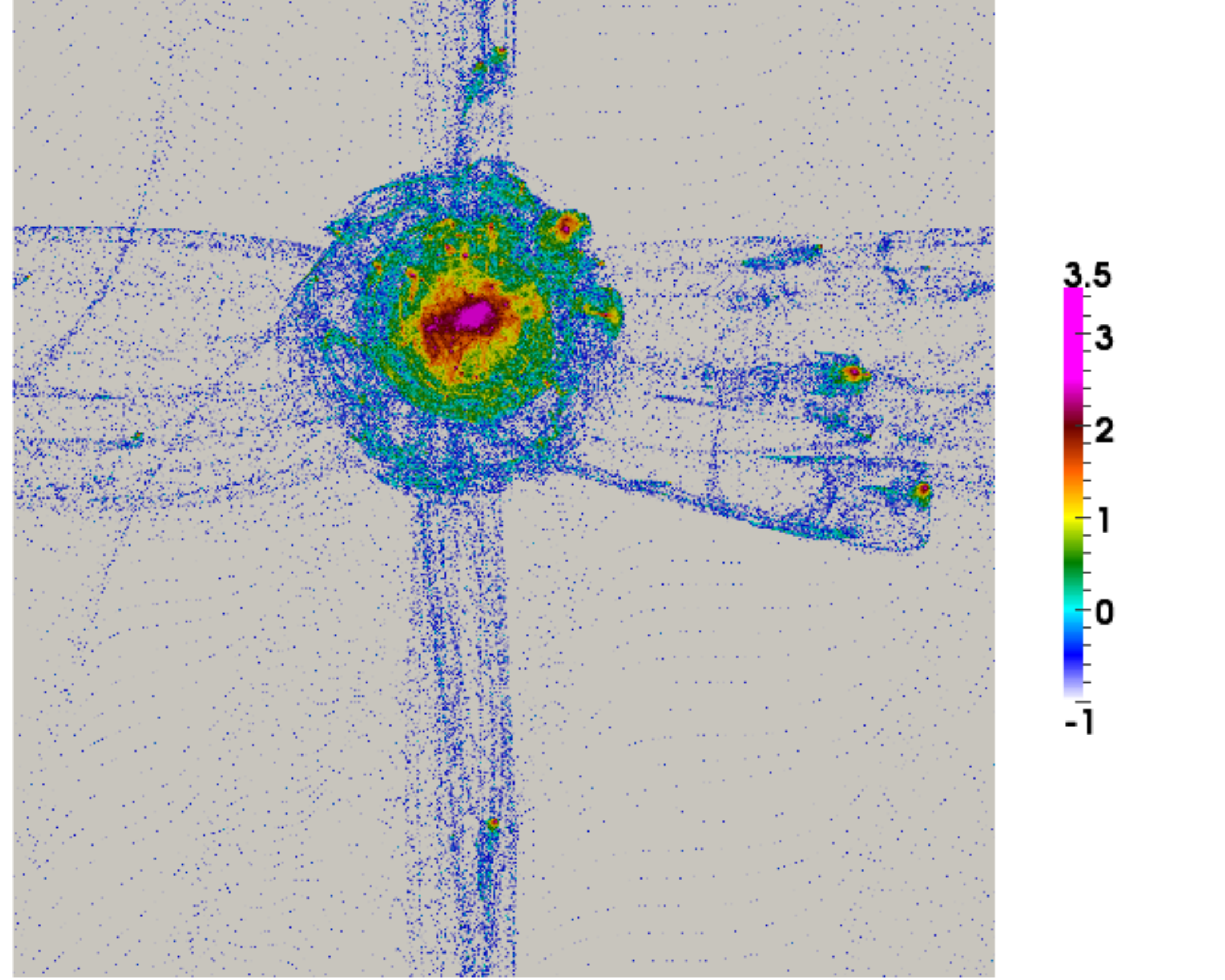}
	                     \includegraphics[scale=0.19]{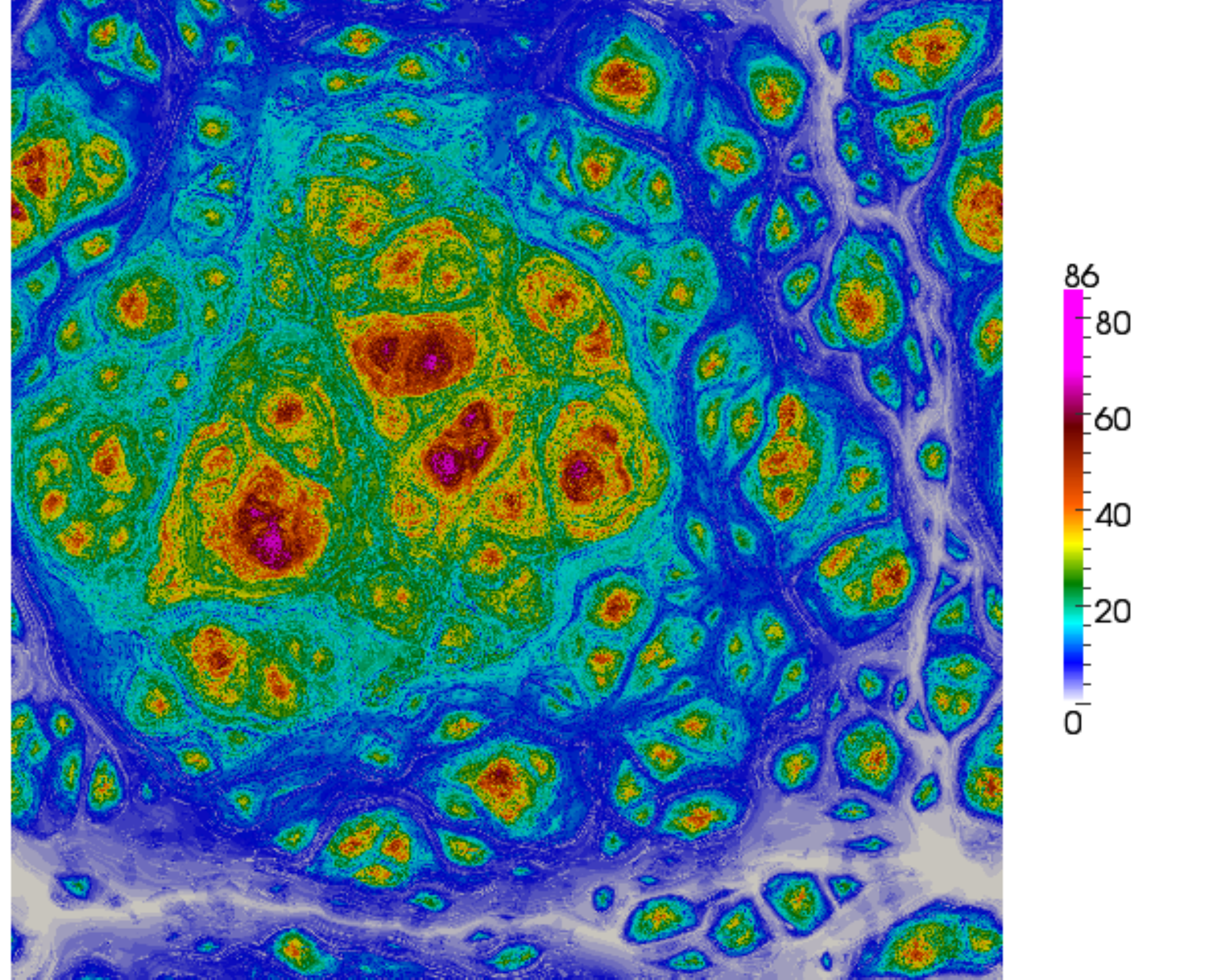}}
	\caption{Evolution of  structure in two-dimensional N-body simulation. Four stages are shown at  
	$a \approx 1.0, ~2.3, ~3.4$ and $58.7$ from
	top to bottom. The density perturbation linearly extrapolated  would result in $\delta_{\rm rms} =1$ at  $a=1$.
	The  CIC density fields in Eulerian space are shown in the left column. The corresponding flip-flop fields in Lagrangian space are shown  on the right.
	The colors in the density plots are given on the logarithmic scale in the range  $-1 < \log_{10}\rho < 3.5 $ from gray to magenta. The colors in the flip-flop
	plots are given on different linear scales in the ranges $0 < n_{\rm ff} < [10, ~21, ~25, ~86]$ in four panels from top to bottom.}
	\label{fig:two-d}
\end{figure} 
 
 The two-dimensional example is based on a simple N-body simulation in the EdS cosmological model 
 with 2048$^2$ particles
 with equivalent mesh for computing CIC density and  gravitational force via FFT. The initial power spectrum 
 was $P \propto k^{-1}$ which corresponds to $P \propto k^{-2}$ in 3D in some important statistical aspects. 
 The initial amplitude was normalized to give linear $\delta_{\rm rms} =1$ at the scale  factor $a=1$.
 The purpose of the simulation was to produce a single halo and evolve it for a long time in order to see
 how the structure originates and evolves in the flip-flop field. 
 In particular how fast it gets erased or smeared by nonlinear processes  or/and numerical noise. 
 
 Figure \ref{fig:two-d} shows four stage in the evolution at $ a \approx 1.0, ~2.3, ~3.4$ and $58.7$.  
 Four panels on the left show the CIC density
 in Eulerian space and the panels on the right show the corresponding flip-flop fields in Lagrangian space.
 Although the colors look similar in all plots they 
 have very different meanings. The density plots use the logarithmic scale with the same range in each  plot 
 $-1 < \log_{10}\rho < 3.5 $ from gray to magenta.
 The flip-flop plots use linear scales with different ranges in all four panels:  $0 < n_{\rm ff} < n_{\rm ff, max} $ where
 $n_{ff, max}  = 10, ~21, ~25$, and $86$ from top to bottom respectively. The pattern in the flip-flop field evolves quite rapidly
 at the beginning of the non-linear stage. It needless to say that before shell crossing the field did not exist or was equal to zero at every point  
 if it is more preferable.
 However,  at the scale factor  approximately between  3 and 4 the landscape gets almost frozen. At later time the  field continues to grow 
 as the increasing range of the color legends show  but the geometrical pattern evolves very little. 
 For instance, all major peaks seen in the bottom panel corresponding to $a \approx 58.7$ could be easily 
 identified in two middle panels   corresponding to $a \approx 2.3$ and $3.4$. 
It is worth stressing that the major haloes shown in the corresponding density plots have not completed merging yet. 
The areas of gray and blue colors in flip-flop field may serve as visual indicators of the mass outside of the web.
Here we do not show the results of the quantitative statistical analysis as  they are similar to that in three-dimensional 
case presented in the following sections.

The paper is organized as follows.
Section 2 explains the method in three dimensions in  detail and Section 3 describes the N-body simulation used 
in the study of the flip-flop field properties.
After having presented the methodology in the previous sections we provide three-dimensional illustrations in Section 4.
Then we discuss a number of  statistical properties of the flip-flop field and in particular its peaks in Section 5.
We also  compare some properties of the flip-flop field  with that  of  density and gravitational 
potential fields in Lagrangian space. In Section 6 we show that AGF haloes contain maxima of flip-flop fields.
We present the results of the study of substructure evolution in the largest halo of the simulation. 
Section 8 is a short summary of the results.
 
 \section{Method}                      
We propose a novel approach to the exploration of the DM web in cosmological N-body simulations. So far most of the studies used either
the particles, or density field in Eulerian space, or the velocity of the particles or gravitational potential field or various combinations of the above 
quantities. We will use the field in Lagrangian space formed by the number of turns inside out experienced by each DM fluid particle
which we call a flip-flop field $n_{\rm ff}({\bf q}; a(t))$.
We  estimate the number of flip flops experienced by each N-body particle by analyzing the mapping ${\bf x} = {\bf x} ({\bf q}; a(t))$ at chosen
times characterized by the value of the scale factor $a(t)$ normalized to the present epoch $a(z=0) =1$.
The particle coordinates ${\bf x}$ and ${\bf q}$ are in Eulerian and Lagrangian  spaces respectively. The Lagrangian coordinates are
the comoving positions of the particles on a regular grid corresponding to the unperturbed initial state.
 Assuming DM to be cold  this mapping, referred to as a Lagrangian 
submanifold,  is a three-dimensional sheet in  the six-dimensional space $({\bf q, x})$. 

The method is based on a concept of a DM sheet ${\bf v} = {\bf v}({\bf x};t)$ in phase space  successfully employed to improve accuracy of the estimates of the density, velocity and other parameters in standard cosmological $N$-body simulations
\citep{Shandarin_etal:12, Abel_etal:12}. 
The major difference between this concept and the conventional one 
lies in a different interpretation of the role of the particles in the simulations.  
Namely how they represent the state and evolution of the continuous DM medium.
In contrast to the common interpretation of particles as carriers of mass, the new approach treats them as massless markers 
of the vertices of a tessellation of the three-dimensional DM sheet  in six-dimentional phase space. 
And the mass is  assumed to be uniformly distributed inside each tetrahedra of the tessellation \citep{Shandarin_etal:12, Abel_etal:12,Hahn_Angulo:16}.
Once the tessellation is built in the initial state of the simulation, it remains intact 
through the entire evolution. 
This requirement results in a significant difference between this approach and
Delaunay tessellation suggested in \cite{Schaap_Weygaert:00} for estimating the density from
particle distributions, where the tessellation must be built at each time.

The particles being the vertices of the tessellation tetrahedra   describe all deformations occurred to the 
geometry of the tessellation. However it remains  continuous in both six-dimensional phase
space (${\bf x}, {\bf v}$) due to the Liouville's theorem as long as the thermal velocities of the DM particles are vanishing. 
In particular, the variations of tetrahedra sizes and volumes result in the corresponding change of the tetrahedra 
densities.  This property is especially valuable because it makes the tessellation self-adaptive to 
the growth of density perturbations with time.

The DM phase-space sheet cannot cross itself in the case of a continuous medium which is an excellent model
for cold DM down to scales of the order of a characteristic DM particle separation.
However the most of  N-body simulations
use particles which are more massive than physical DM particles by many orders of magnitude.  Thus on scales smaller than
the mass resolution scale the simulations are strongly affected by the discreteness effects which emerge at the centers of
DM haloes. One of these effects is selfcrossing of the DM phase space sheet. However the Lagrangian submanifold 
${\bf x} = {\bf x} ({\bf q}; a(t))$ in six-dimensional space never crosses itself. 

Although both (${\bf x}, {\bf v}$) -- and (${\bf q}, {\bf x}$) -- spaces contain all the information about a dynamical system
allowing to compute the whole evolution from ${\bf q}$ to ${\bf x}$ as well as from ${\bf x}$ to ${\bf q}$ (see e.g. \citealt{Landau_Lifshitz:08}) 
they obviously have very different properties. For instance, an advantage of the former consists in ability to calculate the kinetic energy
of the system while an advantage of the latter consists in being a {\it metric} space.
Therefore  the latter is  superior to the former in the analysis of geometry of the structure. 
Moreover, the Lagrangian submanifold ${\bf x=x(q)}$, is a single-valued function, unlike the phase space sheet ${\bf v=v(x)}$ 
or  ${\bf x=x(v)}$ which are multivalued in a projection on arbitrary three-dimensional space formed by any three axes out of six available ($\bf x, v$)
 in the non-linear regime after shell crossing.

Identifying flip-flop events in three dimensions can be done by computing  the Jacobian $J({\bf q},t) = |\partial x_i/\partial q_j|$ on each particle at each time step.
 If the sign of the Jacobian changes, the number of flip-flops for the corresponding particles is increased by one. 
We show that the flip-flop field $n_{\rm ff}({\bf q};a)$ at fixed $a$ exhibits  features in 
generic three-dimensional $N$-body simulation similar to those described in one-dimensional simulation. 
The Lagrangian submanifold technique was implemented in the publicly available cosmological TreePM/SPH 
code GADGET \citep{Springel:05} to compute the flip-flop field. 

\section{N-body simulations}                    
The initial conditions were generated with NGenIC code\footnote{see h-its.org} with the standard $\Lambda$CDM cosmology, $\Omega_m=0.3,~\Omega_\Lambda=0.7,~ \Omega_b=0, ~\sigma_8=0.9,~ h=0.7$ and the initial redshift $z=50$. A set of simulations were carried out with a box  $1 h^{-1}$~Mpc, $M_{\rm b, dm} \approx 1.2\times10^{11} M_{\bigodot}$.
For illustration purposes, we present two  zoomed-in simulations with $128^3$,  $m_{\rm part} \approx 5.7 \times 10^4 M_{\bigodot}$
and $256^3$, $m_{\rm part} \approx 7.1 \times 10^3 M_{\bigodot}$ DM particles in a box with the comoving size of $1 h^{-1}$~Mpc with the force resolution of $1.5 h^{-1}$ and $0.75 h^{-1}$~kpc respectively. The chosen size of the box is obviously too small for the purpose of deriving statistically valid properties of the haloes. However the main purpose of this work is different, namely we would like to demonstrate that the flip-flop field of haloes in a highly nonlinear dynamic  state  retains rich information about the haloes and their substructures as well as about their merging histories.  
We mark the epochs by the values of the scale factor $a$ with $a=1$ corresponding to the present time with $z = 0$. It is worth mentioning that
the common parts of the initial Fourier spaces in two simulations have been generated with the same random numbers.

It is always useful to have a visual concept of any structure when its geometry or/and topology is discussed. 
The visualization of the major object in our analysis the submanifold ${\bf x} = {\bf x} ({\bf q}; a(t))$ 
in six-dimensional (${\bf q}, {\bf x}$) -- space is obviously out of question even in two dimensions.
Its projection on ${\bf x}$ -- space is familiar in the form of ubiquitous dot plots illustrating the results of various N-body simulations.
We are interested in visual illustration of the projections on   ${\bf q}$ -- space which encounters additional problems compared
to the visualization of ${\bf x}$ -- space outlined bellow.

\begin{figure}
	\centering
	\centerline{\includegraphics[scale=0.7]{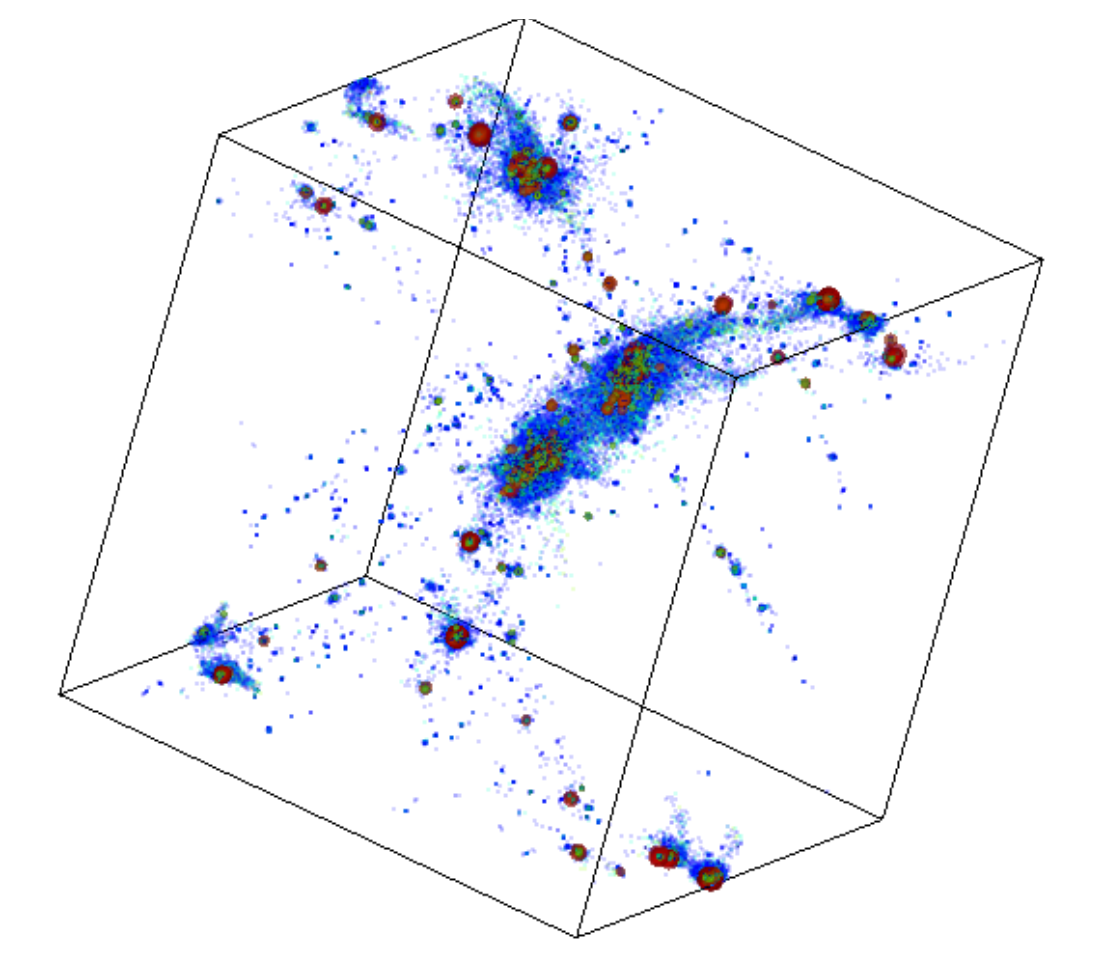}}
	\centerline{\includegraphics[scale=0.7]{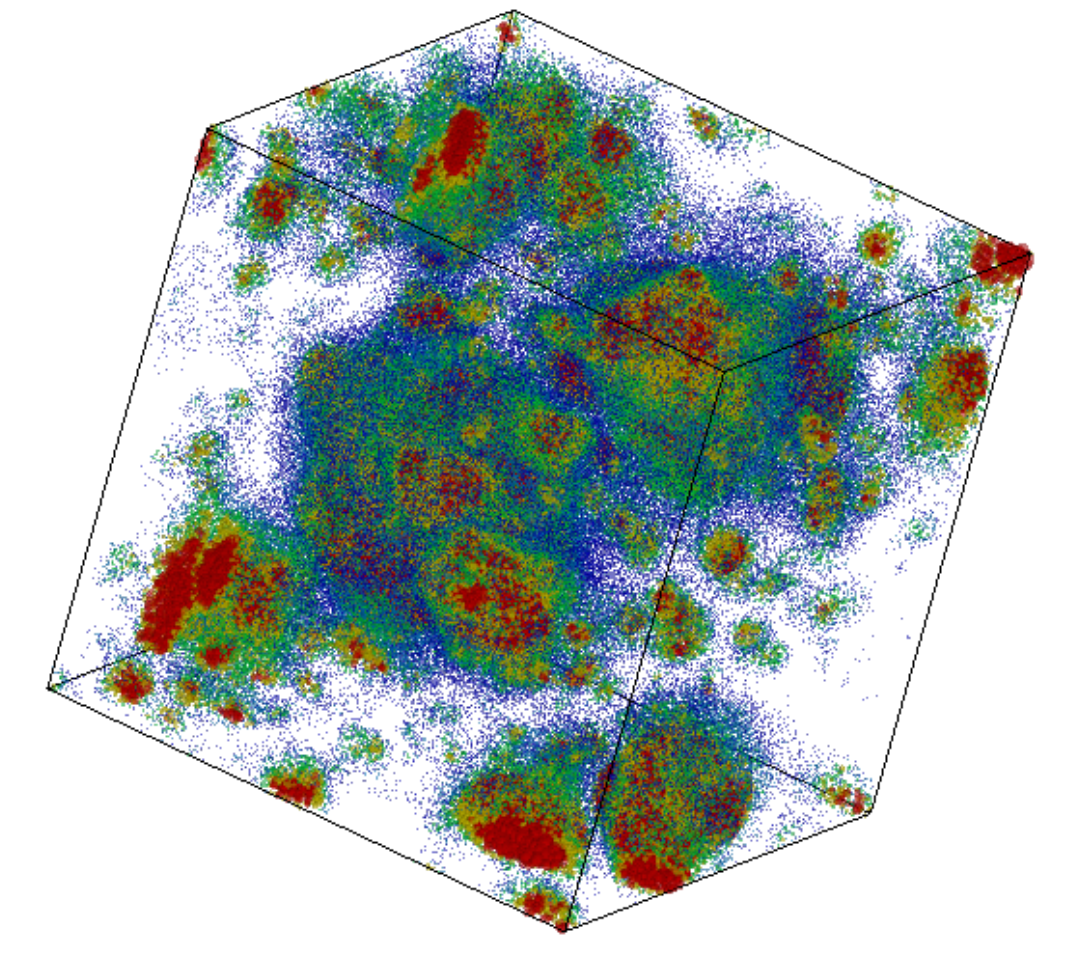}}
	\caption{ Top. The dot plot of a subset of particles with  $n_{\rm ff} \ge 6$ in the simulation of $1/h$~Mpc box in the 
	$\Lambda$CDM cosmology 
	at  $z=0$.  
	The figure illustrates the densest regions of the simulation.
	Bottom. The subset of particles shown in the top panel  is plotted in Lagrangian space. 
	The sizes of dots are scaled  by $n_{\rm ff}$ and the colors change with the growth of   $n_{\rm ff}$ from blue to red.}
	\label{fig:3d_E} 
\end{figure} 
\begin{figure}
	\centering
	\centerline{\includegraphics[scale=0.55]{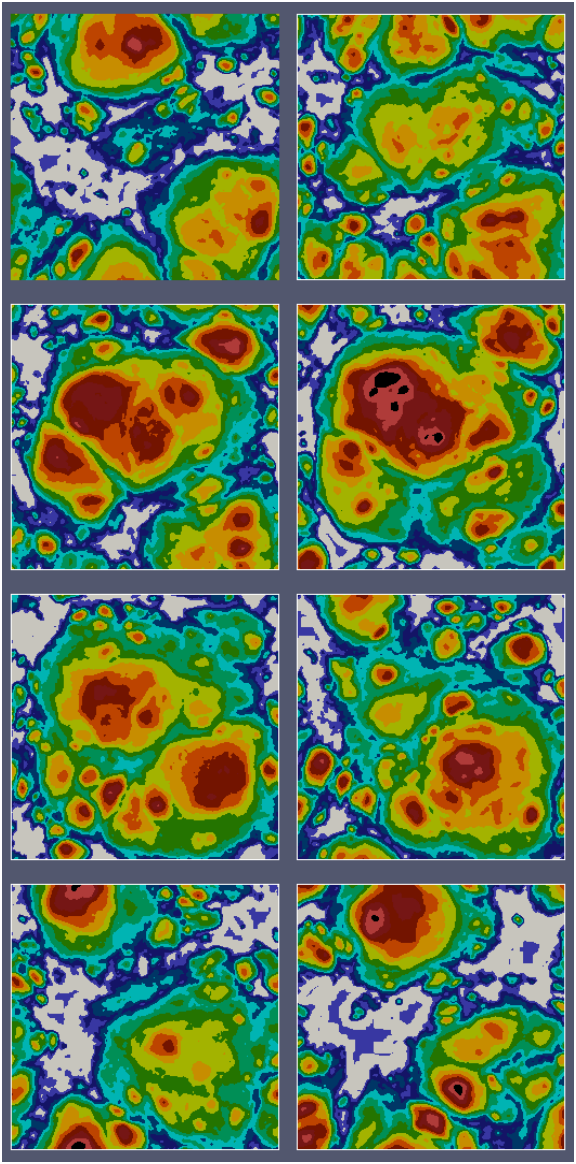}}
	\caption{ Eight panels show $XZ$ -- slices equally spaced along $y$ -- axis. Colors are as follows: $n_{\rm ff} < 1$ -- white
	and $n_{\rm f} > 250 $ -- black, and the range $ 1 \le n_{\rm f} < 250 $ is split in 15 hues from blue to brown in equal
	intervals of  $\log(n_{\rm ff})$.} 
	\label{fig:8z_slices}
\end{figure} 
  \section{Visualization of flip-flop field in three dimensions}
  \subsection{Entire simulation box}
The top of Fig. \ref{fig:3d_E} shows the map of the excursion set $n_{\rm ff}({\bf q},z=0) \ge 6$ in the entire simulation box 
to Eulerian space.
The sizes and colors (from blue to red) of the particles represent the number of flip-flops.
Boosting the sizes of less abundant particles with high flip-flop numbers allows to see them in  crowded regions 
 dominated by numerous particles with low flip-flop values.

Illustrating the flip-flop field in Lagrangian 3D space represents even more difficult problem than the cosmic web in Eulerian space.
This is because the dense regions of the web occupy a small fraction of the volume in $\bf x$ -- space as can be seen in the top of Fig. \ref{fig:3d_E}. 
But the fraction of the volume with $n_{\rm ff} \ge 1$
-- approximately corresponding to the web -- occupies more than 90\% of the volume in $\bf q$ -- space 
as the bottom of  Fig. \ref{fig:3d_E} demonstrates.
It shows the dot plot of the corresponding flip-flop field in $\bf q$ -- space with the same color coding. 
One can see that the flip-flop field has a large number of distinct  peaks occupying the most of  Lagrangian space
however it is not as detailed as in two-dimensional case (Fig.  \ref{fig:two-d}) because  some of the distant peaks are hidden
beneath the nearby structures in the projection on two-dimensional figure.

In order to reveal the much greater richness and complexity of the structure of
subhaloes in the flip-flop field, we also plot a set of two-dimensional slices through Lagrangian box
in Fig. \ref{fig:8z_slices}. This figure shows eight $XZ$ slices equally spaced along $y$-axis through the entire 
256$^3$ simulation box. The sequence of slices is ordered from the top left to bottom right panels. In order to
suppress numerical noise we smoothed $n_{\rm ff}$ field with gaussian filter with
the size being equal to the separation of particles on the Lagrangian grid. 
The colors are explained in the caption.
A complex hierarchy of $n_{\rm ff}$ peaks is revealed in considerably  greater detail however
at the cost of losing a three-dimensional perspective. Unfortunately, this is a typical trade of one for the other.

\begin{figure}
	\centering
	\centerline{\includegraphics[scale=0.39]{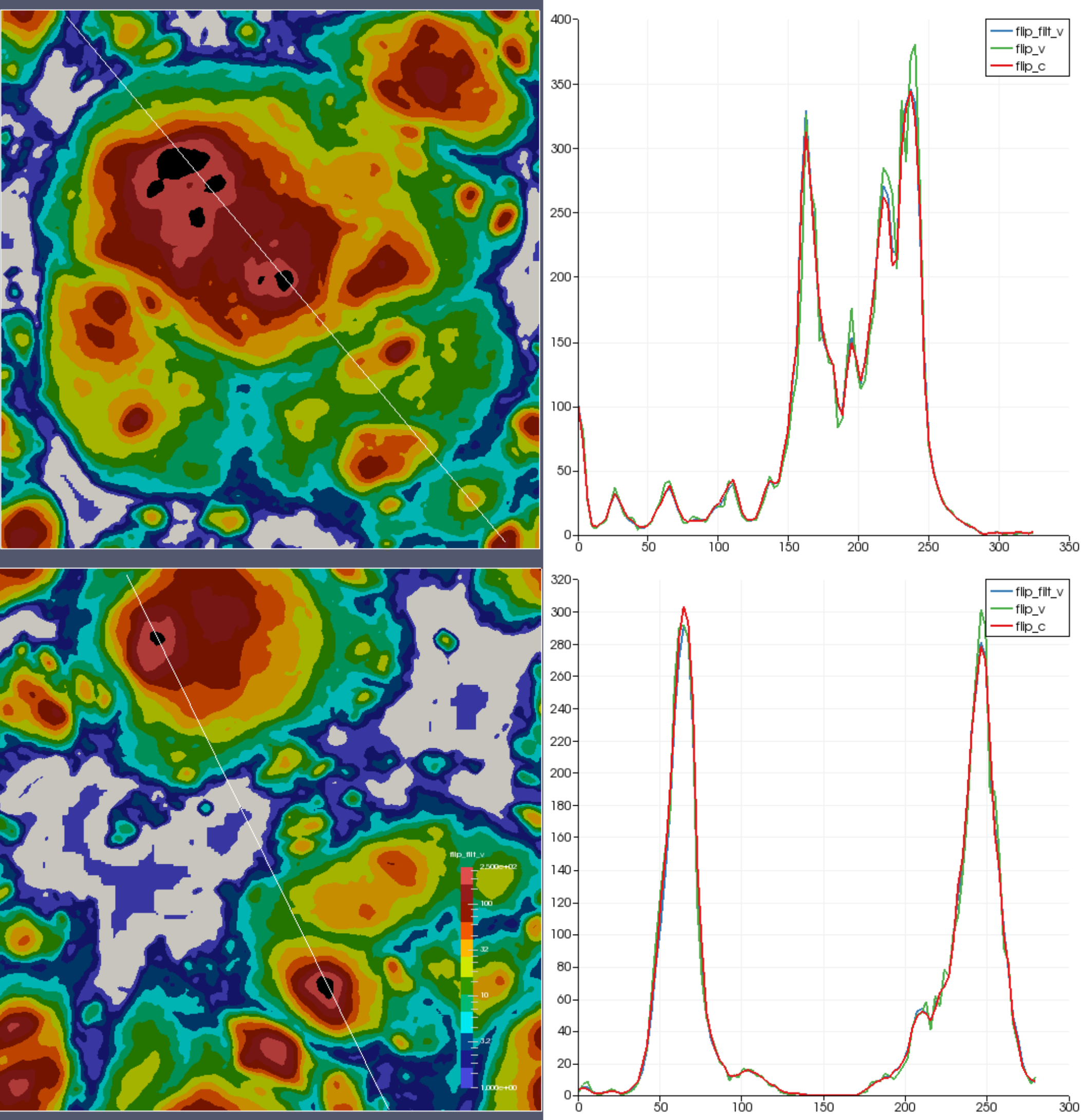}}
	\caption{ Two panels from the right column of Fig. \ref{fig:8z_slices} are shown along with plots over
	lines. The lines are shown in the left hand panels. } 
	\label{fig:2z_slices_wlines}
\end{figure}

Figure \ref{fig:2z_slices_wlines} shows two panels from the right hand column of Fig. \ref{fig:8z_slices} 
 suplemented fy the corresponding plots over lines shown in the left hand panels of Fig. \ref{fig:2z_slices_wlines}.
Each panel on the right hand shows three curves: the green and blue  curves correspond to the raw and filtered fields 
computed on the vertices,  
while the red curves show the number of flip-flops on the  cubic cells.
The flip-flop field on the cells (assumed to be uniform within each cell) equals the mean
of the filtered field at the corresponding eight vertices. Thus the red curves display averaged values.
 Both filtering and averaging play noticeable role only in the vicinities of sharp maxima and narrow minima. 
 This suggests that the numerical noise is not too bad.  However, filtering makes the contours significantly smoother.
\begin{figure}
\centerline{\includegraphics[scale=0.35]{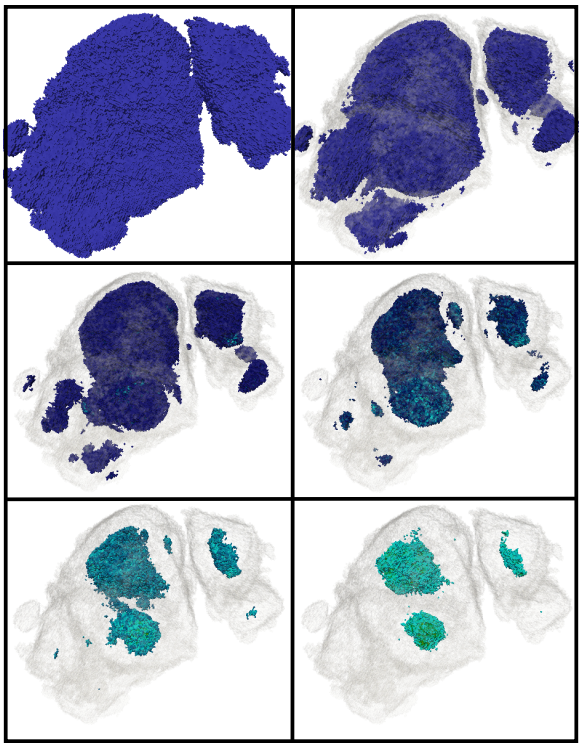}}
\caption{ The substructure in the highest peak of the 256$^3$ simulation with ${\rm max}(n_{\rm ff}) = 616$  is shown
as a set of six isocontours  in 3D. From top left panel to  down right panel color contours are respectively:  
$n_{\rm ff} = 20, 42, 67, 100, 130, 162$.  Five panels show the gray shade of the contour at
$n_{\rm ff} = 20$ shown in the top left panel in blue.}
\label{fig:peak616_a}
\end{figure}

\subsection{An example of substructure in the highest flip-flop peak  }
In this section we focus on the highest flip-flop peak (${\rm max}( n_{\rm ff} )= 616$  at    $ a=1)$ in a set of the peaks
selected by the condition $n_{\rm ff} \ge 20$.  It corresponds to one of  dynamically most evolved haloes.
Figure \ref{fig:peak616_a} shows the  three-dimensional structure of the peak in six panels starting with the contour 
at  $n_{\rm ff} = 20$ in the  top left panel. In each of five remaining panels we plot two contours: one in gray color 
shows the same contour as in the top left  panel and the other at steadily increasing levels 
$n_{\rm ff} = 42, 67, 100, 130, 162$ respectively. One can clearly see a rich nesting structure of the peak.

\section{Statistical properties of the flip-flop field}   
First we briefly discuss some of global statistical properties of the flip-flop field in comparison with density and gravitational potential fields.
 There is a subtlety  in such a comparison. We compute the flip-flop field on particles therefore it is a Lagrangian field.
The density can also be computed on particles  or in tetrahedra of the tessellation of the Lagrangian submanifold as described 
in \citet{Shandarin_etal:12}  and  \citet{Abel_etal:12}, but this would be the density in separate streams not the total density in Eulerian space.
Here we would like to compare the flip-flop field with commonly used density field computed in Eulerian space.   
However the interpolation of the Eulerian density to particles  is not uniquely defined procedure. 
Since our simulations are done with GADGET code we use SPH densities and potentials computed on
particles available in standard outputs of the code.   
We caution that other methods of computing density on particles may produce somewhat different results from reported here
although we do not anticipate a substantial difference.

We start with making a list of obvious differences between three fields in question: (i) conceptually the number of flip-flops
is a uniquely defined field in Lagrangian space but the SPH density or potential on particles in N-body simulations 
of a collisionless medium is only one of many feasible approximate maps of Eulerian fields to Lagrangian space ,
(ii) the flip-flop field is 
a discrete field with  positive integral values  while both density and potential are continuous fields apart from
the discreteness related to the grid, 
(iii) both the flip-flop and density fields are positively defined while potential can be both positive and negative,
(iv) the number of flip-flops  monotonically grows with time at every particle while both the density and potential
do not because some particles can move  back and forth between  high and low density environments
as well between regions with high and low potentials.
\begin{figure}[ht]
	\begin{center}
	\advance\leftskip-3cm
	\includegraphics[scale=0.45]{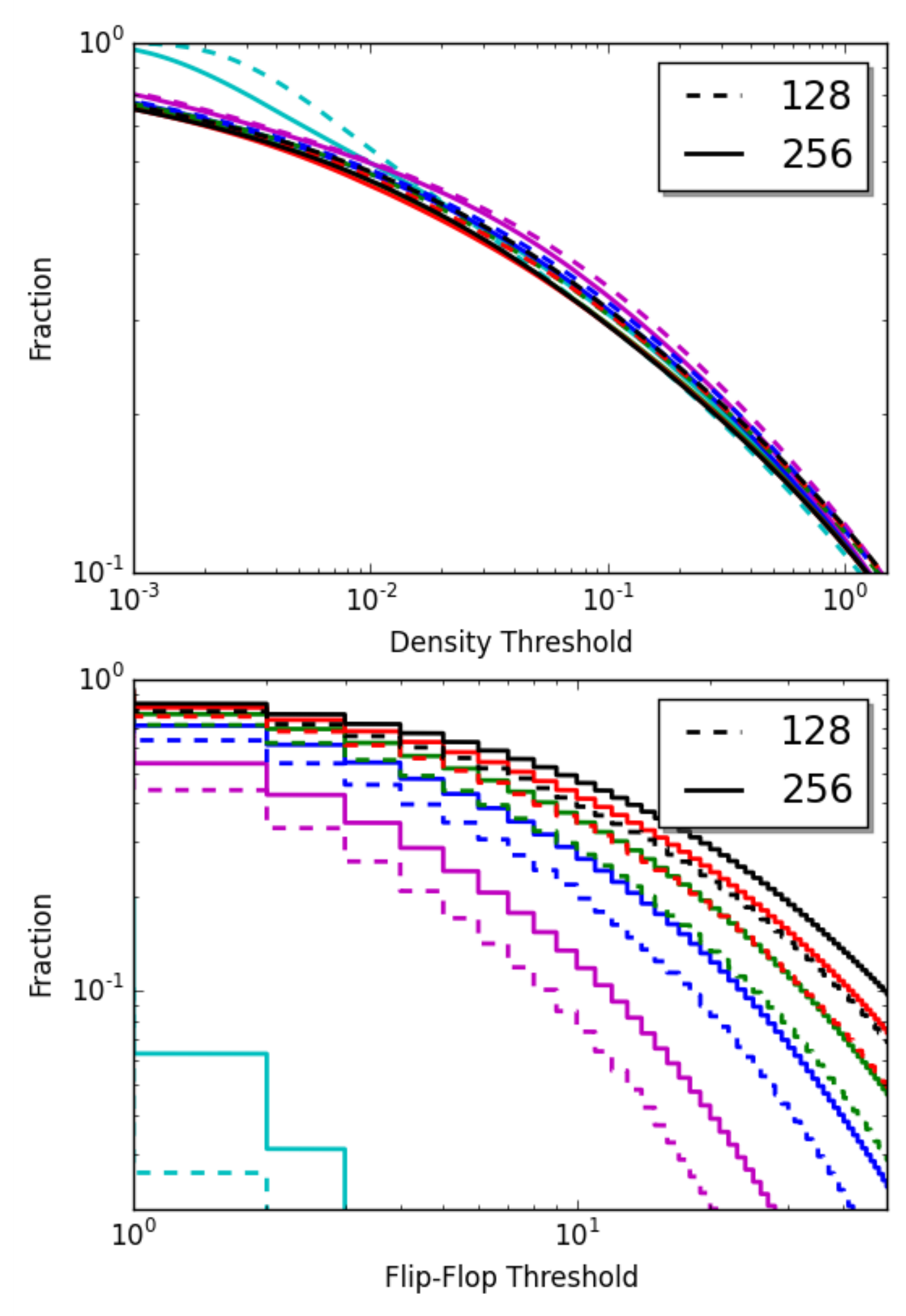}           
	\caption{ The cumulative probability functions of the  lowest  90\% of density (top)  and flip-flop (bottom) fields  are shown at six epochs: 
	$a=0.026, 0.058, 0.129, 0.242, 0.493, 1.000$ in $128^3$ and $256^3$ simulations. 
	Colors in the order of epochs are: cyan, magenta, blue, green, red and black.}
	\label{fig:cpf_rho_flip_bulk}
\end{center}\end{figure} 
\begin{figure}
	\centering
	\centerline{\includegraphics[scale=0.45]{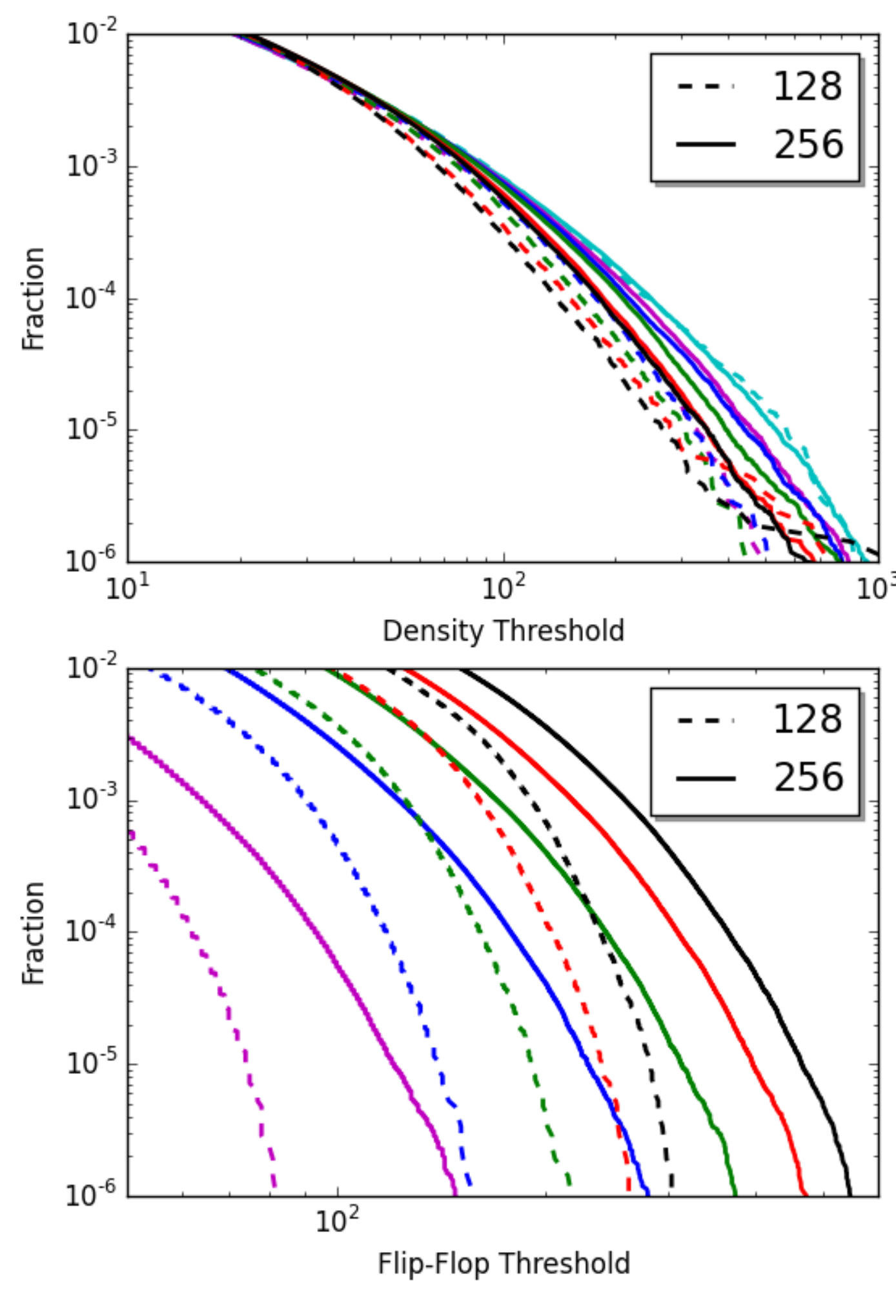}}   
	\caption{ Same as Fig. \ref{fig:cpf_rho_flip_bulk} except that  these are the cpfs of the highest one percent of the values of the fields are shown.}
	\label{fig:cpf_rho_flip_tail}
\end{figure} 
\subsection{Cumulative probability functions}
Figure \ref{fig:cpf_rho_flip_bulk}   shows the fractions of particles with densities above the threshold $\rho_{\rm th}$  in the range
of $0.001 \ge \rho_{\rm th}  \ge15$ and  the fractions of particles with the number of flip-flops above the threshold $n_{\rm th}$ in the range $1 \ge n_{\rm th} \ge 50$ in the top and bottom panels respectively.
The ranges correspond approximately 90\% of lowest values of the fields in both plots. The plots of the cpfs showing the highest
1\% of the values are shown in Fig. \ref{fig:cpf_rho_flip_tail}. The cpf is shown for both 128$^3$ and 256$^3$ simulations.

The figures demonstrate that the cpf of the flip-flop field is considerably 
more regular function of the cosmological epoch and the size of the simulation than the density in Lagrangian space.
The flip-flop cpf monotonically increases with time and with the mass resolution  of the simulation.  
The density cpf seems do not show clear dependence on either the size of the grid or the epoch.

\subsection{The growth of the web mass}    
\begin{figure}
	\centering
	\centerline{\includegraphics[scale=0.45]{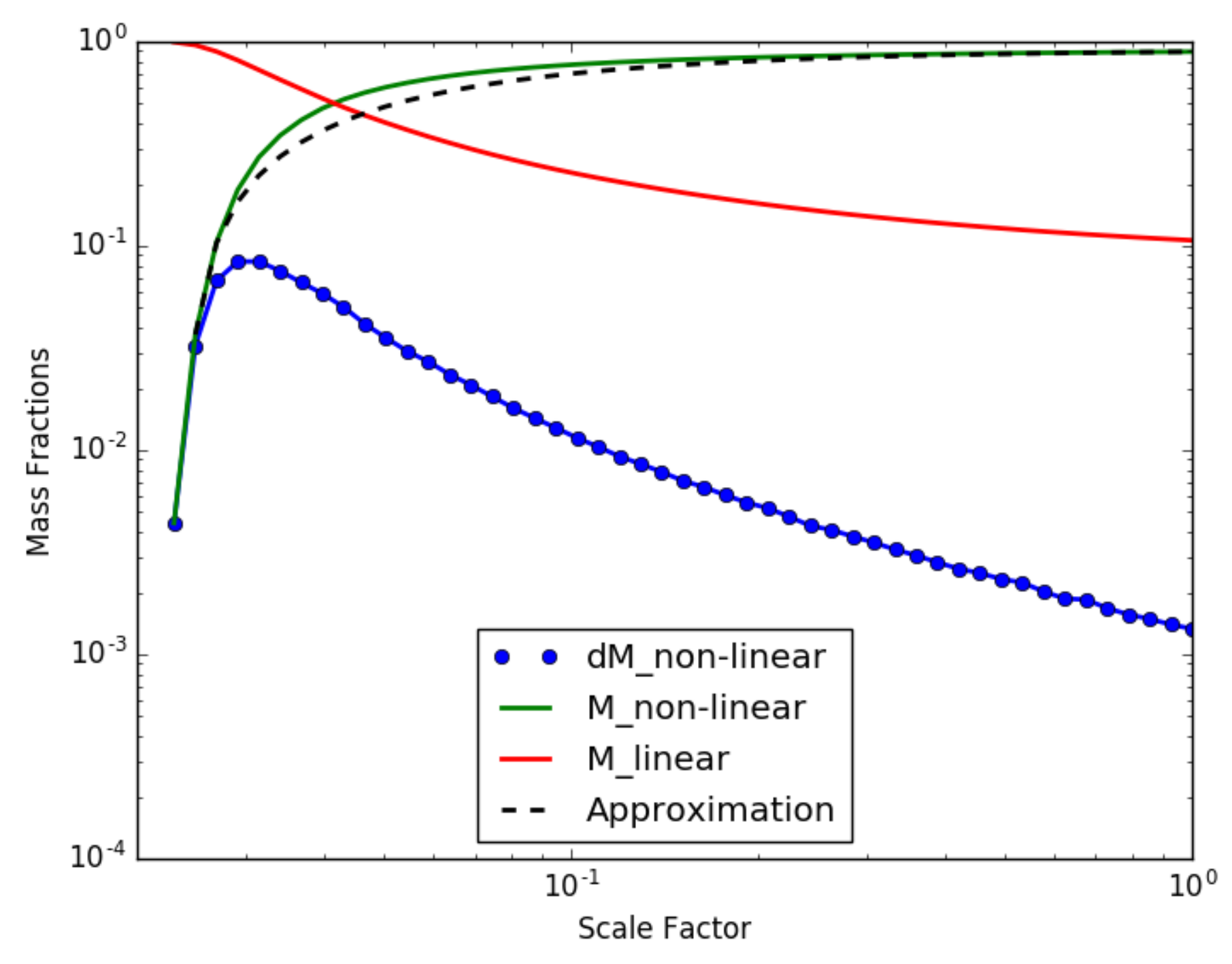}}     
	\caption{ 128$^3$ simulation. 
	A blue  line with dots shows the fractions of mass experienced the first flip-flop between  (i-1)-th and   i-th
	outputs of the N-body code as a function of $a_i$. The green curve  shows the  accumulation of mass
	experienced at least one flip-flop by $a_i$ while the red curve demonstrates the mass fraction which have not experienced flip-flops
	at all  by $a_i$. The dashed line in black shows a crude analytical approximation to the green line. }
	\label{fig:mass_fracs_128}
\end{figure}

\begin{figure}
	\centering
	\centerline{\includegraphics[scale=0.45]{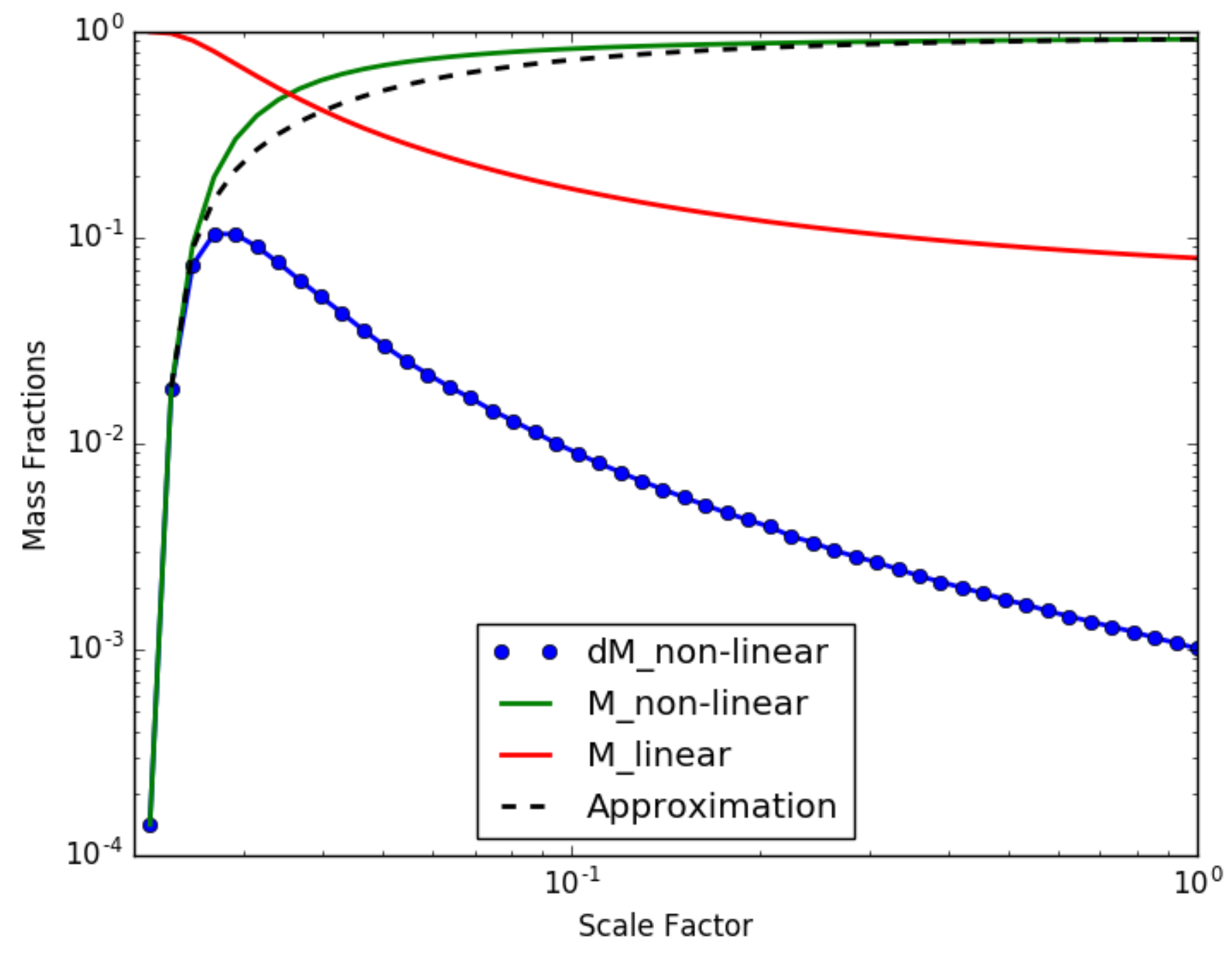}} 
	\caption{ 256$^3$ simulation. Notations are the same as in Fig. \ref{fig:mass_fracs_128}}
	\label{fig:mass_fracs_256}
\end{figure}

The fraction of mass experienced the strongest non-linear event -- flip-flop -- monotonically increases with time.
The slope of the power spectrum of the linear density perturbations in the range from the Nyquist wavelength
$L_{\rm Ny} \approx   16\, h^{-1}$ kpc or $\approx   8\, h^{-1}$ kpc  in 128$^3$ or 256$^3$ simulation respectively 
to the size of the simulation box  is quite steep. Therefore the fraction
of mass reached a strong non-linear regime when a fluid element experiences a flip-flops grow very fast.
The blue curves in Fig. \ref{fig:mass_fracs_128} and \ref{fig:mass_fracs_256} show the fractions of mass experienced the first flip-flop between the output times
equally spaced on  the $\log_{\rm 10}{(a)}$  scale from approximately 0.023 to 1. The green  lines show the growth 
of the mass fraction  experienced at least one flip-flop by $a_{\rm i}$ while the red lines show the decrease of the mass 
fraction that did not experience even a single flip-flop. 
 After reaching about a third of the total mass
at $a\approx 0.03$ the growth of the mass in the non-linear regime is steadily reducing reaching
about 0.1\% by the present time ($a=1$).  The total mass in the particles never experienced flip-flops has dropped
to about 10\% or 8\% in  the 128$^3$ or 256$^3$ simulations respectively.
The black dashed line is a crude analytic approximation of the growth of the mass fraction reached
non-linearity by a function with two free parameters
$$
f(a) = C_1 - {C_2 \over a}.
$$
The parameters are chosen to make the approximation curve to pass  through the second and last points of the green curve 
obtained in the N-body simulations. 
The approximation obviously is far from perfect and is not intended for further use.
It still may be useful because it demonstrates a very fast growth of the mass in the  dark matter web,
which is probably faster than exponential rate between $a=0.023$ and $a=0.03$.
In the 128$^3$ simulation $C_i = (0.022, 0.92)$ and in the 256$^3$ simulation the are $C_i = (0.021, 0.94)$ showing
that the difference between two simulation is noticeable but quite small.

\subsection{Correlation properties }    
\subsubsection{ Correlations between the pairs of different fields at the same epoch}   
\begin{figure}
	\centering
	\centerline{\includegraphics[scale=0.45]{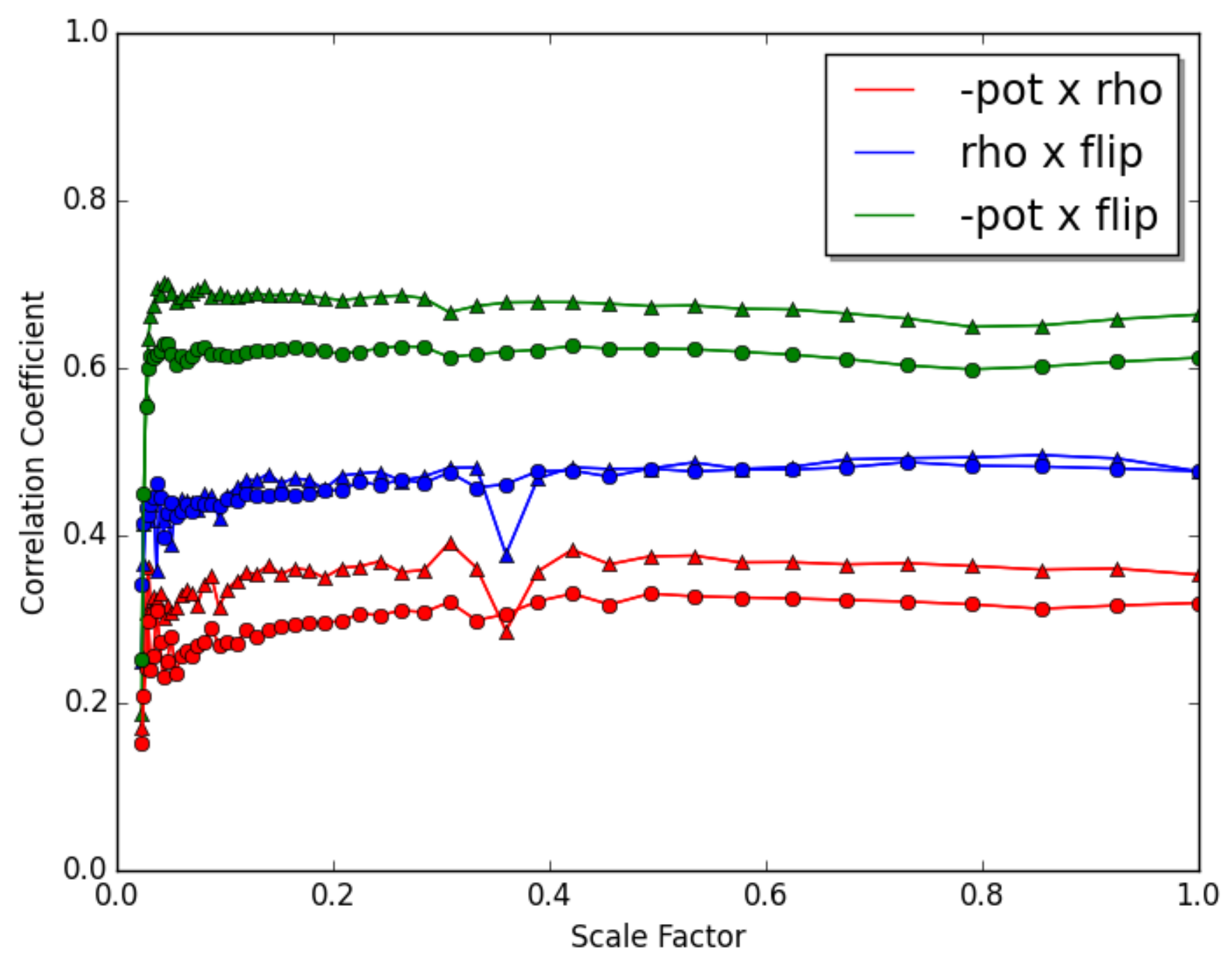}}     
	\caption{ Correlation coefficients of  all pairs of three Lagrangian fields in Lagrangian space: negative potential (- pot), density (rho) and flip-flop (flip)
	as a function of the scale factor. Triangles and circles correspond to the simulations with $128^3$ and $256^3$ particles respectively.}
	\label{fig:corcoefs_p_r_f}
\end{figure}

It is expected that the  density, potential and flip-flop fields correlate with each other since
the higher densities and higher numbers of flip-flops tend to be more frequent in the regions of negative potential. 
In order to quantify these anticipations we evaluate the correlation coefficients of the fields in question
at a number of  epochs in both $128^3$ an $256^3$ simulations.
Figure \ref{fig:corcoefs_p_r_f} shows each correlation coefficient as a function of a scale factor.
They exhibit a remarkably weak evolution if any with time.
The highest correlation is between the potential and  and flip-flop fields, while the lowest
between potential and density fields. Both become weaker with the growth of the mass resolution
while the correlation coefficient of the density and flip-flop fields seems to be very similar in
both simulations. 
 The  correlations between the density and flip-flop fields is intermediate  for these simulations
 and do not show any dependence on the size of the simulation. Its magnitude being around 0.45
 suggests that the fields seem to  resemble each other but they are not very similar.
Both dependences on the size of the simulation looks natural. Adding  small scale power to the initial perturbations 
in the $256^3$ simulation results in  reducing the scale of the both  density  and flip-flop fields
without a significant  effect on the scale of the potential. 
\subsubsection{ Correlations between the pairs of  the same field taken at different epochs}  
\begin{figure}
	\centering
	\centerline{\includegraphics[scale=0.45]{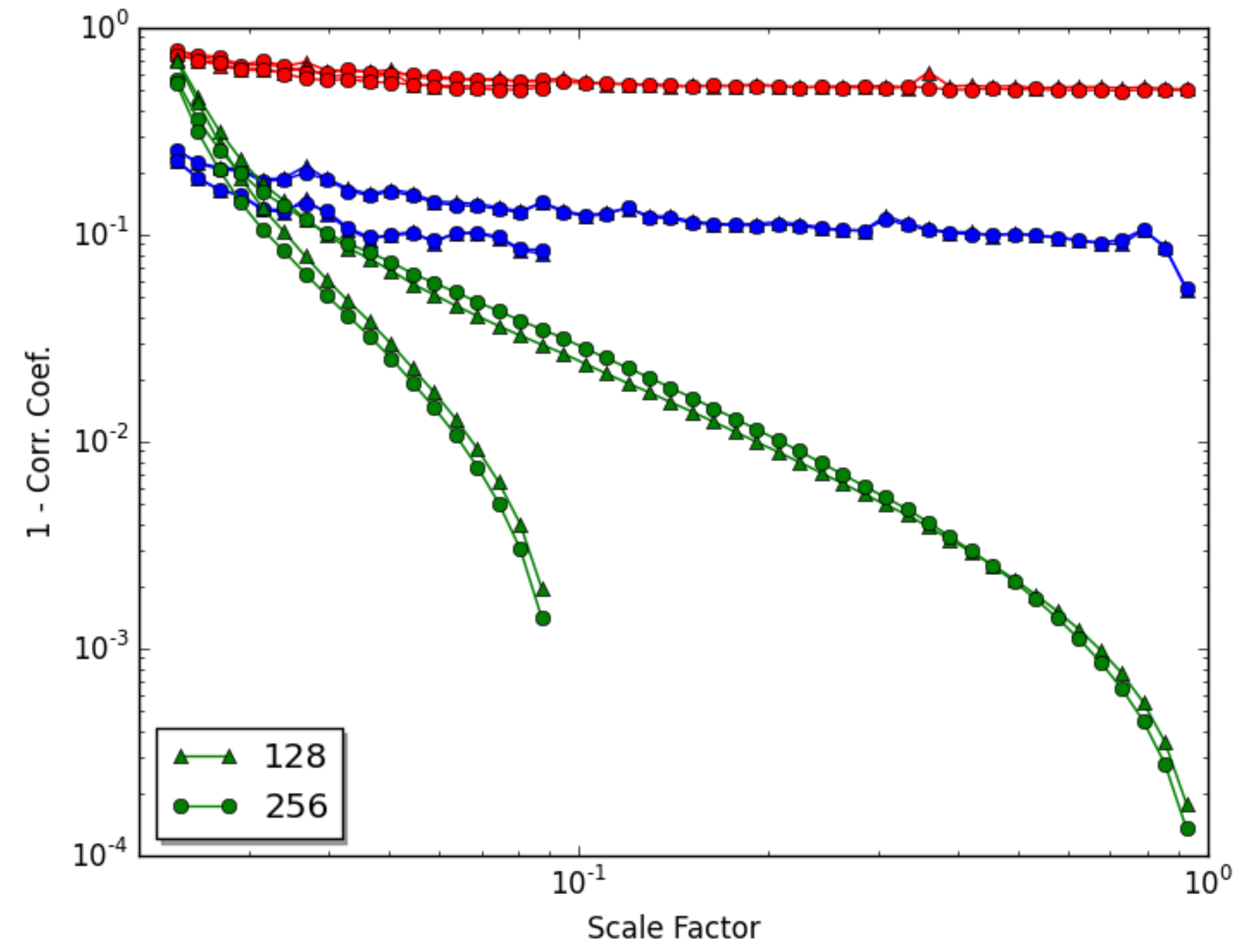}}    
	\caption{ Correlation coefficient  of the density $\xi_{\rho\rho}$ (red), potential $\xi_{\varphi\varphi}$ (blue) and flip-flop 
	$\xi_{ff \cdot ff}$ fields at $a=1$ (long curves) and $a=0.1$ 
	(short curves) with corresponding fields at all previous stages.   In order to see how close to unity
	$\xi_{ff \cdot ff}$ is we plot the logarithm of its difference from unity. 
	The curves are shown for both $N_p = 128$ (triangles) and $256$  (circles) simulations.}
	\label{fig:corcoef}
\end{figure}

Here we present the correlation coefficient of the same field taken at two different epochs: ${\rm corcoef}[f(a_i), f(a_{\rm ref})]$
where $f = [\rho, \phi, n_{\rm ff}]$ \ie it is either density or potential of flip-flop field.
We selected two reference epochs:  $a_{\rm ref\,1}=1$ and $ a_{\rm ref\,0.1} = 0.1$.
In both cases we compute the correlation coefficients with all previous stages $a_i < a_{ \rm ref}$.
In order to emphasize  how close to unity the correlation coefficient of the flip-flop field  is we plot  
$ \log{(1- {\rm corcoef}(f_i,f_k))}$  instead of $ \log{({\rm corcoef}(f_i,f_k))}$ in Fig. \ref{fig:corcoef}. 

It is no surprise  that three fields correlate stronger
when two epochs get closer to each other.  However the strength of the correlation 
as well as dependence on the epoch $a_i$  are substantially different between all three fields.
The density correlation coefficient shown in red is the lowest and the most stably  growing from about 0.2 
at the largest separation of the epochs to about 0.5 for the closest epochs. 
There is practically no difference between $128^3$ and $256^3$ simulations and barely noticeable difference
between two reference epochs $a=1$ and $a=0.1$. The potential correlation coefficient  shown in blue is significantly 
higher with the range from about 0.75 to 0.95 with distinct difference between two reference epochs $a=0.1$ and $a=1$. 

The overall highest  correlation coefficient 
and the greatest difference between two reference epochs is displayed by the flip-flop field. Although it is similar to the density
correlation coefficient at the largest separation of the epochs it grows very quickly above 0.9 at $a_i \approx 0.03$ and then
above 0.99 at $a_i \approx 0.05$ and at $a_i \approx 0.2$ for the reference epochs $a=0.1$ and $a=1$ respectively.
Such a small difference between the correlation coefficient and unity suggests that the difference  of the fields 
at the corresponding epochs is mostly due to a constant factor. It means that the geometry of the flip-flop field does 
not evolve much after some epoch which means in turn that the peak structure in the flip-flop field keeps the record
of the formation of the haloes even after they experienced  mergers with other haloes. 

\subsection{Comparison of the flip-flop, density and potential fields 
 in Lagrangian space}
\begin{figure}
	\centering
	\centerline{\includegraphics[scale=0.6]{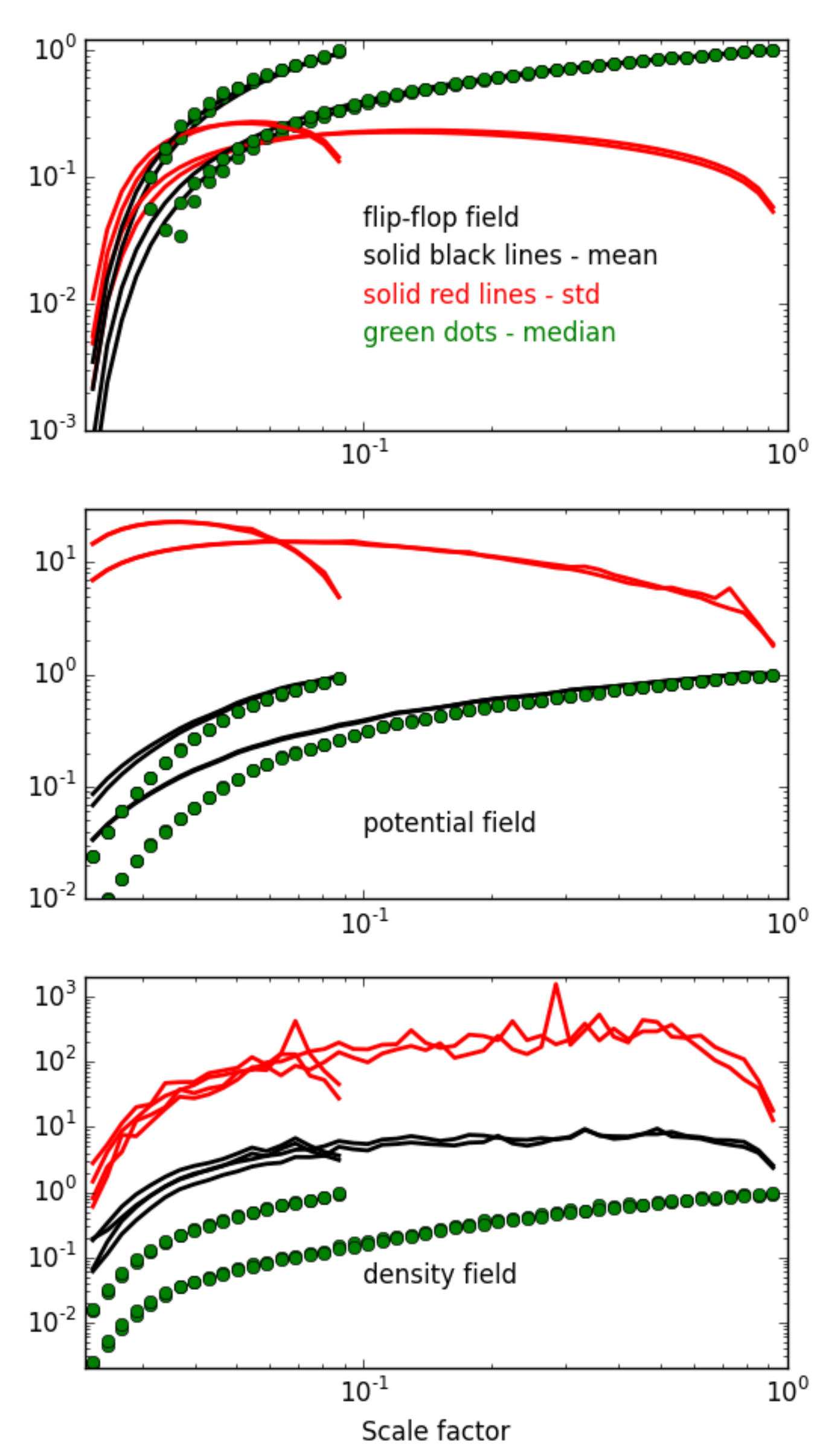}}    
	\caption{The mean, std and median of the ratios of the field at the scale factors shown on the horizontal to 
	the same field at the reference scale factor. Two sets of curves correspond to two reference scale factors $a=0.1$ (short curves)
	and $a=1$ (long curves).  The curves corresponding to $128^3$ and $256^3$  simulations noticeably different only
	for mean and std values of the density field shown in the bottom panel.   }
	\label{fig:ratios}
\end{figure}

We computed the ratios of a field at chosen epoch $a_i$ to it at a reference epoch  $a_{\rm ref} > a_i$:
$R^{(F)}_{i,  {\rm ref}}=F({\bf q}, a_i) / F({\bf q}, a_{\rm ref})$, where $F$ is either the flip-flop or density or potential  field and
$a_{\rm ref}$ is either 0.1 or 1.
Then we estimated the mean, median and standard deviations of the ratio fields.  
Figure \ref{fig:ratios} show the results
for all three fields  at 47 scale factors in the range from 0.02 to 0.92 when the reference 
field was chosen at $a=1$ and at 18 scale factors in the range from 0.02 to 0.087 when the reference field was chosen at $a=0.1$.
The mean and std values are shown in black and red colors respectively in three panels: flip-flop field - in the top, 
potential - in the middle, and density - in the bottom panels of Fig. \ref{fig:ratios}. The median values of  $R^{(F)}_{i,  {\rm ref}}$
are shown by green dots.
\subsubsection{Flip-flop field (top panel of Fig. \ref{fig:ratios})} 
 For the majority of  the epochs $a_i$ the values of $\sigma(R^{\rm (ff)}_{i,  {\rm ref}}$) is significantly smaller than the mean and median 
 values which are practically the same: 
 $\sigma(R^{\rm (ff)}_{i,  {\rm ref}}) \ll {\rm mean}(R^{\rm (ff)}_{i,  {\rm ref}} ) = {\rm median}(R^{\rm (ff)}_{i,  {\rm ref}}) $. 
 It demonstrates that the flip-flop field grows in very orderly manner with
 stochastic effects being quite small.
 
\subsubsection{Potential field (middle panel  of Fig. \ref{fig:ratios})}  
The difference between ${\rm mean}(R^{(\varphi)}_{i,  {\rm ref}} )$  and ${\rm median}(R^{(\varphi)}_{i,  {\rm ref}} )$ 
are more distinct at large separations of the epochs and significantly diminishes
as the epochs get closer. The most conspicuous difference of $R^{(\varphi)}_{i,  {\rm ref}}$ from $R^{\rm (ff)}_{i,  {\rm ref}}$
is a large value of std  at all epochs 
$\sigma(R^{(\varphi)}_{i,  {\rm ref}}) \gg {\rm mean}(R^{(\varphi)}_{i,  {\rm ref}} )\approx  {\rm median}(R^{(\varphi)}_{i,  {\rm ref}}) $
with the lowest value
of about twice of the mean at the smallest separation from $a=1$. 

\subsubsection{Density field (bottom panel  of Fig. \ref{fig:ratios})}  
The evolution of  $R^{(\rho)}_{i,  {\rm ref}}$ is essentially stochastic: 
$\sigma(R^{(\rho)}_{i,  {\rm ref}}) \sim 100 \times{\rm mean}(R^{(\rho)}_{i,  {\rm ref}})$ 
and $\sigma(R^{(\rho)}_{i,  {\rm ref}}) \sim 1000 \times{\rm median}(R^{(\rho)}_{i,  {\rm ref}} )$ making both  rather useless.
 
Summarizing Section 6.3 we would like to stress the qualitative difference between evolutions of the flip-flop and SPH density
 and gravitational potential fields evaluated on particles. The flip-flop evolves in a remarkably orderly manner
 with mean and median of the ratios $n_{\rm ff}({\bf q}, a_i) /n_{\rm ff}({\bf q}, a_{\rm ref})$ being almost exactly 
 equal and std being much smaller.
It is a strong evidence that the geometry of the flip-flop field in Lagrangian space evolves very little.
On the contrary the evolution of  the potential and especially density field is almost  stochastic:
while the both the  mean and median of the ratios of the both fields monotonically increase with time the standard deviations
are considerably greater at all epochs making the mean and median useless.
\begin{figure}
	\centering
	\centerline{\includegraphics[scale=0.6]{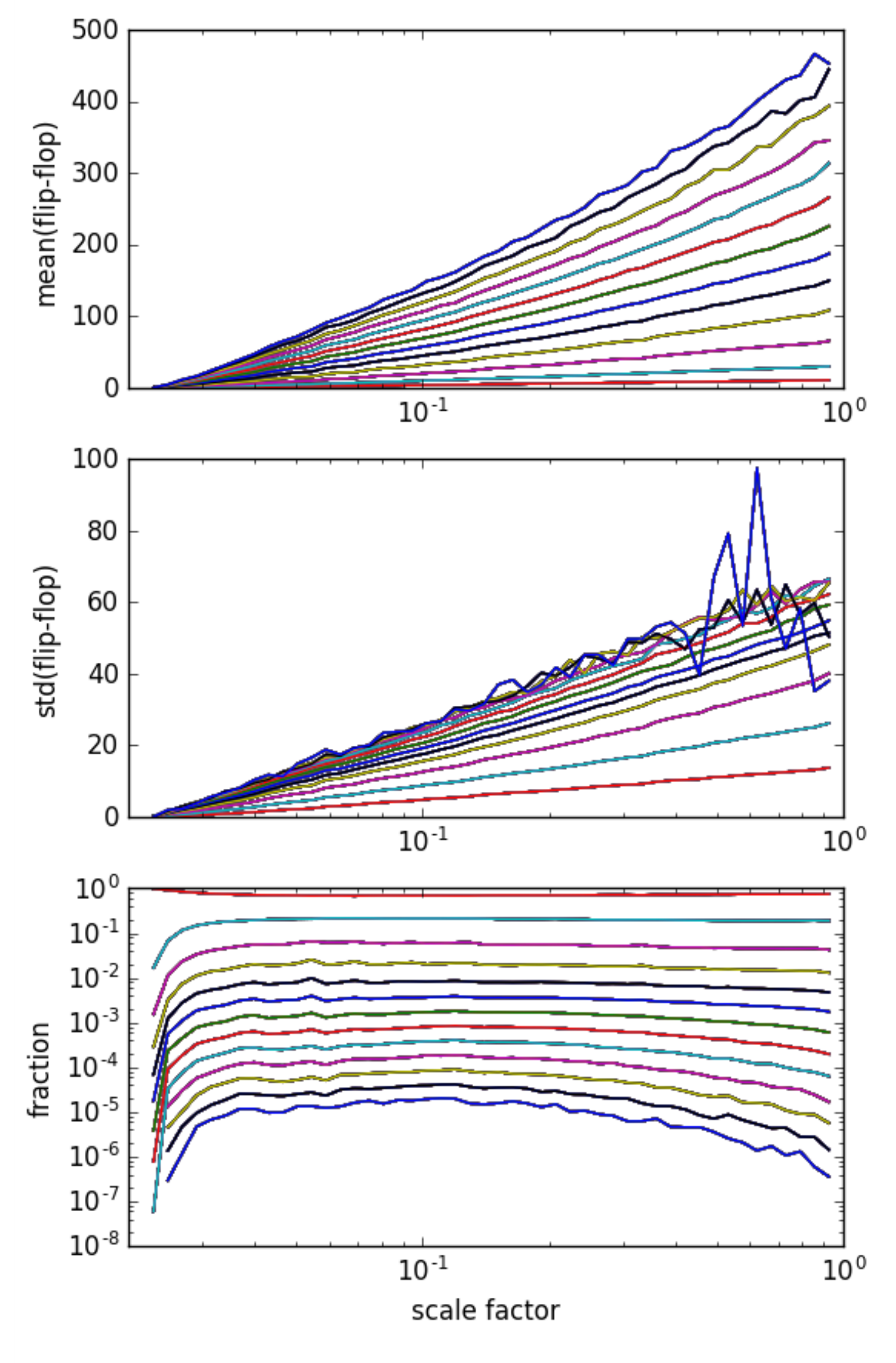}}   
	\caption{Conditional statistics of the flip-flop field as a function of the scale factor. The particles are binned according to the change of the number of flip-flops 
	 between the outputs $\Delta_{\rm ff} = n_{\rm ff}(a_{i+1}) - n_{\rm ff}(a_i) $. The curves are in  the range $0 \le \Delta_{\rm ff} \le 12$ from the bottom to top in top two panels 
	and in the reverse order in the bottom panel. The panels show the mean, std and fraction of the particles in each bin from top to bottom}
	\label{fig:growth_ff}
\end{figure} 
\begin{figure}
	\centering
	\centerline{\includegraphics[scale=0.45]{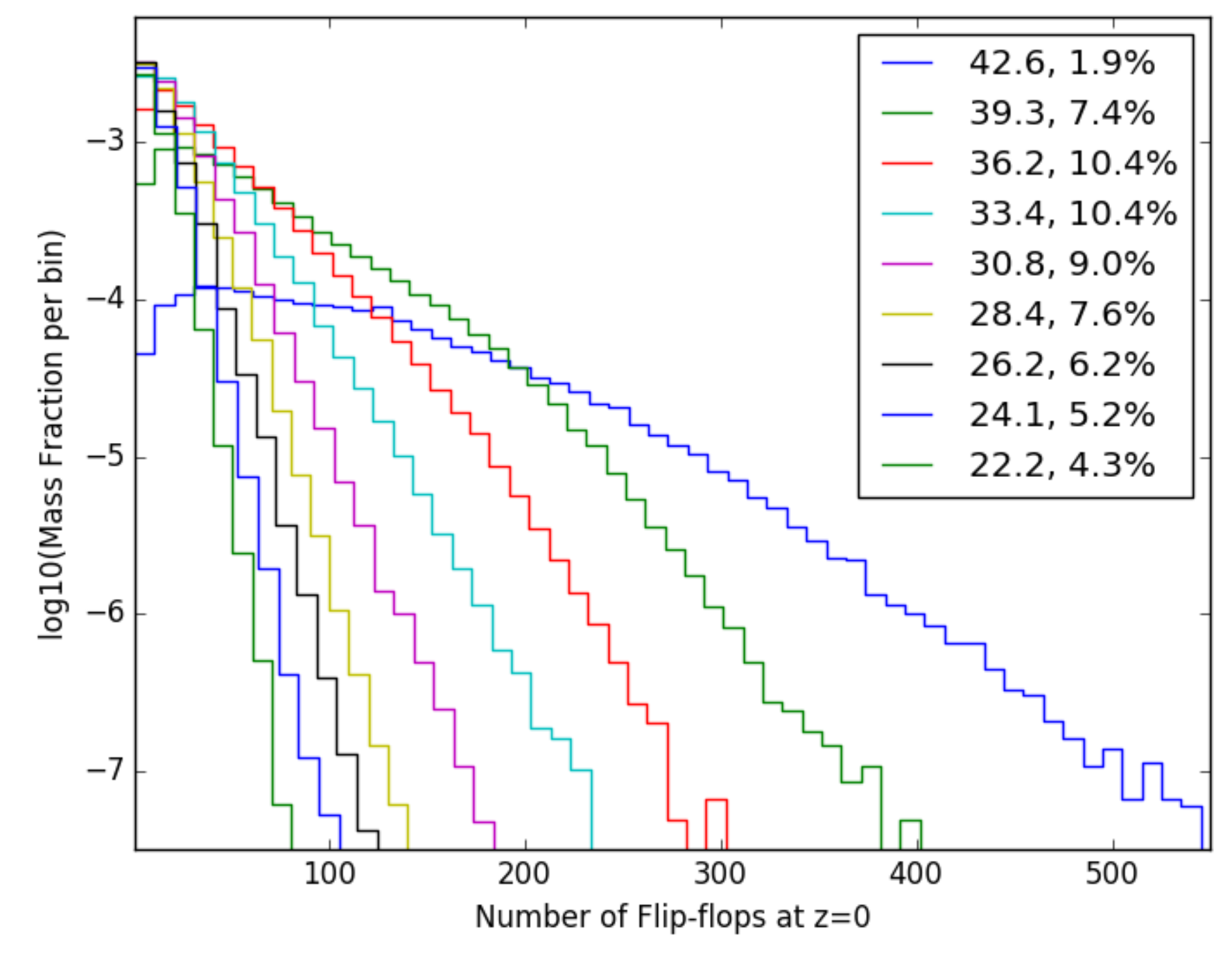}}    
	\caption{The logarithm of the fraction of mass in bins with the number of flip-flops at $z=0$ shown on the horizontal. 
	Nine  curves correspond to the particles  experienced first flip-flopping before redshift shown in the first column of the legend
	except those experienced flip-flopping before the previous redshift. 
	For example, the second (green) curve from the right corresponds
	to the particles that did not experience a single flip-flop before $z=42.6$ but experienced some by $z=39.3$.
	The second column in the legend  shows the total fractions of the mass in the particles satisfying the above
	criterion.}
	\label{fig:mass_frac}
\end{figure} 

\subsection{A unique feature of the evolution of the flip-flop field}  
Discussing a simple  one-dimensional halo  in  Introduction we argued that the closer 
the orbit of a fluid element to the center of the halo the shorter its characteristic time.
It means that the earlier the fluid element collapsed for the first time the greater its counts of flip-flops
at later epochs.
A similar rule is also valid in generic haloes with substructures in three-dimensions although it is not
exact but valid in statistical sense as we describe bellow. 
Figure \ref{fig:growth_ff} provides a quantitative illustration of this assertion.

For the plot we selected and binned particles according to their gains in flip-flops counts
$\Delta n_{\rm ff} = n_{\rm ff}(a_{i+1}) - n_{\rm ff}(a_i)$  where $a_i$ mark the output stages  of the 256$^3$ simulation. 
Then  we computed 
the mean number of flip-flops $n_{\rm ff}(a_i)$, standard deviation
$\sigma(n_{\rm ff}(a_i))$ and the fractions of the particles in thirteen bins with $0 \le \Delta n_{\rm ff} \le 12$.
Thirteen lines in the top panel show  ${\rm mean}[{n}_{\rm ff}(a_i)]$ for each bin in ascending order
of $\Delta n_{\rm ff}$ from the bottom to top. The middle panel shows $\sigma(n_{\rm ff}(a_i))$ for all bins, the order of bins
is same as in the top panel\footnote{Large fluctuations of $\sigma(n_{\rm ff}(a_i))$ for a couple of the highest bins
is caused by very limited statistics as the bottom panel shows.}.
The bottom panel shows the fraction of particles in every bin. Obviously the  lines in this panel  are in the reversed order 
of two top panels: the largest fraction of particles gets zero raise at all times (red line).
 
 Although the particles do not always stay in the same bin their mobility between bins is rather limited.
 The bins reasonably well separated  as $\sigma(n_{\rm ff}(a_i))$  in the middle panel show.
 In addition the particles in the higher bin have on average higher ${\rm mean}[{n}_{\rm ff}(a_i)]$ and higher 
 raise $\Delta n_{\rm ff}$.
 Thus the particles that gained superiority in the number of flip-flops at early nonlinear stages 
on average remain among the leaders at later times.  
Therefore we  conclude that the contrast between peaks of different
heights and valleys in the flip-flop landscape on average does not diminish with time.
However in N-body simulations some saddle points in the $n_{\rm ff}$  landscape may disappear  due to
discreteness effect and other sources of numerical noise.

Figure \ref{fig:mass_frac} provides an additional evidence for the conclusion of the previous paragraph. It shows 
the histogram of mass fractions in the bins of the final counts of flip-flops at $z=0$ in nine sets of particles. The first
set comprising about 1.9\% of all particles are particles experienced first flip-flops at $z >42.6$.  The corresponding
histogram is shown by blue curve on the right demonstrates the distribution of the flip-flop counts at $z=0$. 
The next histogram in green shows the final distribution of the numbers of flip-flops on particles that entered 
the nonlinearity, i.e. experienced flip-flops, at $39.3 <  z  < 42.6$, comprising 7.4\% of all mass an so forth until the group of
particles experienced first flip-flop at $22.2 < z < 24.1$ 
The corresponding redshifts and amounts are shown  in the figure legend in two columns.
The figure is clearly demonstrates that at $z=0$ the high counts of flip-flops, say $n_{\rm ff} > 50$, 
are dominated by the
particles experienced  first flip-flops at $z >36.2$. They make less than 20\% of the total mass or less than 22\% of
all particles experienced at least one flip-flop as Fig. \ref{fig:mass_frac} shows. These particles  obviously
make the tips of the highest peaks of the flip-flop field.

\subsection{The number of peaks}  
\begin{figure}
	\centering
	\centerline{\includegraphics[scale=0.45]{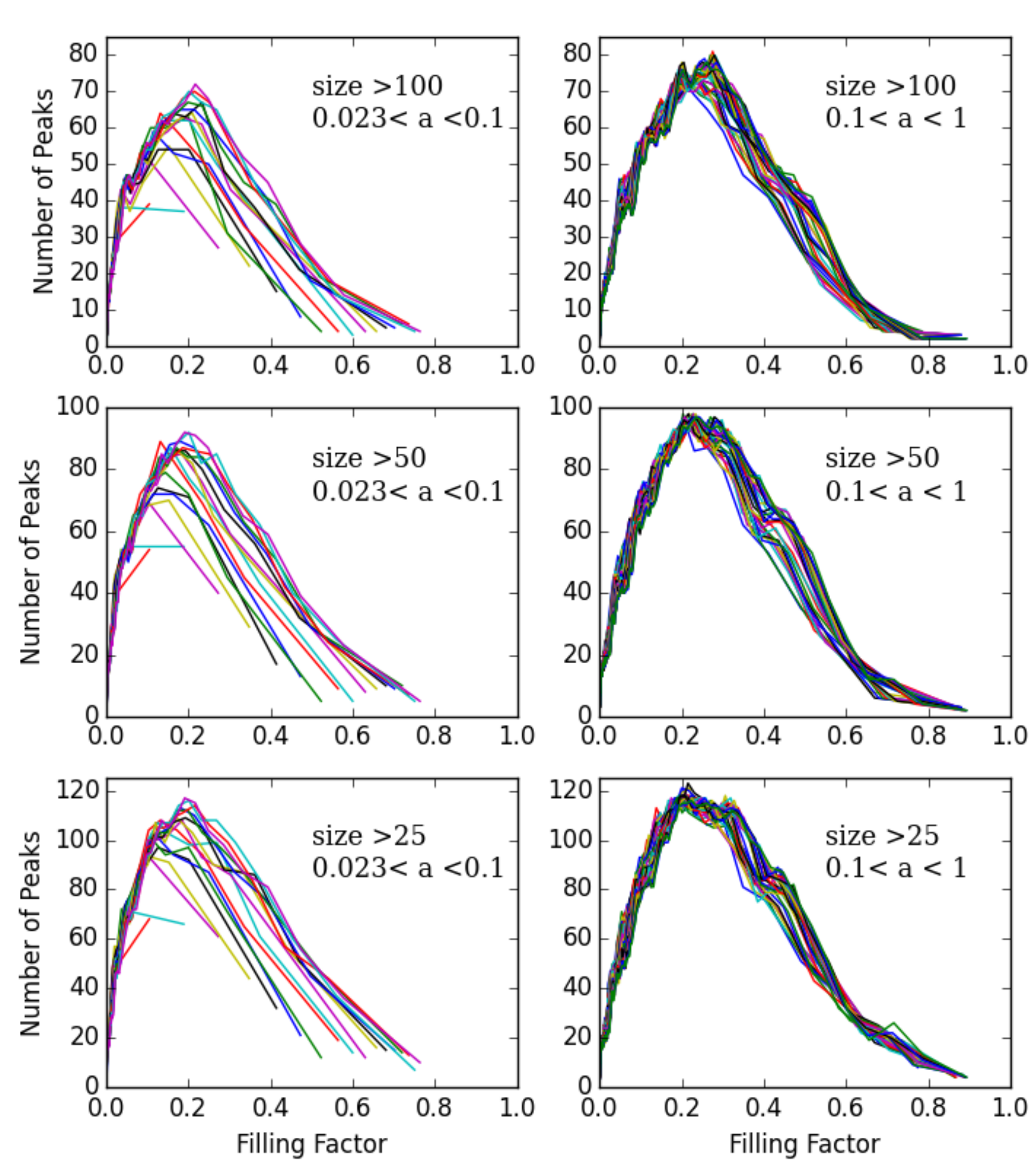}}
	\caption{ Number of peaks as a function of filling factor is plotted for every output of the simulation with $128^3$ particles.
	The counts for outputs at early stages $a< 0.1$ are shown on the left and the rest are shown on the right. 
	The counts are made for three thresholds of the size as shown in the panels.}
	\label{fig:num_peaks_128}
\end{figure} 

Now we will turn to the issue of the number of peaks in the flip-flop field. It is worth stressing that here we do not mean 
the points of  maxima of $ n_{\rm ff}$ in Lagrangian space 
but rather the regions of the field with the number of flip-flops above a certain level.  More exactly we will look at the
evolution of the excursion set in the flip-flop field at all fifty output epochs. The peaks and there structures 
represent the major interest because they are associated with the haloes and subhaloes.

\begin{figure}
	\centering
	\centerline{\includegraphics[scale=0.45]{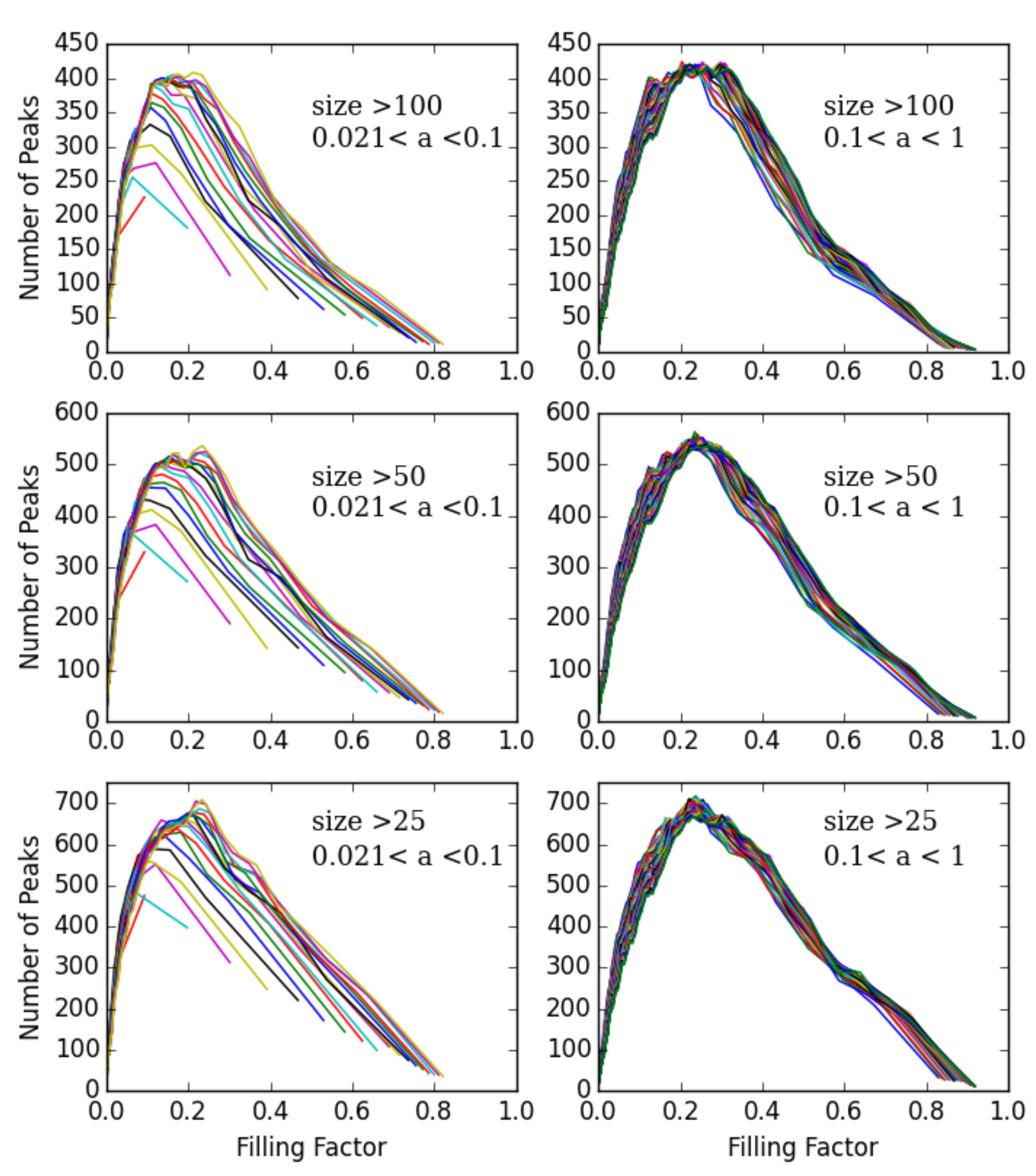}}
	\caption{ Same as in Fig. \ref{fig:num_peaks_128} but for the simulate with $256^3$ particles.}
	\label{fig:num_peaks_256}
\end{figure}

Figures \ref{fig:num_peaks_128}
and \ref{fig:num_peaks_256} show the number of peaks as a function of the filling factor at every output epoch in the $128^3$
and $256^3$ N-body simulations. We count peaks with three different sizes: $>100$, $>50$ and $>25$ particles. 
The left panels show earlier stages $a < 0.1$ characterized by relatively fast evolution 
and the right panels show later stages with $0.1 \le a \le 1$ when the evolution is considerably slower.  
The number of  peaks increases with the growth
of the filling factor as the threshold is decreasing and reaches a broad maximum around $FF \approx 0.2$. 
Then at greater filling factors 
many  peaks begin to  merge with each other and therefore their number decreases. 

It is remarkable that at $a > 0.1$
the curves corresponding to different scale factors $a$ are packed quite tightly forming a relatively narrow strip. 
This behavior is quite different from that at earlier stages $a <0.1$. 
There is no much difference between  the 128$^3$ and 256$^3$ simulations apart  from the number of peaks,
which is expected. Generally it is in agreement with the results described in previous sections and confirms
that the geometry of the flip-flop landscape does not evolve much at $a > 0.1$ 


\section{Comparison with  AHF haloes}  
\subsection{Illustration}  
\begin{figure}
\centerline{\includegraphics[scale=0.55]{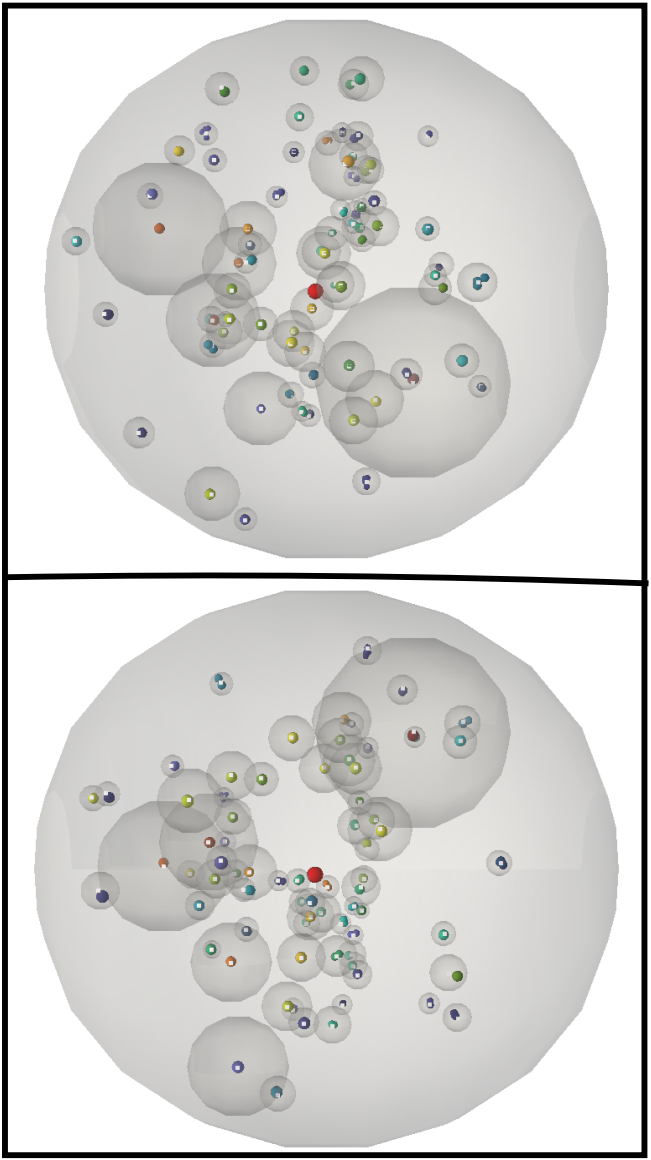}}
\caption{ Two mutually orthogonal views of one of the largest halo along with its subhaloes in the simulation found by the
AMF code. The haloes are represented by the spheres of the virial radii. Small colored spheres mark the particles
with the maximum value of the flip-flop field and white dots show the centers of the sphere, some of which may be
obscured by the respective colored spheres.
 }
\label{fig:amf_halo_subhaloes}
\end{figure}

As we indicated in Introduction there are many different methods and techniques used to identify and study
haloes, in particular DM haloes. We  compare the positions of the maxima of the flip-flop field a publicly available halo finder called Amiga Halo Finder or AHF
\citep{Knollmann_Knebe:09,Gill_etal:04} for comparison of some flip-flop peak properties
with AHF haloes. We are arguing that the peaks of the flip-flop field computed in Lagrangian space are directly related with
the DM haloes. If it is true then the maxima of the flip-flop field must be inside  of the AHF haloes.
In order check this we find the maxima of the flip-flop field in Lagrangian space and map them in Eulerian space.
Figure \ref{fig:amf_halo_subhaloes} shows two orthogonal projections of one of the largest haloes found by the AHF
in our simulation. The halo is shown as a sphere of the viral radius. It also shows the sabhaloes with more than a hundred
particles.  The centers of the virial spheres are shown as white dots. The particles with the maxima of the flip-flop field
are shown by small colorful spheres where both the radii and colors reflect the magnitude of the flip-flop maxima.
Colors from dark blue to red correspond to the ascending order of the heights of the maxima. Generally both the
colorful spheres and white dots are clearly seen in one or the other projection. However in a few cases the spheres 
obscure the centers of the virial spheres in both projections. The largest red sphere shows the particle with the
maximum value of flip-flops  corresponding to the virial center of the halo itself.

\subsection{Statistics}     
\begin{figure}
\centerline{\includegraphics[scale=0.45]{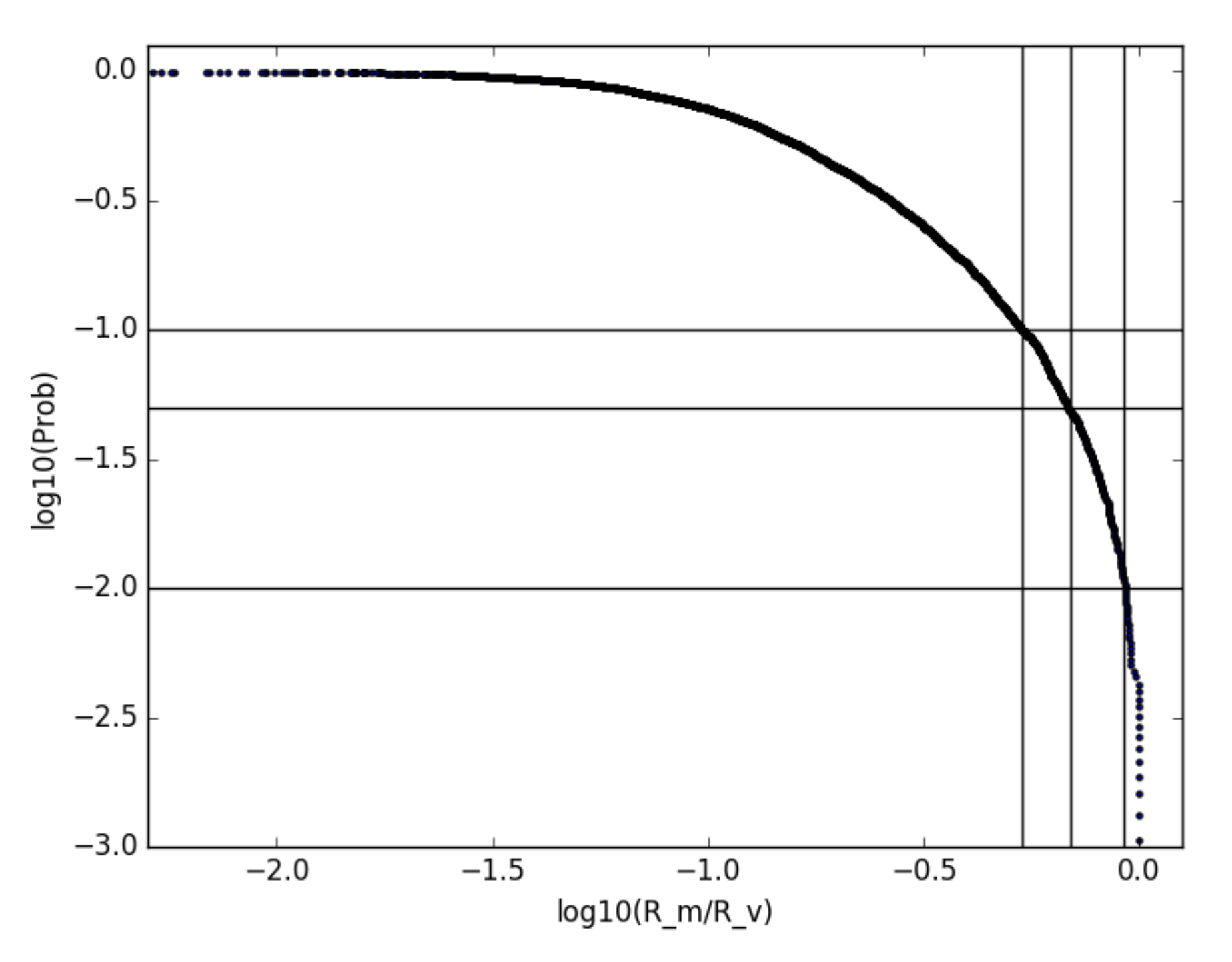}}
\caption{ Cumulative probability function of the ratio $R_{\rm max}/R_{\rm vir}$. Three horizontal lines show 1, 5 and 10\%
levels and the vertical lines show  the corresponding  values  of $R_{\rm max}/R_{\rm vir}$ which are
0.92, 0.69 and 0.54 respectively,  \ie for example, in 90\% of haloes the distances of the flip-flop maxima are
less than 0.54 of the virial radius. 
 }
\label{fig:prob_ratio}
\end{figure}
In order to quantitatively address the question of  how close are the maxima of the flip-flop field to the virial centers of the AHF
haloes we computed the cumulative probability function of the ratios $R_{\rm max}/R_{\rm vir}$ for haloes or subhaloes
with more than thirty particles.
First we selected all the particles with the largest value of flip-flop, i.e. the global maximum in each set of particles 
comprising a halo or subhalo identified by the AFH  algorithm. 
In most of haloes it was just a single particle, however since the number of flip-flops is integral on rare occasions
-- mostly in haloes with fewer than a hundred particles -- more than one particle had the largest count of flip-flops.
Then for each particle with the maximal count of flip-flops we computed the distance from the virial center of the halo or subhalo. 
The  cumulative probability function of the ratios $R_{\rm max}/R_{\rm vir}$ is displayed in Fig. \ref{fig:prob_ratio}.
For convenience  we show three horizontal lines marking 1, 5 and 10\% of particles  with highest values of the
ratio $R_{\rm max}/R_{\rm vir}$. The vertical lines show the corresponding values of the ratio. 
Thus  $R_{\rm max}/R_{\rm vir} < 0.54, 0.69, 0.92$ in 90\%, 95\% and 99\% of the haloes respectively.

\begin{figure}
\centerline{\includegraphics[scale=0.5]{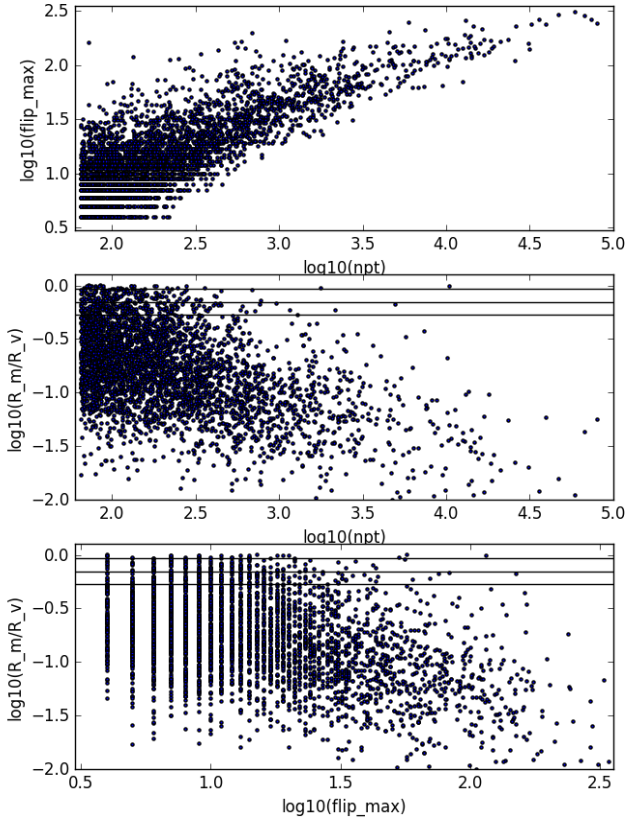}}
\caption{ Two top panels show the flip-flop maxima  and the ratio $R_{\rm max}/R_{\rm vir}$ respectively vs the number of 
particles  in selected AMF haloes. The bottom panel shows the ratio $R_{\rm max}/R_{\rm vir}$ vs the maximum of the
flip-flop field in the haloes. Horizontal lines in two lower panels correspond to the vertical lines in Fig. \ref{fig:prob_ratio}, i.e. 
$R_{\rm max}/R_{\rm vir} = 0.92, 0.69, 0.54$
 }
\label{fig:npt-flip-max-ratio}
\end{figure}

In order to show how the positions of the maxima of flip-flops, their  values and masses of haloes correlate we provide the
scatter plots of three pairs of these parameters in Fig. \ref{fig:npt-flip-max-ratio}. The top panel shows the correlation of 
the  maxima with the number of particles in the halo. For the haloes with more than about 200 particles the tendency 
that the larger the halo the higher maximum of the flip-flop in the halo is unambiguously clear.  The tendency
of the ratio $R_{\rm max}/R_{\rm vir}$ to decrease with the growth of the halo mass is also quite clear shown in
the middle panel. The bottom panel exhibits similar tendency with the growth of the flip-flop maximum of the halo at
$ff_{\rm max} \gtrsim 20$, which qualitatively follows  from the correlations displayed in the top and middle panels.
For haloes with $N_p \lesssim 200$ or/and $ff_{\rm max} \lesssim 20$ these correlations are not observed.

The horizontal lines in the middle and bottom panels correspond to the vertical lines in Fig. \ref{fig:prob_ratio}.
Although they show that the largest ratios of $R_{\rm max}/R_{\rm vir}$ are characteristic for small haloes with low maxima of flip-flops but the most of them have a clear maximum of the flip-flop field within a sphere centered on the virial center
and the radius less than about a half of the virial radius.
It is worth stressing that the haloes with fewer than a hundred particles are probably seriously affected by numerical noise
in our simulations.

\section{The evolution of the highest flip-flop peak}  

\subsection{Statistics of subpeaks in one of the largest peaks} 
First we look at the dependence of the number of subpeaks defined as peaks within a parent peak.
The parent  peak is selected as  a compact region at some threshold $n_{\rm p, th}$.
Then we continuously elevate the threshold $n_{\rm th}$ up to the highest maximum  within the peak
and identify distinct subpeaks with sizes greater than selected threshold at each level $n_{\rm th}$. 
We repeat this procedure  three times for three size threshods: 100, 50 and 25 particles.
The bottom panel of Fig. \ref{fig:num_subs_maxs} shows this dependence
 in the $256^3$ simulation. Three lines show the effect of the size threshold as indicated by the legend. 
 For comparison and as an illustration of the effect of the small scale cutoff in the power spectrum 
 of the initial density perturbations the top panel shows the substructure
 in the $128^3$ simulation\footnote{We are reminding that the common part of the Fourier amplitudes was the same in both simulations}.
As expected the number of substructures increases with the size of the simulation box and decreases with the growth 
of the size threshold. This is in a qualitative agreement with Fig. \ref{fig:num_peaks_128} and \ref{fig:num_peaks_256}.

\subsection{Selection of subpeaks}  
\begin{figure}
\includegraphics[scale=0.55]{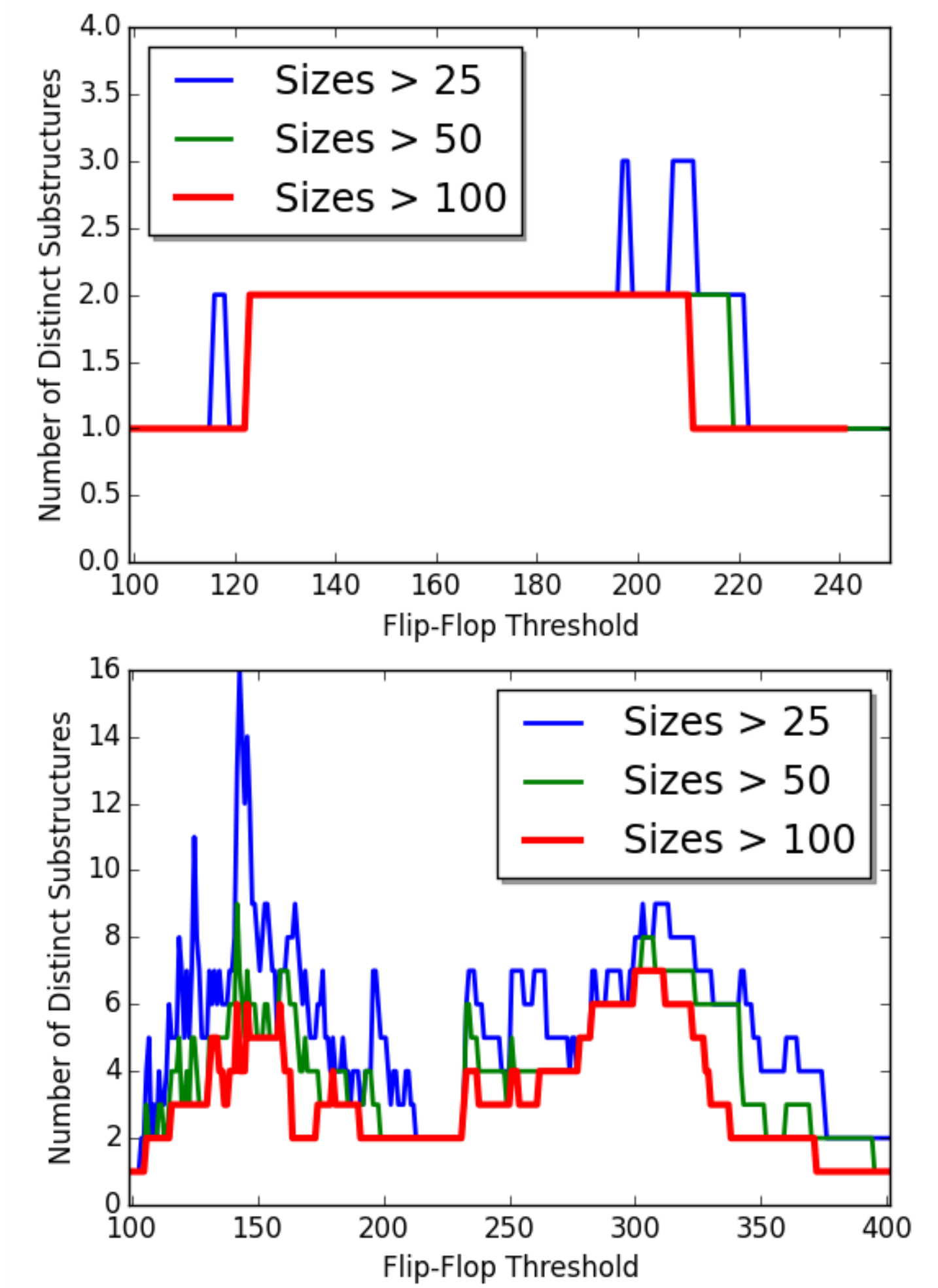}
\caption{ Number of subpeaks within the largest flip-flop peak as a function of the threshold. 
The top and bottom panels correspond to the $128^3$ and  $256^3$ simulations respectively. 
The lines of different colors show the dependence on the size threshold as indicated by the legend.}
\label{fig:num_subs_maxs}
\end{figure}

In order to make an unambiguous  case  we consider only substractures selected
by  two conditions: one by their sizes in Lagrangian space that can be quantified by the number of   
particles $n_{\rm ff} \ge n_{\rm ff, thr} = 100$   and the other by their heights $n_{\rm p} \ge n_{\rm p, th} = 100$. 
Therefore we identify individual substructure within the parent peak at four  thresholds: $n_{\rm ff} = 150, ~240, ~270, ~{\rm and} ~ 300$.
The choice is based on  the red curve in the bottom panel of   Fig. \ref{fig:num_subs_maxs}. 

\begin{figure}
\includegraphics[scale=0.4]{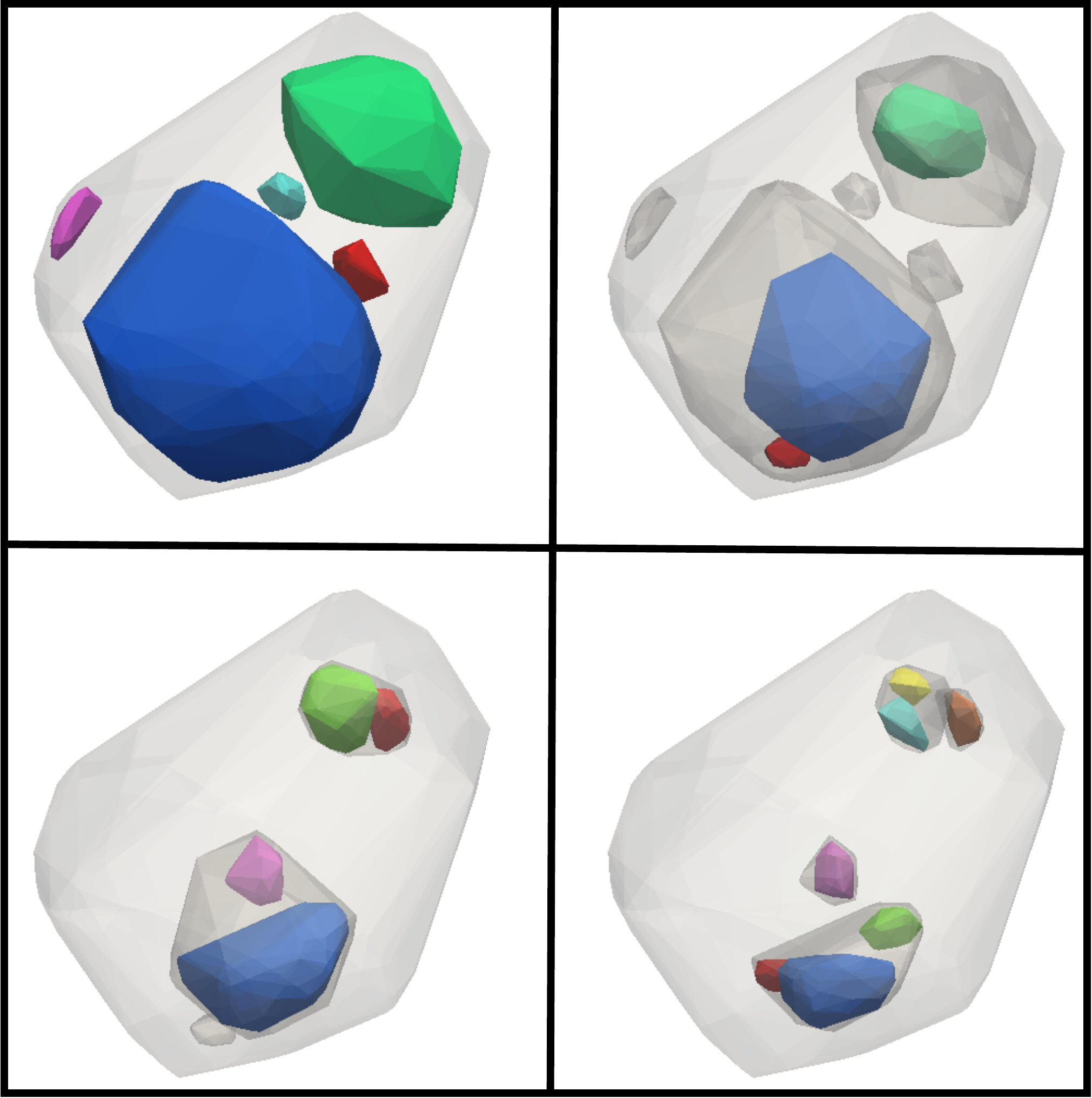}    
\caption{ \small Contours of the flip-flop field show five levels of hierarchical structure of the largest flip-flop peak 
in the $256^3$ simulation. The top left panel shows the convex hull of the peak identified as a connected region
with  $n_{\rm ff} \ge100$ in gray and five distinct substructures at $n_{\rm ff} \ge150$ in color.
In the remaining three panels the contour $n_{\rm ff} =100$ is displayed in light gray.
	The top right panel displays the convex hulls at  $n_{\rm ff} \ge 150$ in light pink that are shown in the previous panel
	 in color and  three substructures at  $n_{\rm ff} \ge240$ in color. The bottom subplots show three contours 
	 at $n_{\rm ff} = 240$ in light pink and four  at $n_{\rm ff} = 270$ in color on the left 
	 and four contolurs at $n_{\rm ff} = 270$ in light pink and seven  at $n_{\rm ff} = 300$ in color.
	Colors in order blue, dark cyan, green, light brown, and dark brown correspond to the mass of the substructures
	in descending order . }
\label{fig:substr3d_L}
\end{figure}

The flip-flop subpeaks in Lagrangian space are displayed in Fig. \ref{fig:substr3d_L}. 
The parent peak in three-dimensional space represents a nesting structure 
resembling a generalized Russian doll or 'matreshka'-doll.
 Combining all five levels of  this substructure in one plot significantly obscures the complex geometrical 
pattern of the largest peak. Therefore we show the substructure in four steps each of which displays only two levels
of the nesting structure: the low level of  substructure is shown as  gray surfaces and the higher level as 
color surfaces. Surfaces are the convex hulls of the peaks. 
Color scheme is same in every step: the colors (blue, magenta, cyan, green, yellow, and red)
correspond to the sizes/masses  of subpeaks in descending order. Thus the top left portion of  Fig. \ref{fig:substr3d_L}
shows levels (1,2), top right -- levels (2,3), bottom left -- levels (3,4), and bottom right - levels (4,5). The sizes and
masses of the peaks at every level are given in Tables 1 and 2.

It is worth stressing that  Fig. \ref{fig:substr3d_L} does not show the entire substructure of the largest peak.
Four thresholds  were selected on the basis of the red line in the bottom panel of Fig. \ref{fig:num_subs_maxs}
as a few representative examples of substructure.
The three-dimensional illustration is shown in Fig. \ref{fig:substr3d_L} and their parameters are given in Table 1 and 2
which  display the number of particles  and the masses respectively

\begin{table}{ Table 1. Number of particles in substructures shown in Fig. \ref{fig:substr3d_L}}\\[1ex]
\begin{tabular} {crrrrrrr}
\hline
flip-flop &$N_1$ &$N_2$ &$N_3$ &$N_4$ &$N_5$ &$N_6$ &$N_7$\\
threshold & \\
\hline
100 &206670 & & & & \\
150 & 61960 & 17035&   513&   308&   265 \\
240 & 8698 & 3960 & 211 \\
270 &4363& 1474&  753&  521 \\
300&1976 & 416&  405&  381&  286&  267&  178 \\
\hline
\end{tabular}
\end{table}
\begin{table}{  Table 2. Approximate masses of  substructures shown in Fig. \ref{fig:substr3d_L} in  units of  $10^6 M_{\bigodot}$}\\[1ex]
\begin{tabular} {crrrrrrr}
\hline
flip-flop &$M_1$ &$M_2$ &$M_3$ &$M_4$ &$M_5$ &$M_6$ &$M_7$\\
threshold & \\
\hline
100 &1464. & & & & \\
150 & 439. & 121.&   3.63&   2.18&   1.87 \\
240 & 61.6 & 28.1 & 1.50 \\
270 & 30.9  & 10.4&  5.34&  3.70 \\
300& 14.0 & 2.94&  2.87&  2.70&  2.03&  1.89&  1.26 \\
\hline
\end{tabular}
\end{table}

\subsection{Illustration of the evolution of the largest  flip-flop peak}  
\begin{figure}
\centerline{\includegraphics[scale=0.24]{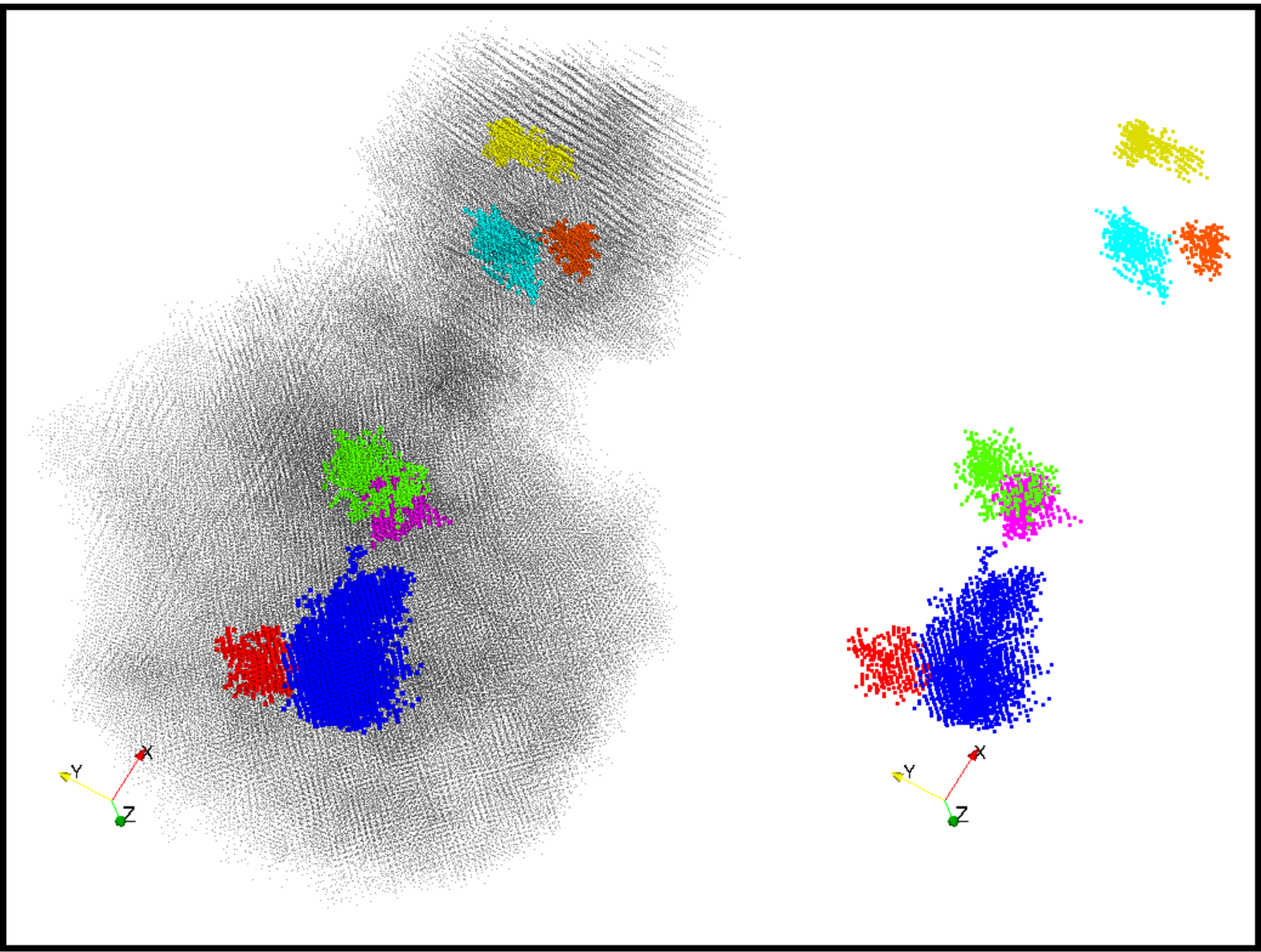}}
\centerline{\includegraphics[scale=0.24]{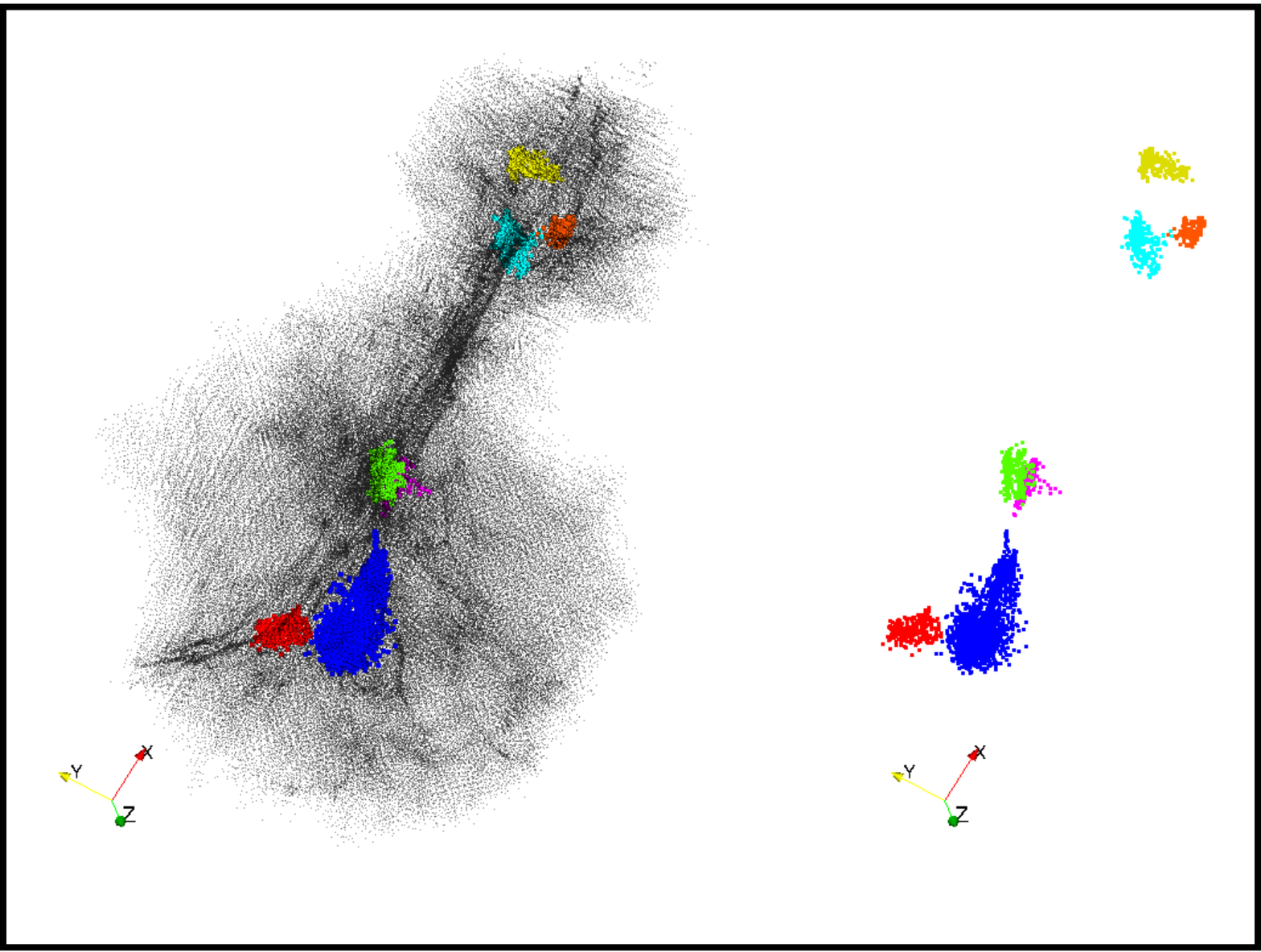}}
\centerline{\includegraphics[scale=0.24]{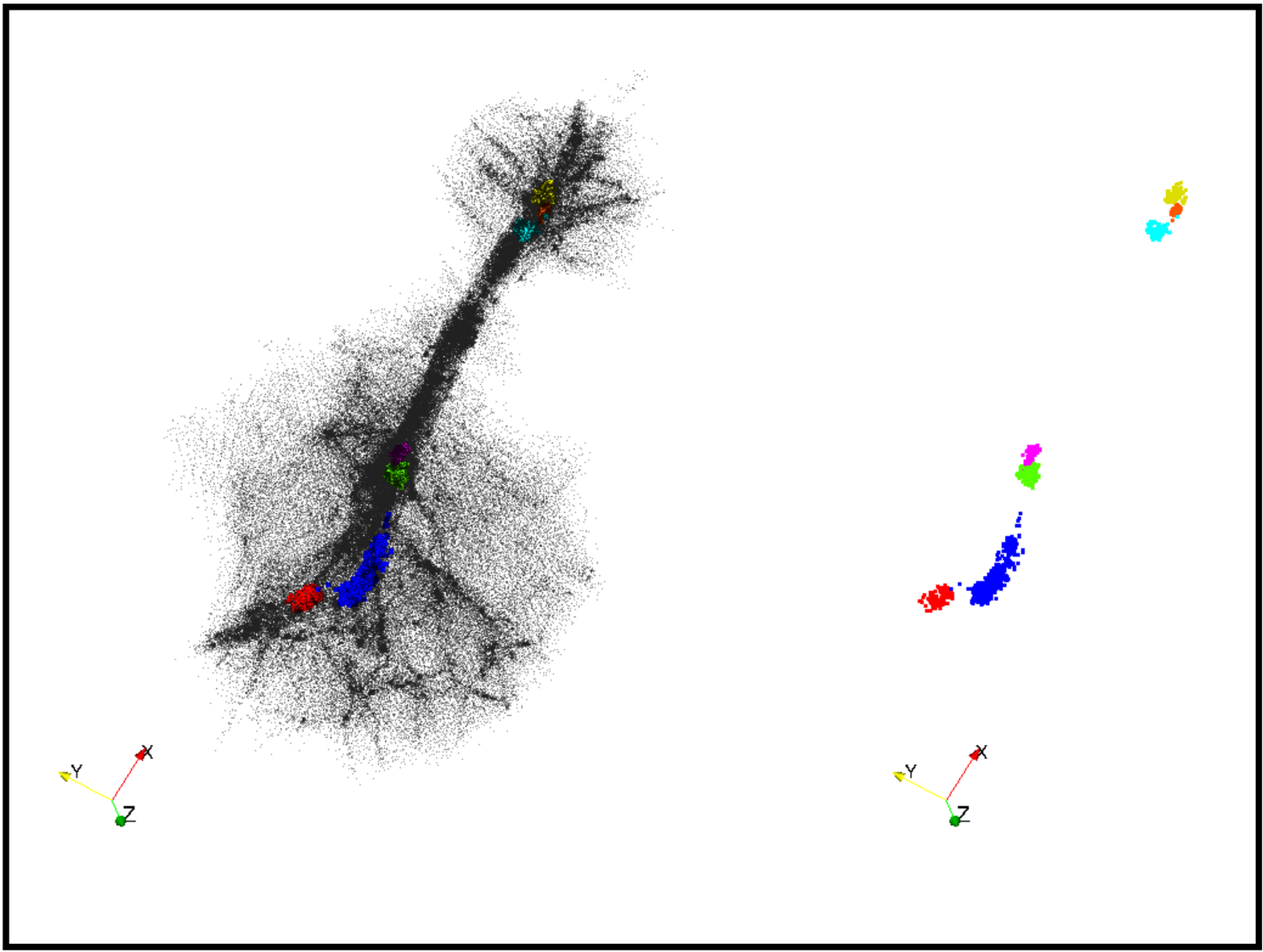}}
\centerline{\includegraphics[scale=0.24]{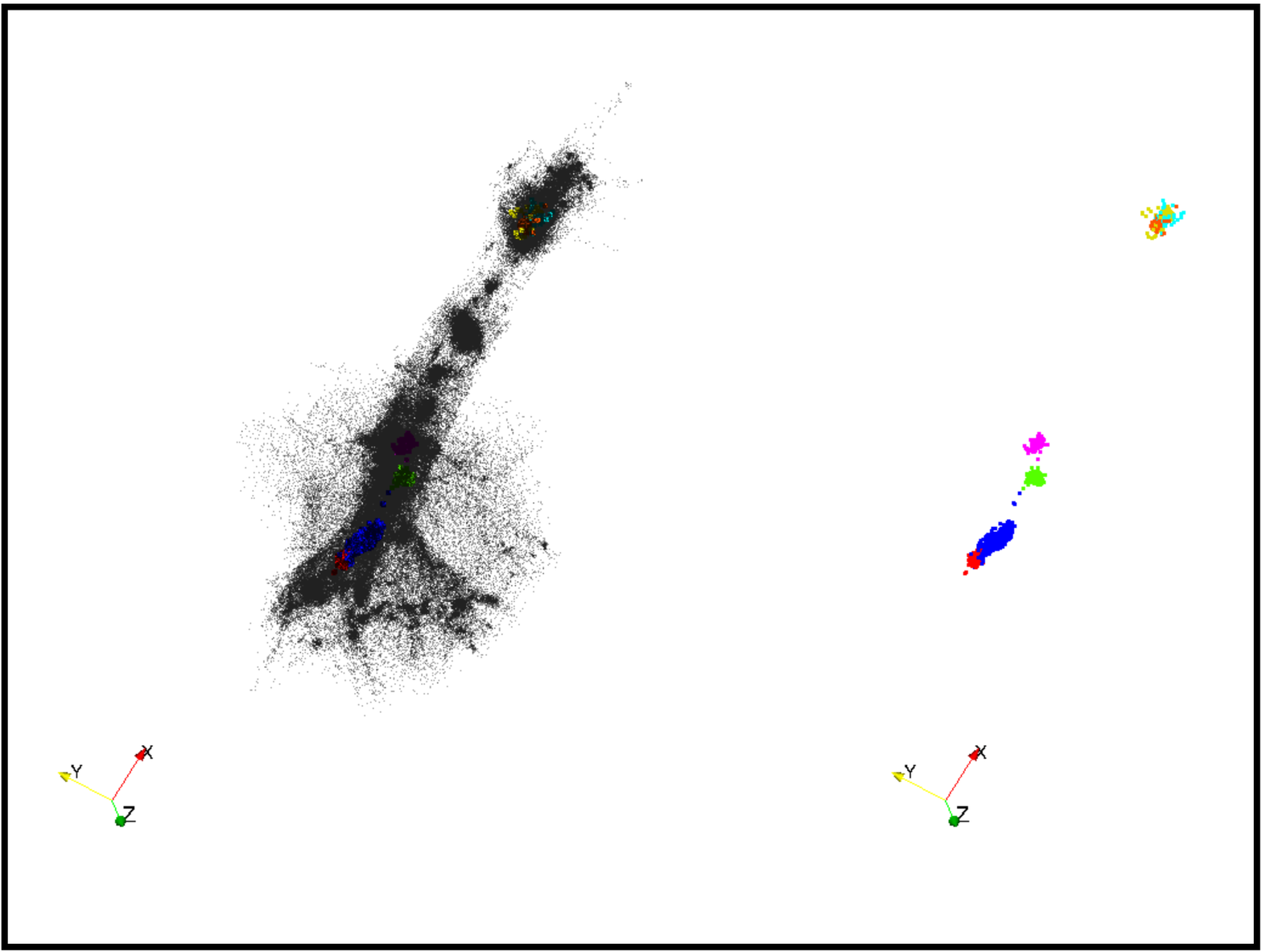}}
\caption{The figure illustrates the evolution of the whole halo and its substructure shown in color in the bottom right panel of 
Fig. \ref{fig:substr3d_L}. The plots on the left show all particles in the halo in gray and the particles of the substructure are shown
in  color on both sides. The stages from the top to bottom corresponds to  $z= 50,~ 42.6,~ 39.3,~ 36.2$ or respectively
$a = 0.020,~ 0.023,~ 0.025,~ 0.027$. All plots have the same scale.}
\label{fig:substr_evol_1}
\end{figure}

\begin{figure}
\centerline{\includegraphics[scale=0.25]{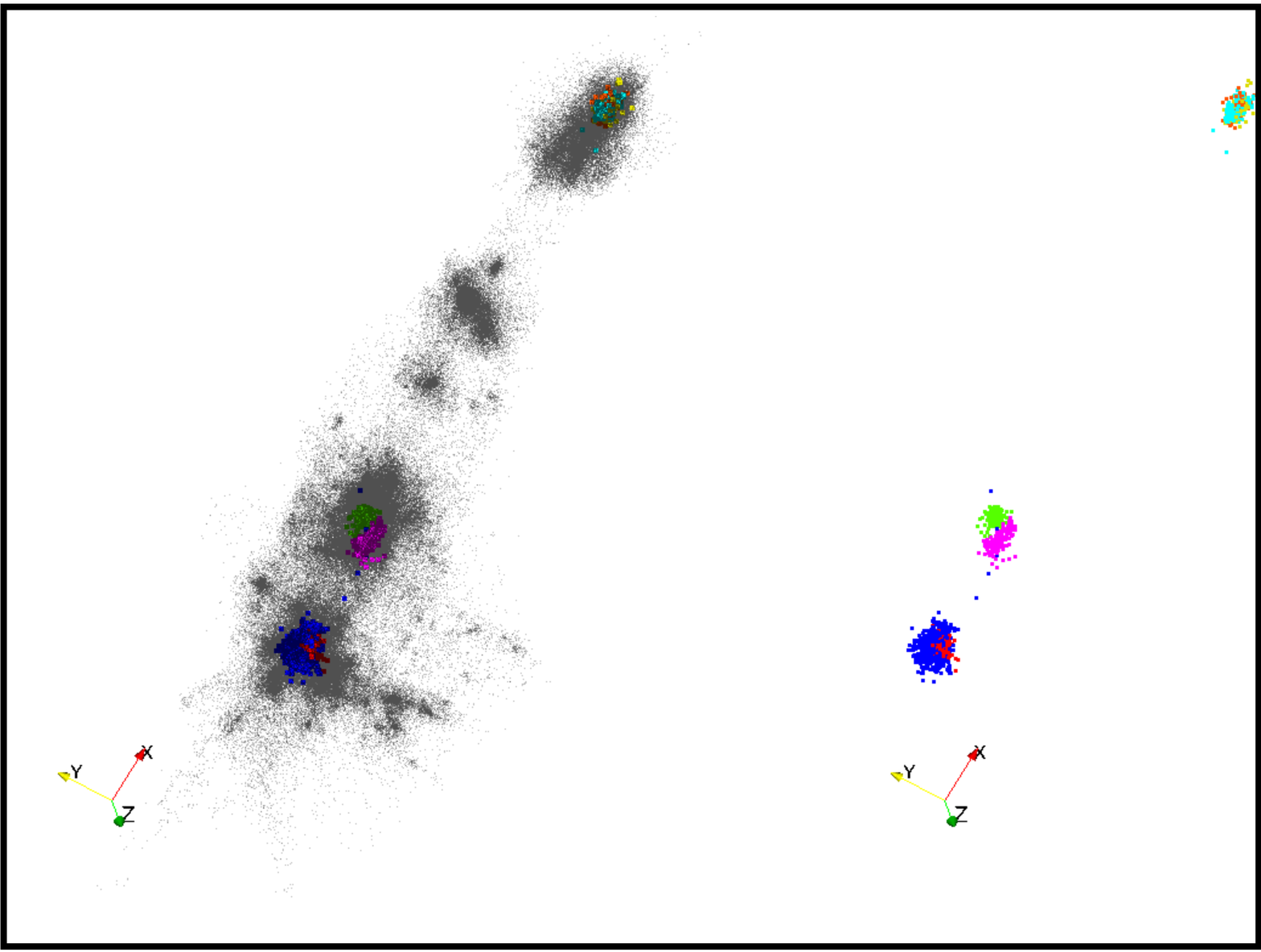}}
\centerline{\includegraphics[scale=0.25]{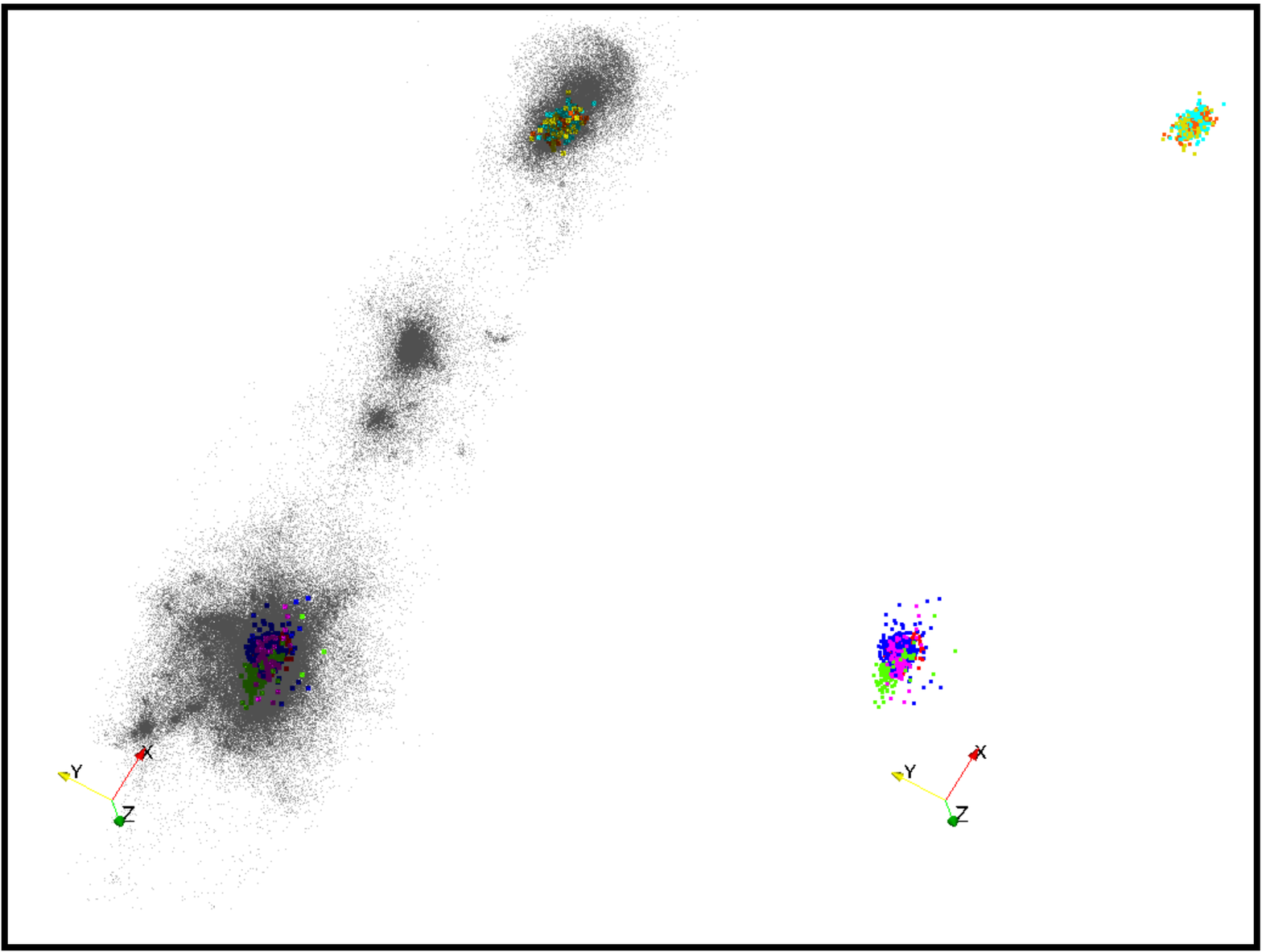}}
\centerline{\includegraphics[scale=0.25]{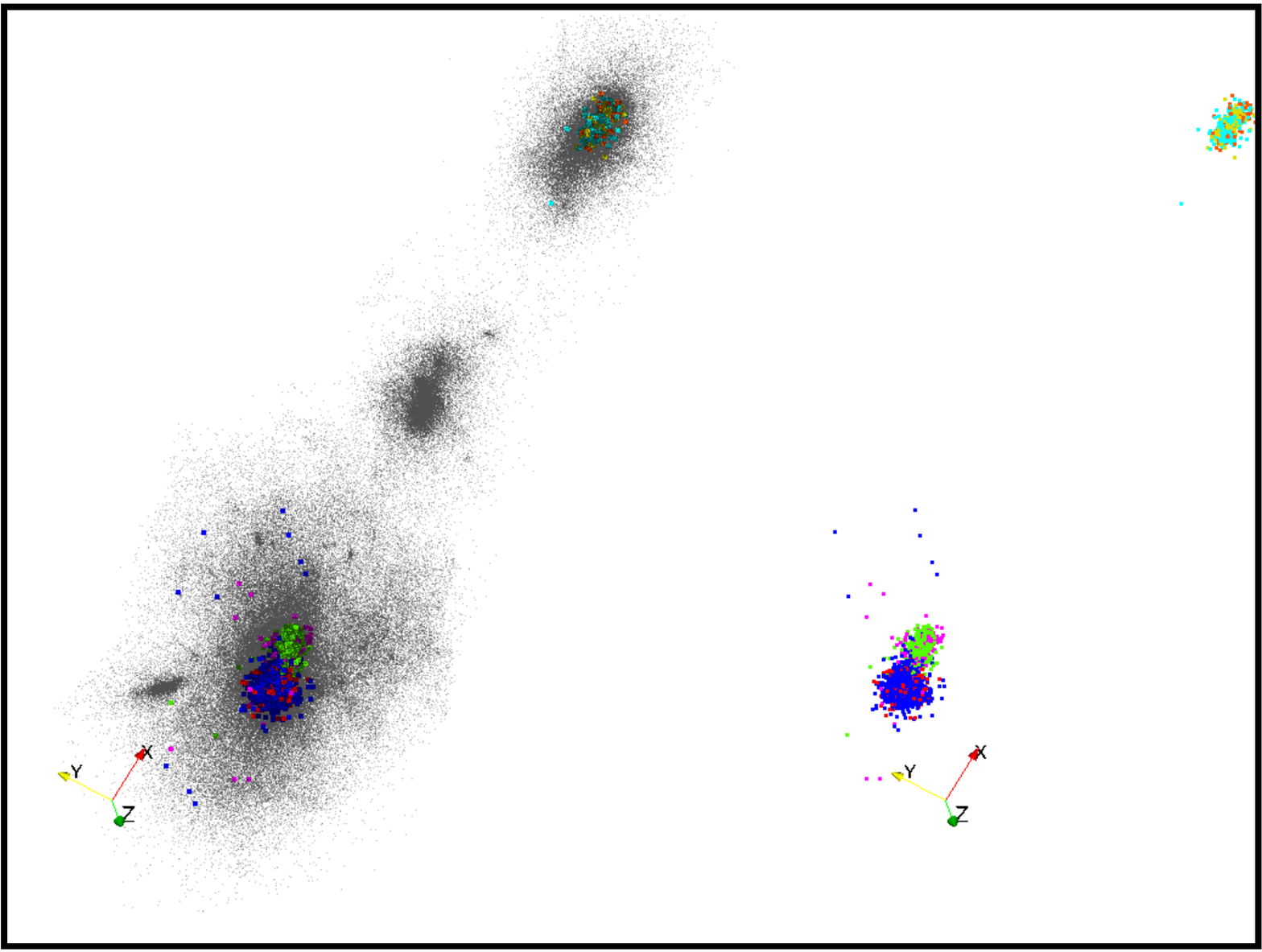}}
\centerline{\includegraphics[scale=0.25]{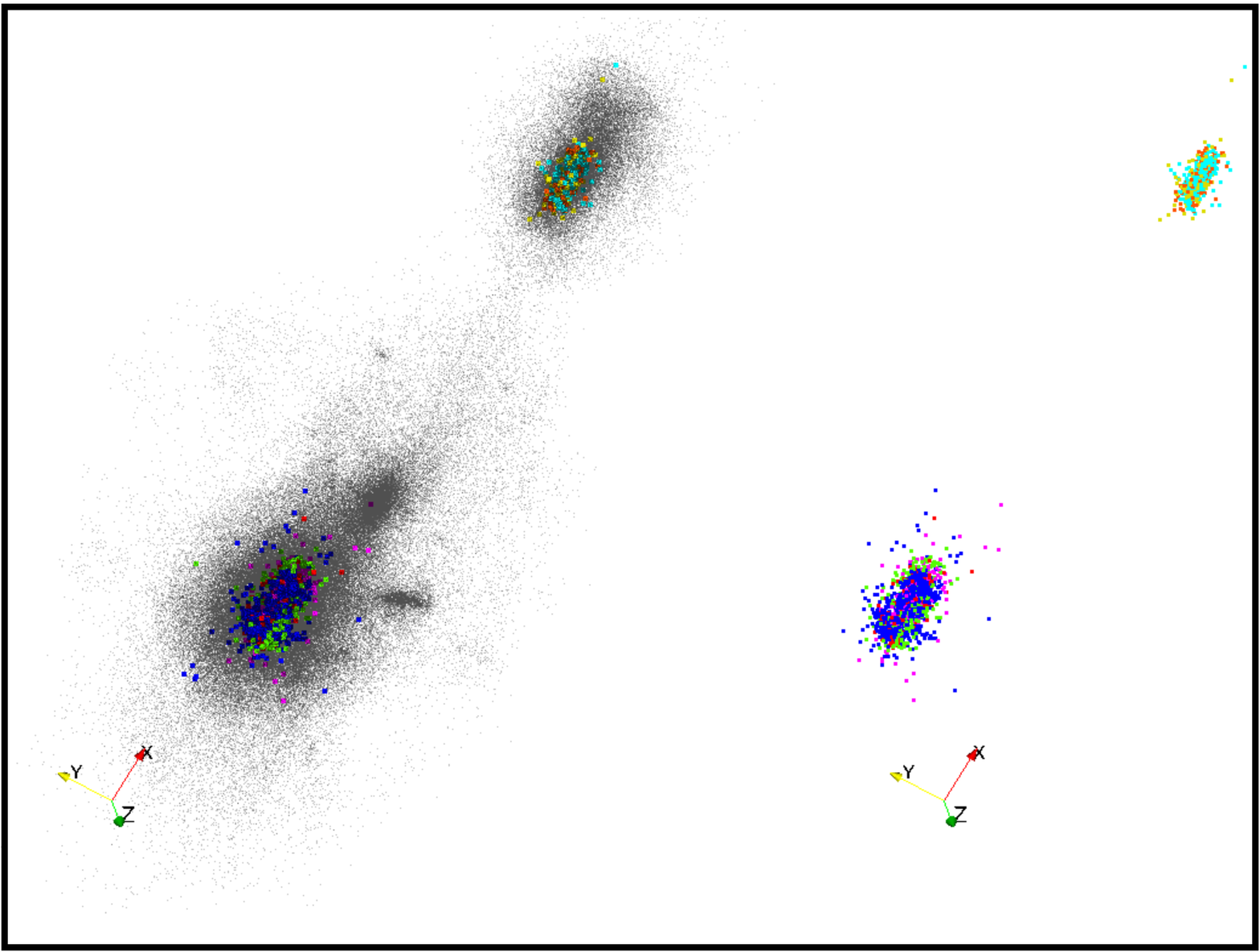}}
\caption{Same as Fig. \ref{fig:substr_evol_1} except the stages  are  $z= 33.4,~ 30.8,~ 28.4,~ 26.2 $
or respectively $a = 0.029,~ 0.031,~ 0.034,~ 0.037$.
In order to improve the visibility of subhaloes the plots are progressively zoomed in with time.}
\label{fig:substr_evol_2}
\end{figure}

\begin{figure}
\centerline{\includegraphics[scale=0.26]{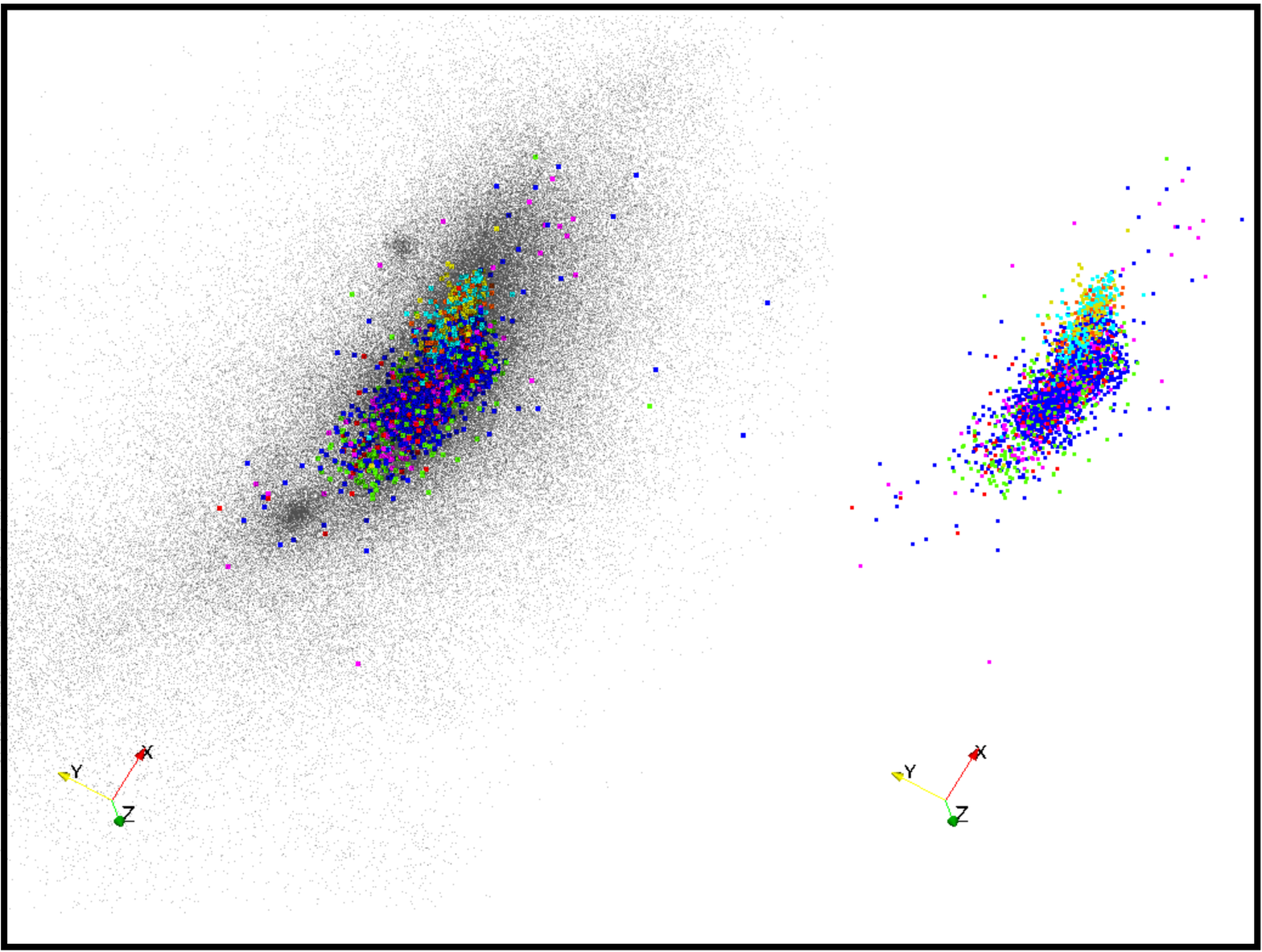}}
\centerline{\includegraphics[scale=0.26]{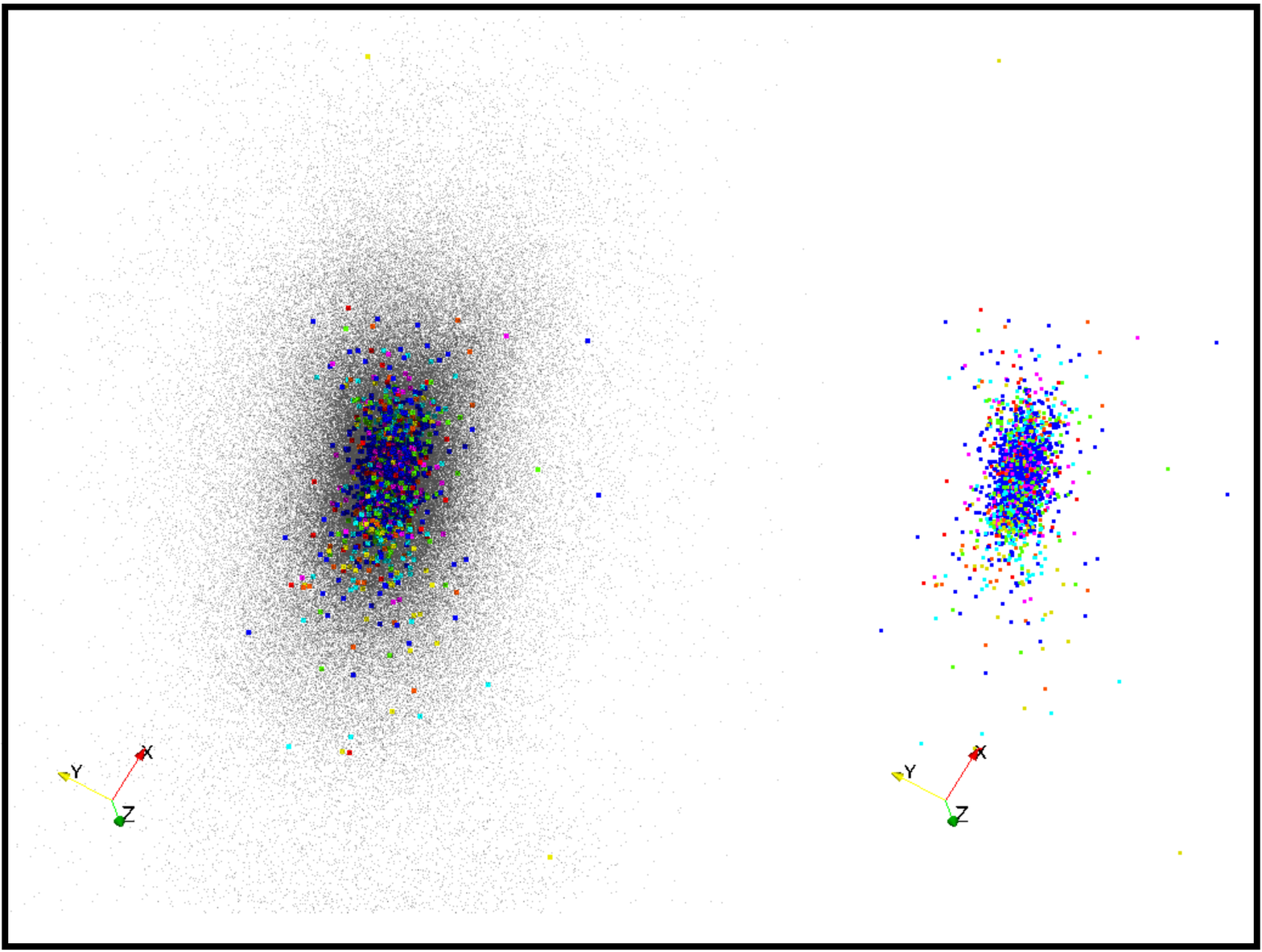}}
\centerline{\includegraphics[scale=0.26]{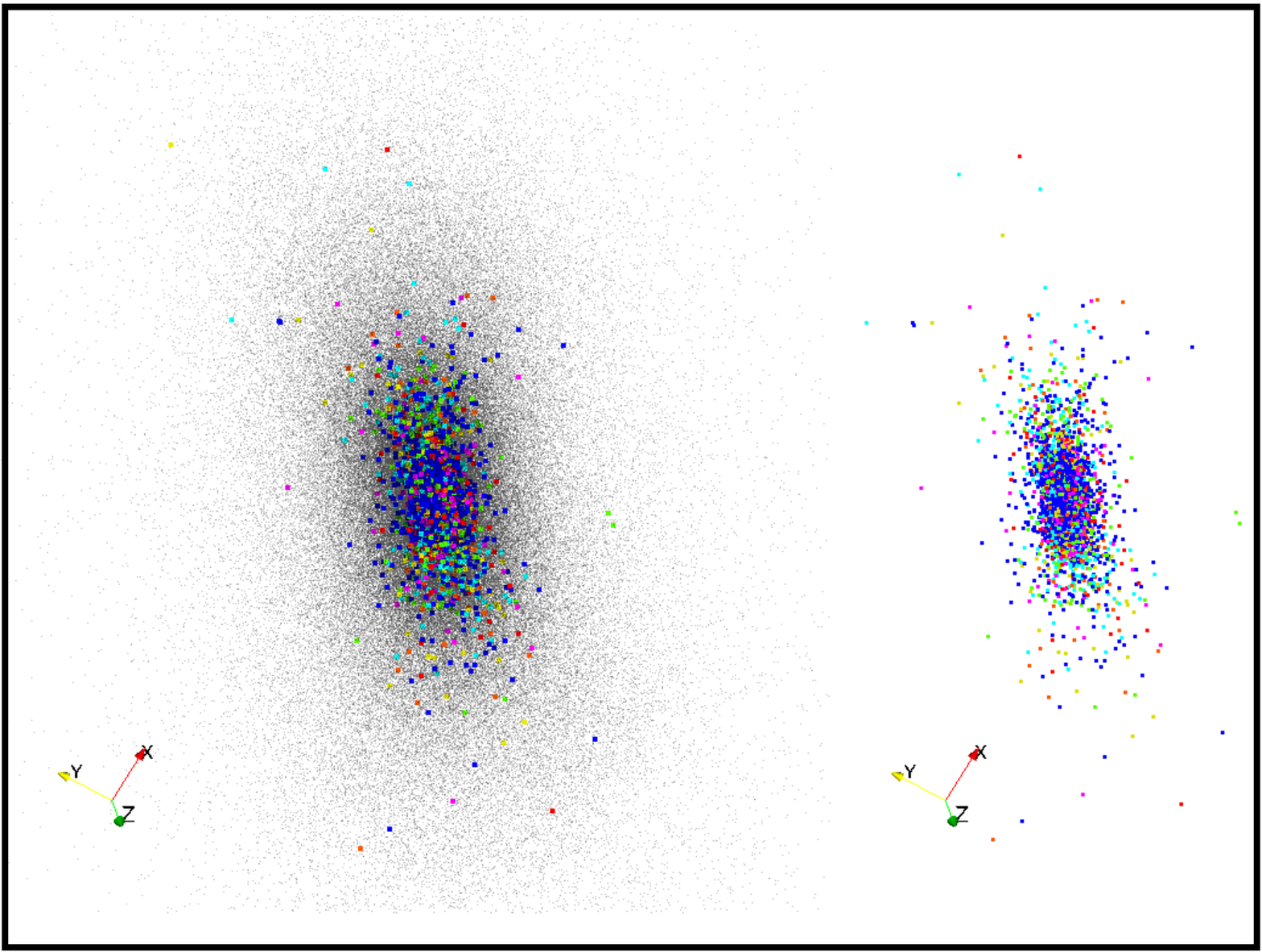}}
\centerline{\includegraphics[scale=0.26]{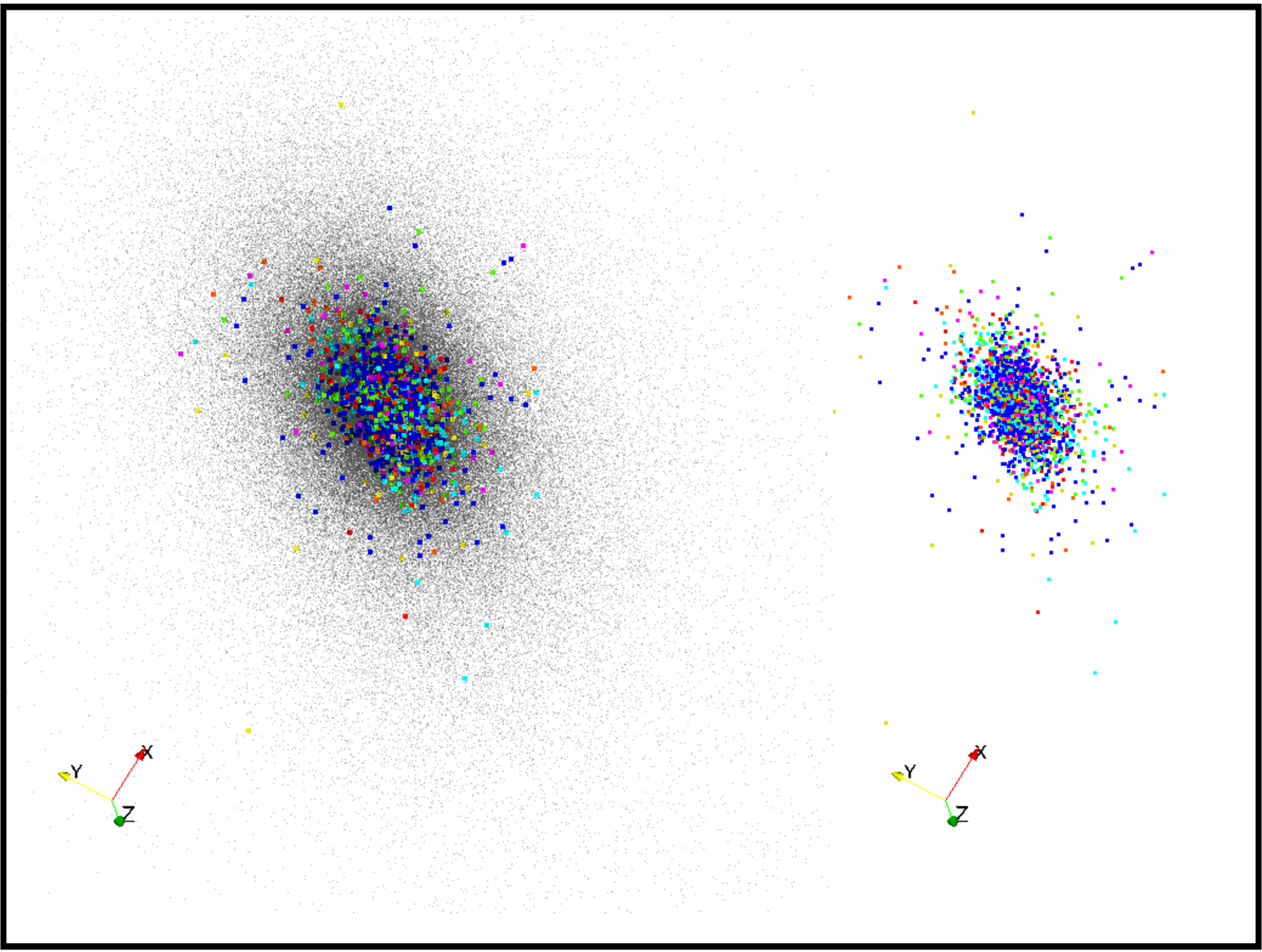}}
\caption{Same as Fig. \ref{fig:substr_evol_1}  except the stages  are  $z= 24.1,~ 9.58,~ 3.82,~ 0 $
or respectively  $a = 0.040,~ 0.094,~ 0.21,~ 1.0$.
In order to show subhaloes the plots are arbitrary zoomed in.}
\label{fig:substr_evol_3}
\end{figure}

In this section we will closely follow the evolution of seven highest peaks selected by the condition
$n_{\rm ff}\ge 300$ at $z=0$.
We demonstrate that these peaks of the flip-flop field become  small  individual haloes at early times.
Then we  show how they merge with each other in a hierarchical process of assembling the parent halo.
Finally we illustrate how they evolve after they all reside  in the central region of the cloud formed by the particles of the 
parent peak selected  by the condition $n_{\rm ff}\ge 100$. It is worth stressing that
only particles corresponding to the bottom right panel of Fig. \ref{fig:substr3d_L} retain their colors
in the plots described in this section
(i.e. Figs. \ref{fig:substr_evol_1}, \ref{fig:substr_evol_2} and \ref{fig:substr_evol_3}).
 All the rest particles  are in black.
According to Table 1 the total number of particles in these peaks  is only 3,909 which makes
less than 2\% of the parent peak comprised of  206,670 particles. 
Thus we consider a tiny fraction of the collapsed region within a gray surface in the top left corner of Fig. \ref{fig:substr3d_L},
or more accurately as a cloud of particles in the top left panel of  Fig. \ref{fig:substr_evol_1}\footnote{For the purpose of
better visualization the boundary surfaces of the peaks in Fig. \ref{fig:substr3d_L}  are approximated by the convex hulls.
No calculation used this approximation.}.
It is worth emphasizing that it makes only the central part of the largest halo.
The whole halo is even more massive since there are particles with $n_{\rm ff} < 100$ in the region in question.

 We concentrate on this small set of particles because 
they reside in the  highest density environment and thus have experienced the most vigorous dynamical evolution.
The major question we are interested in is what is the final state of  particles in the selected seven peaks 
of the flip-flop field. In particular is it feasible to identify them as distinct structures in Eulerian space
at $z=0$. 

Figures \ref{fig:substr_evol_1}, \ref{fig:substr_evol_2} and \ref{fig:substr_evol_3} show a sequence of dot plots 
illustrating the evolution of selected particles from $z=50$ to $z=0$ or respectively from $a=0.02$ to $a=1$.  
Each figure shows the structure formed by these particles at four redshifts indicated in the captions. 
It is no surprise that the colorful particles find themselves in the central parts of the cloud of black particles
and thus become obscured very quickly. In order to see the structures formed by them more clearly
each of four panels in every figure consists of two parts: all particles are plotted on the left hand side and
only colorful particles are shown on the right hand side. Please note that the orientation in Fig. \ref{fig:substr3d_L}
 is different from that in figs.\ref{fig:substr_evol_1}, \ref{fig:substr_evol_2} and \ref{fig:substr_evol_3} where it 
 is the same. The orientations in Lagrangian and Eulerian spaces were chosen with the goal of minimizing 
 obscuration due to projection. 

The top panel of Fig. \ref{fig:substr_evol_1} shows the initial state at $z=50$. The shape of the cloud of black particles in 
Fig. \ref{fig:substr_evol_1} is more accurate than the gray surface in the top left corner of Fig. \ref{fig:substr3d_L} because
the latter is the convex hull of the cloud in Fig. \ref{fig:substr_evol_1}. Except for this particular case other peaks are of remarkably 
convex shapes and thus approximating them by convex hulls in Fig. \ref{fig:substr3d_L} is quite accurate. Convex hulls 
require fewer surface elements and thus improve visibility of nesting surfaces depicted in Fig. \ref{fig:substr3d_L}.

The top panel in Fig. \ref{fig:substr_evol_1} shows Lagrangian space. The second from the top panel  shows 
the first output ($z=42.6$) when the flip-flop field on some particles differs from zero.
By $z=39.3$ shown in the second from the bottom panel of Fig. \ref{fig:substr_evol_1} the colorful particles collapsed 
in small haloes incorporated into a complicated filament formed by black particles. 

The  top panel of  Fig. \ref{fig:substr_evol_2} 
shows that at $z=33.4$ seven  colorful haloes merged into three larger haloes approximately corresponding to
 the bottom left  panel in Fig. \ref{fig:substr3d_L}. One can see four or even five haloes on the left hand side of this panel
 but only three of them contain colorful particles. 
 The flip-flop peaks corresponding  to two smaller haloes without colorful particles emerged from peaks lower than $n_{\rm ff}=300$
 on which the colorful particles have been selected. The rest panels of Fig. \ref{fig:substr_evol_2} illustrate the hierarchical 
 process of merging  of colorful particles.   
 The bottom panel of Fig. \ref{fig:substr_evol_2} shows that all colorful particles merged in two haloes. 
  It is worth noting that the colorful particles reside inside bigger haloes of black particles.
  
  Finally the top panel of 
Fig. \ref{fig:substr_evol_3}   displays the state when all colorful particles merged into a single halo at $z= 24.1$.
The remaining three panels show three states at $z=9.58,~ 3.82$  and finally at $z=0$. The cloud of colorful
particles is well mixed and remains intact  after $z\approx 10$.
 
In order to see the central part of the halo it has been  zoomed in and the colorful particles are plotted with
greater sizes than black particles. In addition the opacity of black particles has been reduced. 
Without this trickery the colorful particles were hardly being seen at all.

The outputs of the simulation are equally spaced in logarithm of the scale factor $a$.
Starting from the second from top panel in Fig. \ref{fig:substr_evol_1} until the top panel of Fig. \ref{fig:substr_evol_3}
every output is displayed.  Two middle panels of Fig. \ref{fig:substr_evol_3} show just two states in a long evolution
from $z=24.1$ shown at the top to $z=0$ shown at the bottom.
There are probably two remarkable features in this evolution. First, the orientation of both the central part of the halo
depicted by the colorful particles and fifty times more massive part of the halo shown by black particles substantially 
and synchronously have changed their orientation. This obviously happened because they are only 
small central parts of a much more massive halo experienced  considerable accretion of mass.
Second, the final state shown by colored particles looks like a dynamically relaxed configuration with
all seven subpeaks marked by different colors well mixed. Nevertheless they are easily identified 
as distinct peaks of the flip-flop field $n_{\rm ff}({\bf q}, z=0)$ in Fig. \ref{fig:substr3d_L}.

\section{Summary}   
We explored the properties of flip-flop field in the N-body simulations designed to simulate the formation of DM haloes 
at very early stage of the evolution of the universe. We considered various correlation properties of the flip-flop field
estimated at fifty epochs and compared them with that of density and gravitational potential fields computed on
the simulation particle by the GADGET code. We conclude that 
the flip-flop field $n_{\rm ff}({\bf q},z)$  carries a wealth of information 
about the substructures in the collisionless Cosmic Web.
In particular the peaks  of  the flip-flop field at the final stage $n_{\rm ff}({\bf q},z=0)$  store substantial
information on the merging process.  

The final flip-flop field looks like a set of bulky regions well isolated by narrow valleys,  see Fig. \ref{fig:1d_example},
 \ref{fig:two-d} and  \ref{fig:8z_slices} for one-, two- and three-dimensional examples respectively.
The peaks typically have a highly nested structure consisting of several smaller by sizes peaks which in turn 
may consist of even smaller peaks in Lagrangian space. 
The hierarchy can be very extensive and 
these properties are universal for collisionless self gravitating media in 1D,
2D and 3D. The topography of the flip-flop field is very different from a Gaussian field in all cases.

We selected a halo the central part of which has developed 
at least by $z \approx 10$ or even as early as $z\approx 20$ as Fig. \ref{fig:substr_evol_3} suggests. 
We used seven highest peaks in the flip-flop field $n_{\rm ff}({\bf q}, z=0)$ within the largest peak in the simulation 
with 256$^3$ particles for a study of the robustness of the flip-flop field as an indicator of substructure existed
in the halo. We demonstrated that the selected seven peaks were initially individual small haloes
(Fig. \ref{fig:substr_evol_1}) then they experienced multiple mergers (Fig. \ref{fig:substr_evol_2}) and
finally merged in a single approximately ellipsoidal cloud at the center of the halo (Fig. \ref{fig:substr_evol_3}). 
During the following evolution from at least $z=10$  or  from even earlier epoch the merged seven peaks 
have remained at the center all the time. 
Although the  orientation of the ellipsoid significantly changed but it remained intact until $z=0$.
It is remarkable  that despite a very vigorous mixing in Eulerian configuration space and the lack of significant
differences between mean and std  velocities of the components the flip-flop field retains a 'record' of the merging
tree in the form of easily identified isolated peaks, see Fig. \ref{fig:num_subs_maxs} and \ref{fig:substr3d_L}.

The topography of the 
flip-flop landscape  evolves rapidly after the onset of non-perturbative nonlinearity marked by the origin of
the first regions with $n_{\rm ff}({\bf q},z_{\rm nl}) > 0$. However relatively soon its  evolution is
considerably impeded (see Fig. \ref{fig:corcoef}, \ref{fig:ratios} almost to complete freeze  while  the heights of  peaks continue to grow. 
The latter indicates ongoing rapid dynamics inside the haloes and subhaloes themselves while the former suggests
a remarkable stability of the flip-flop landscape. 

We are suggesting the following explanation why  only the flip-flop field has these unique characteristics.
The period of time between sequential flip-flops of a fluid element is a characteristic time between its most significant
dynamical metamorphoses. The dynamical significance of each flip-flop may be attributed to passing of
the fluid element through the state of infinite density. 
Thus counting the flip-flops of a fluid element can be considered as counting ticks of its own dynamical clock.
We demonstrated that there is a sort of dynamical instability:  the greater the current counts of flip-flop the more likely 
the next one happens,  see Fig. \ref{fig:growth_ff}.  This results in the remarkable stability of the topography of the flip-flop field,
because the particles with lower counts of flip-flops have little chance to overtake  their neighbors 
with currently higher flip-flop counts. In other words  it is very unlikely that the valleys of the flip-flop field will become 
peaks, see Fig. \ref{fig:two-d}. 

The flip-flop field approximates a very complicated structure of dark matter haloes and subhaloes in six-dimensional phase
space as a much simpler set of  tree-lilke structures in only four dimensions made by three Lagrangian axes and
the number of flip-flops axis. Thus, the simplification is not only due to reduction of the number of dimensions
but more importantly by the fact that the flip-flop distribution in Lagrangian space is a single valued function
i.e. a field in three dimensions.

The discovered properties of the flip-flop field and easiness of its computing in cosmological N-body simulations 
make it a good  candidate for a valuable addition to the suite of various techniques suggested
for studies of substructures in the dark matter Cosmic Web.
Neither  density nor potential computed on particles by GADGET code possess such  properties.

\section*{Acknowledgements}
SSh acknowledges support  by the Templeton Foundation and  sabbatical support at Kapteyn Astronomical Institute at the University of Groningen The Netherlands and by Argonne  National Labs where the significant part of the work was done. SSh also thanks S. Habib for fruitful discussions. The authors are grateful to S.D.M. White for useful critical comments on the first draft of the paper.
MM acknowledges partial support by DOE and NSF via grants DE-FG02-07ER54940 and AST-1209665

\bibliographystyle{mn2e}
\bibliography{bibliography}

\end{document}